\documentclass{article}

\usepackage[preprint]{neurips_2024}

\usepackage[utf8]{inputenc} % allow utf-8 input
\usepackage[T1]{fontenc}    % use 8-bit T1 fonts
\usepackage{hyperref}       % hyperlinks
\usepackage{url}            % simple URL typesetting
\usepackage{booktabs}       % professional-quality tables
\usepackage{amsfonts}       % blackboard math symbols
\usepackage{nicefrac}       % compact symbols for 1/2, etc.
\usepackage{microtype}      % microtypography
\usepackage{xcolor}         % colors
\usepackage{graphicx}
\usepackage{array}
\usepackage{multirow}
\usepackage{footnote}
\usepackage{enumitem}
\usepackage{graphicx}
\usepackage{caption}
\usepackage{subcaption}
\makesavenoteenv{table}
\usepackage{hyperref}

\hypersetup{
    colorlinks=true,
    urlcolor=magenta     % For URLs
}

\title{Sprite Sheet Diffusion: Generate Game Character for Animation}

\author{%
  Cheng-An Hsieh* \\
  \texttt{chengan2@andrew.cmu.edu} 
  \And
  Jing Zhang*\\
  \texttt{jingzha4@andrew.cmu.edu} 
  \And
  Ava Yan*\\
  \texttt{lany2@andrew.cmu.edu} 
  \And
\url{https://chenganhsieh.github.io/spritesheet-diffusion/}
}

\begin{document}
\maketitle
\def\thefootnote{*}\footnotetext{These authors contributed equally to this work}
% \begin{abstract}
%   ...
% \end{abstract}

\section{Introduction}

Creating character animations is a cornerstone of 2D game development, requiring illustrators to design a main character image and then manually draw multiple keyframes depicting actions such as running or jumping. This process is both time-consuming and labor-intensive, as maintaining consistency in style, proportions, and details across frames demands meticulous effort (Figure~\ref{fig:side_by_side}).

Existing generative AI methods primarily focus on single image generation or video synthesis, overlooking the specific challenges of conditional image sequence generation for sprites. Additionally, this task involves conditioning on both a reference image for appearance and a pose sequence for intended actions, significantly increasing its complexity. Furthermore, while related methods have shown success in human-centered domains, they rarely address the sprite domain, where limited high-quality training data further complicates the problem. In this work, we make an attempt to address these challenges and our contributions are summarized as follows:

\begin{enumerate}
    \item \textbf{Propose a New Task}: We define the sprite sheet generation task, introducing evaluation methods to benchmark models on consistency, pose alignment, and sequence quality.

    \item \textbf{Curate a Specialized Dataset}: We create a high-quality sprite-domain dataset with \textbf{150+} paired reference images, pose sequences, and target action sequences.

    \item \textbf{Explore Two Baselines and Propose a Method}: We explore using two existing methods in conditional image and video generation domain respectively and creatively adapt one of them, Animate Anyone \citep{hu2024animate}.

    \item \textbf{Obtain a Capable Model}: We successfully train a model that generates faithful, consistent, and high-quality sprite action sequences and conduct extensive experiments to analyze key factors affecting its performance.
\end{enumerate}

This work demonstrates how generative diffusion models can streamline 2D game development by reducing manual effort and enabling efficient sprite animation. Beyond games, the framework extends to virtual avatars, storytelling, and education, offering tools and insights for advancing creative industries and interactive media.

\section{Dataset and Task}
\label{dataset_and_task}
The proposed task involves generating an \textbf{action sequence} of a game character conditioned on an initial \textbf{game character reference image} (Figure~\ref{fig:ref_image}), a specified \textbf{pose sequence} (Figure~\ref{fig:pose_sequence}), and the resulting \textbf{action sequence} (Figure~\ref{fig:action_sequence}). The reference image defines the character's appearance, while the pose sequence outlines the intended actions. The goal is to create a sequence of images that depict the character performing these actions, maintaining consistency with the reference image. Formally, let $C$ denote the reference image of the game character, $P = \{p_1, p_2, \dots, p_n\}$ denote the pose sequence, where each $p_i$ represents a specific pose in the sequence, and $\hat{I} = \{\hat{i}_1, \hat{i}_2, \dots, \hat{i}_n\}$ denote the generated image sequence, where $\hat{i}_i$ corresponds to the character in pose $p_i$. The model learns a mapping $f: (C, P) \to \hat{I}$ such that each $\hat{i}_i = f(C, p_i)$.

We curate a sprite dataset by collecting sprite sheets from two sources: (1) GameArt2D\footnote{GameArt2D: \url{https://www.gameart2d.com/.}}, which provides high-quality and uniform sprites with limited diversity, and (2) SpriteDatabase\footnote{SpriteDatabase: \url{https://spritedatabase.net/.}}, which offers diverse styles and poses but exhibits inconsistent quality. From each sprite, one frame was selected as the reference image, while the poses for the remaining frames were annotated using pose detection models or manual labeling. Each sprite typically includes 3 to 8 frames per action sequence.  See Figure~\ref{fig:data_example} for an example. We split the dataset into training, validation, and two test sets (in-sample and out-sample) at the action sequence level. In-sample test set includes unseen action sequences of characters present in the training set, while out-sample test set contains motion sequences from entirely unseen characters. In total, the dataset comprises 152 paired action sequences and 916 paired frame sequences. See Table~\ref{tab:dataset_summary} for dataset statistics.

  % Since this is a novel problem, no existing datasets fully meet the requirements, but we found publicly available sprite sheet datasets, and applied pose detection algorithms or manual pose annotations to generate corresponding pose sequences. We curated the dataset from two sources: GameArt2D\footnote{\url{https://www.gameart2d.com/freebies.html}} and SpriteDatabase\footnote{\url{https://spritedatabase.net/system/pc}}. The GameArt2D dataset contains high-quality, pre-cropped, and uniformly formatted sprite sequences. From this source, we collected xxx (reference image, pose image, target image) pairs across xxx actions from 16 characters. These pairs were split into training and testing sets at the character level, with 11 characters used for training and 5 for testing. While the GameArt2D sprites are of high quality, they exhibit limited diversity in terms of style and pose. To address this limitation, we augmented the dataset with 30 (reference image, pose image, target image) pairs extracted from SpriteDatabase.

We evaluate the generated results with both qualitative and quantitative assessments. For qualitative analysis, we manually evaluate four examples to verify alignment with conditioned images and detail consistency across frames. For quantitative analysis, we evaluate the generated motion frames based on: (1) similarity to ground truth action sequence; and (2) subject consistency within a generated sequence. For (1), we use Structural Similarity Index Measure (SSIM) \citep{1284395} to assess structural similarity in luminance and contrast, Peak Signal-to-Noise Ratio (PSNR) \citep{5596999} to evaluate the pixel-wise difference, and Learned Perceptual Image Patch Similarity (LPIPS) \citep{zhang2018unreasonable} to measure perceptual differences aligned with human judgment. For (2), we utilize the subject consistency score proposed by \citet{huang2024vbench}, which is calculated based on the DINO feature similarity \citep{ruiz2023dreambooth}.

\section{Related Work}

\subsection{Image-to-Pose: Foundation for Pose Conditioning}

Image-to-pose techniques are fundamental for sprite generation, providing the necessary pose information from reference images and enabling pose constraints during synthesis. 2D human pose estimation (HPE) is a well-studied task focusing on predicting keypoint coordinates from images \citep{zheng2023survey}. Regression-based methods, such as DeepPose \citep{toshev2014deeppose}, directly predict joint positions, while heatmap-based approaches like DW-Pose \citep{dwpose2021} improve spatial precision through probability maps. Multi-person estimators, such as OpenPose \citep{cao2017openpose}, handle more complex scenarios. However, existing pose estimation models struggle with game sprites, as game sprites often feature exaggerated proportions, occluding costumes, and non-standard poses. The transfer learning approach by \citet{chen2022bizarre} adapts human pose models for illustrations but is limited in generalizability. Given these challenges, we manually annotate poses for difficult cases to ensure data reliability.

\subsection{Pose-to-Image: Conditional Image Generation}

Our task can be treated as a conditional image generation problem, where each frame is conditioned on both a reference image for appearance and a specific pose for structure. Diffusion-based models, such as Latent Diffusion \citep{rombach2022high}, are widely used in image generation tasks \citep{dhariwal2021diffusion}. For pose conditioning, models like ControlNet \citep{zhang2023adding} enable spatial guidance by integrating pose maps, while methods like IP-Adapter \citep{ye2023ip} ensure style and content consistency with reference images. These techniques excel at generating single images aligned with pose or style constraints. However, our task adds the complexity of producing multiple frames that maintain coherence in appearance and motion. While single-frame conditional models offer effective solutions for isolated generations, they lack mechanisms to ensure consistency across sequential outputs. 

\subsection{Pose-to-Video: Temporal Consistency in Animation}

Pose-to-video synthesis extends pose-to-image tasks by introducing temporal coherence, generating realistic video sequences from pose inputs. This task is central to applications like animation, virtual reality, and game development, where smooth and consistent motion is essential. \citet{hu2024animate} proposed Animate Anyone, a diffusion-based framework that leverages ReferenceNet for appearance conditioning, a Pose Guider for structural alignment, and temporal modeling to produce smooth transitions between frames. Similarly, \citet{wei2024aniportrait} developed AniPortrait, which integrates audio and reference portraits to generate temporally consistent facial animations. 
Although these approaches emphasize sequence-level temporal smoothness, our task differs as it involves discrete frames. Moreover, their results often exhibit noticeable artifacts when applied to anime that significantly differs from human figures.

\section{Methods}

\subsection{Baseline Approaches}

\textbf{Stable Diffusion with ControlNet and IP-Adapter Integration (SD-IPCN)}:
First, we employ a method from the domain of conditional image generation, leveraging pose and character images as prompts to guide the process. We follow IP-Adapter \citep{ye2023ip}'s method to integrate the appearance adaptor with SD-v1.5 \citep{rombach2022high} and ControlNet \citep{zhang2023adding}. For testing with IP-Adapter's given weights, we generate all frames of an action sequence as a single concatenated image to improve frame-to-frame consistency of this baseline. For fine-tuning, we adapt IP-Adapter's given training code to fine-tune with ControlNet.

\textbf{Animate Anyone}:
Second, we refer to Animate Anyone \citep{hu2024animate}, which differs from IP-Adapter \citep{ye2023ip} in two key aspects. First, Animate Anyone focuses on conditional video generation, introducing an additional layer of consistency to image generation. Our approach adapts Animate Anyone's framework to address sequential image generation by treating it as a video generation problem. Second, Animate Anyone employs a more complex image encoder compared to IP-Adapter, which only leverages the semantics of the reference image. Building on Animate Anyone \citep{hu2024animate}, we further extend it to develop our main method.

\subsection{Main Method}
\begin{figure}[h]
\centering
\includegraphics[width=0.7\textwidth]{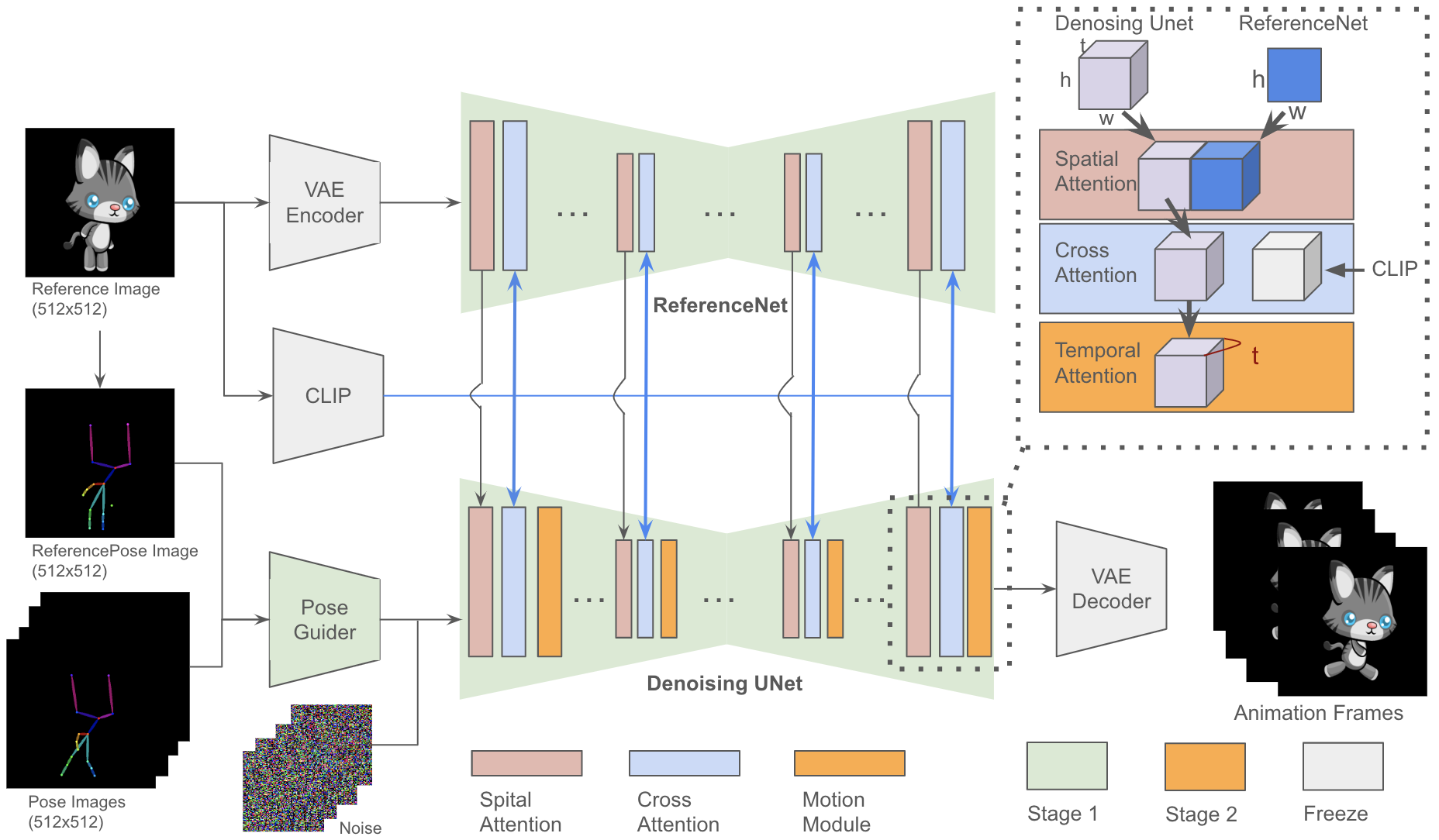}
\caption{Framework}
\label{fig:framework}
\end{figure}

We adapt the framework from \citet{wei2024aniportrait} for portrait animation, enhancing Animate Anyone with a Pose Guider that adds pose features to the noise latent during downsampling. The pipeline of our method is illustrated in Figure~\ref{fig:framework} and comprises three key components: ReferenceNet, Pose Guider, and Motion Module. ReferenceNet encodes appearance features using an SD model with spatial-attention layers, while IP-Adapter \citep{ye2023ip} uses a CLIP image encoder with limited resolution (224x224). Cross-attention, driven by the CLIP image encoder, enhances feature integration between ReferenceNet and the denoising network. The Pose Guider encodes motion information using four convolutional layers to align the pose image to the same resolution as the noise latent, which is then added to the noisy latent before being input into the denoising network. To ensure smooth frame transitions, the Motion Module is embedded within the Res-Trans block after attention layers.

The training process consists of two stages aimed at optimizing both pose conditioning and temporal consistency. In Stage 1 (Pose-to-Image), ReferenceNet, the denoising network, and the Pose Guider are trained to generate individual character images based on input poses. ReferenceNet encodes the reference image, while the Pose Guider encodes pose information, allowing the denoising network to integrate these features and produce accurate character representations. In Stage 2 (Pose-to-Sprite), the Motion Module is trained for temporal consistency while the weights of ReferenceNet, the denoising network, and the Pose Guider are frozen from Stage 1. This stage focuses on achieving smooth and coherent transitions between animation frames.

%%%%%%%%%%%%%%Experiments
\section{Experiments}
% To demonstrate the superiority of our fine-tuned Animate Anyone model, we evaluate its performance on crafted in-sample and out-sample test datasets, showcasing its ability to handle seen and unseen characters. For comparison, we use three baselines: Animate Anyone with vanilla weights, SD-IPCN, and a fine-tuned SD-IPCN trained on our dataset. Results are analyzed using metrics from Section \ref{dataset_and_task}, focusing on image quality and subject consistency.
To showcase the superiority of our fine-tuned Animate Anyone model, we assess its performance on the in-sample and out-sample test dataset we crafted, aiming to demonstrate its ability to handle both seen and unseen characters. We also establish three baseline methods for comparison: Animate Anyone with its vanilla weights, SD-IPCN, and a fine-tuned version of SD-IPCN trained on our training set. Results are analyzed using the metrics described in Section \ref{dataset_and_task}, focusing on the quality of generated images as well as their subject consistency.

\subsection{Implementation Details}
We trained IP-Adapter model for 40 epochs (approximately 5,000 steps) on an NVIDIA A10G GPU (24GB), taking around 2 hours. The training loss gradually saturates after 10 epochs, and we report results at epoch 20 to avoid overfitting.

Fine-tuned Animate Anyone model has two training stages. In the first stage, we train the PoseGuider, ReferenceNet, and DenoisingNet, which require over 30GB of GPU memory. Therefore, we use an NVIDIA L40S (42GB) GPU for 30,000 steps (10 hours). In the second stage, only the temporal layers are trained, allowing us to use a single NVIDIA 4090 GPU (24GB) for 20,000 steps (2 hours).

% Up to the Midway Report deadline, we have conducted experiments on the dataset we created so far. For the Animate Anyone fine-tuned model, we ran stage 1 training on an NVIDIA L40S for 10 hours and stage 2 training on an NVIDIA 4090 GPU for 2 hours. For inference, both with the pretrained Animate Anyone model and the fine-tuned version, we used the NVIDIA 4090. For running the SD-IPCN pipeline, we used an NVIDIA A10G.

\subsection{Qualitative Comparison}
% TODO: Jing - could be further simplified, overlaps a bit with 5.3 :)

% We randomly sampled four examples from our test set: two from the in-sample and two from the out-sample, to manually evaluate the quality of the generated motion frames, as shown in Appendix \ref{app:qualitative}. Our fine-tuned Animate Anyone model successfully recreates the appearance of the reference image and follows the corresponding pose image to generate appropriate motion frames, while also generalizing to unseen characters. However, it struggles with finer details such as hairstyles (see Figure \ref{fig: QA1}) and props (see Figures \ref{fig: QA2} and \ref{fig: QA3}). In comparison, the Animate Anyone model with vanilla weights performs significantly worse, producing garbled results that fail to capture the reference image or pose accurately. For the SD-IPCN, the vanilla weights version achieves acceptable results, with fine-tuning offering only minimal improvement. Nonetheless, both models fail to accurately replicate the reference image and instead generate outputs that are merely stylistically similar to the reference.

We randomly sampled four test set examples: two in-sample and two out-sample, to manually evaluate the generated motion frames (Appendix \ref{app:qualitative}). Our fine-tuned Animate Anyone model successfully recreates the reference image's appearance and follows the corresponding pose image to generate appropriate motion frames, while also generalizing to unseen characters. However, it is still underperforming with finer details like hairstyles (Figure \ref{fig: QA1}) and props (Figures \ref{fig: QA2}, \ref{fig: QA3}). The vanilla-weight version performs worse, producing garbled outputs failing to match reference or pose. For SD-IPCN, the vanilla model generates acceptable results, with fine-tuning offering limited improvement. Nonetheless, both models fail to accurately replicate the reference image, producing only stylistically similar outputs.

\subsection{Quantitative Comparison}

Tables \ref{tab:result:in-sample} and \ref{tab:result:out-sample} present the quantitative comparison between our fine-tuned Animate Anyone model and baseline approaches. Our fine-tuned model demonstrates remarkable improvements in SSIM, PSNR, and LPIPS compared to other baseline methods, which aligns with our quantitative analysis and highlights its effectiveness in maintaining fidelity to reference images. Notably, it achieved significant improvements over the version with vanilla weights. This is because the original Animate Anyone model was trained to generate realistic animated videos, where the training dataset consisted of characters with clear, human-like pose structures. Sprite images, however, lack such clear pose structures, making it challenging for the original model to generate accurate sequences. Through the fine-tuning of Animate Anyone, our tailored model effectively learns the relationship between these specialized sprite images and their corresponding pose images, enabling it to produce results that are highly aligned with the reference images and follow the given pose image structure.

The SD-IPCN model with vanilla weights achieves a similar Subject Consistency score as fine-tuned Animate Anyone. After fine-tuning, SD-IPCN surpasses fine-tuned Animate Anyone in subject consistency and shows an improved SSIM score over its vanilla-weight counterpart, indicating enhanced subject identity preservation. However, its SSIM, PSNR, and LPIPS scores remain lower, suggesting that it struggles to align with the detailed appearance of the reference images. The primary difference lies in how these methods approach image condition modeling. SD-IPCN relies solely on CLIP embeddings, which can preserve some image similarity but fail to effectively transfer fine-grained details. In contrast, the ReferenceNet in our fine-tuned Animate Anyone model leverages a UNet structure to capture the spatial details of the reference images, enabling superior preservation of appearance details.

\begin{table}[!ht]
\centering
\resizebox{0.9\textwidth}{!}{ % Resize to fit within the text width
\begin{tabular}{lcccc}
\toprule
& SSIM$\uparrow$ & PSNR$\uparrow$ & LPIPS$\downarrow$ & Subject Consistency$\uparrow$ \\ 
\midrule
SD-IPCN & 0.294 ± 0.074 & 10.752 ± 2.187 & 0.412 ± 0.127 & 0.885 ± 0.044 \\
SD-IPCN (finetuned) & 0.421 ± 0.157 & 10.060 ± 2.984 & 0.401 ± 0.108 & \textbf{0.910 ± 0.039} \\
Animate Anyone & 0.330 ± 0.143 & 9.786 ± 2.159 & 0.557 ± 0.184 & - \footnotemark \\
Animate Anyone (finetuned) & \textbf{0.659 ± 0.250} & \textbf{18.405 ± 5.280} & \textbf{0.125 ± 0.088} & 0.901 ± 0.064 \\
\bottomrule
\end{tabular}
}
\footnotetext{Due to Animate Anyone's limited pose adaptation capability without fine-tuning, the generated frames show minimal variation. Therefore, subject consistency evaluation was not conducted for this baseline method.}
\caption{Quantitative Comparison (In-Sample).}
\label{tab:result:in-sample}
\end{table}
\begin{table}[!ht]
\centering
\resizebox{0.9\textwidth}{!}{ % Resize to fit within the text width
\begin{tabular}{lcccc}
\toprule
& SSIM$ \uparrow$ & PSNR$ \uparrow$ & LPIPS$ \downarrow$ & Subject Consistency$ \uparrow$ \\ 
\midrule
SD-IPCN & 0.308 ± 0.098 & 10.919 ± 2.316 & 0.386 ± 0.079 & 0.892 ± 0.038 \\
SD-IPCN (finetuned) & 0.385 ± 0.124 & 10.298 ± 3.045 & 0.393 ± 0.111 & \textbf{0.932 ± 0.025} \\
Animate Anyone & 0.236 ± 0.055 & 8.957 ± 1.908 & 0.694 ± 0.083 & -\footnotemark  \\
Animate Anyone (finetuned) & \textbf{0.655 ± 0.195} & \textbf{18.809 ± 4.806} & \textbf{0.139 ± 0.090} & 0.893 ± 0.038 \\
\bottomrule
\end{tabular}
}
\caption{Quantitative Comparison (Out-Sample).}
\label{tab:result:out-sample}
\end{table}

% The SD-IPCN model with vanilla weights achieves a similar Subject Consistency score compared to the fine-tuned Animate Anyone. After fine-tuning, the SD-IPCN pipeline can even surpass the fine-tuned Animate Anyone in subject consistency and shows an improved SSIM score over its vanilla-weight counterpart, indicating its enhanced ability to maintain subject identity. However, its SSIM, PSNR, and LPIPS scores remain significantly lower than the fine-tuned Animate Anyone model, suggesting that it struggles to align with the detailed appearance of the reference images. The primary difference lies in how these methods approach image condition modeling. SD-IPCN relies solely on CLIP embeddings, which can preserve some image similarity but fail to effectively transfer fine-grained details. In contrast, the ReferenceNet in our fine-tuned Animate Anyone model leverages a UNet structure to capture the spatial details of the reference images, enabling superior preservation of appearance details.

\subsection{Ablation Study}
\label{ablation_study}
% In order to verify which component being fine-tuned plays a significant role in adapting the Animate Anyone model to the sprite sheet generation task, we conduct an ablation study by fine-tuning specific components in the framework while freezing others during training. The experiments include: (1) training only the pose guider in stage 1 while freezing all other components (Pose Guider Only), (2) training only the pose guider and the denoising Unet in stage 1 while freezing all other components (Pose Guider + Denoising Unet), (3) training only the stage 1 components, including the pose guider, reference net, and denoising net (Stage 1 Only), (4) training both stages 1 and 2, which is the same as the fine-tuned Animate Anyone mentioned earlier (Fully Fine-Tuned).

To identify which fine-tuned components are crucial for adapting the Animate Anyone model to sprite sheet generation, we conducted an ablation study by selectively fine-tuning components while freezing others. The experiments include: (1) fine-tuning only the Pose Guider (Pose Guider Only), (2) fine-tuning the Pose Guider and denoising UNet (Pose Guider + Denoising UNet), (3) fine-tuning all Stage 1 components (Stage 1 Only), and (4) fine-tuning both Stage 1 and Stage 2 components, which is the same as the fine-tuned Animate Anyone mentioned earlier (Fully Fine-Tuned).

Tables \ref{tab:ablation:in-sample} and \ref{tab:ablation:out-sample} in the Appendix \ref{ablation:quantitative} show the quantitative comparison of the performance across different configurations. We also randomly sampled four examples for manual evaluation (see Appendix \ref{ablation:quantitative}). Results show that fine-tuning only the Pose Guider allows the model to reconstruct most appearance details, while incorporating additional components during Stage 1 training, such as the reference network and denoising UNet, consistently improves the model's performance across all evaluation metrics. However, the fully fine-tuned model underperforms "Stage 1 Only" on some metrics, revealing the need to adjust Stage 2 training configuration. While Stage 2 is critical for ensuring the generated character’s poses align closely with the given pose images, overfitting at this stage can potentially lead to the loss of important details. For example, as shown in Figure \ref{fig-abl: QA3}, the character's gun disappears in the output generated by the fully fine-tuned model.

% Tables \ref{tab:ablation:in-sample} and \ref{tab:ablation:out-sample} in the Appendix \ref{ablation:quantitative} show the quantitative comparison of the performance across different configurations. We also randomly sampled four examples for manual evaluation (see Appendix \ref{ablation:quantitative}). From the results, it can be observed that by fine-tuning only the pose guider, the model is able to reconstruct most of the appearance details from the reference image. By incorporating additional components during Stage 1 training, such as the reference network and denoising UNet, model's performance is able to consistently improve across all evaluation metrics. Notably, the fully fine-tuned model (Stages 1 and 2) fails to outperform the "Stage 1 Only" configuration on certain metrics, which reveals the need for careful adjustment of the Stage 2 training step. While Stage 2 is critical for ensuring the generated character’s poses align closely with the given pose images, overfitting at this stage can potentially lead to the loss of important details. For example, as shown in Figure \ref{fig-abl: QA3}, the character's gun disappears in the output generated by the fully fine-tuned model.

% \footnote{Due to limited computational resources, we were not able to run experiments to determine the optimal training step for Stage 2.}

%%%%%%%%%%
\section{Code Overview}
% \subsection{Dataset Creation}
\textbf{Dataset Creation}: Our data pipeline consists of the following steps: (1) for uncropped sprite action sequences, we split the sprite sheet into separate rows, adjust the height, define the width of each frame, save cropped frames (2) label each frame manually or use OpenPose/DW-Pose. The detailed code is available on Appendix \ref{sec:pipeline}.

\textbf{IP-Adapter}: For testing with the IP-Adapter \citep{ye2023ip} repo's provided weights, we adapt dataset loading part of IP-Adapter repo's provided \texttt{ip\_adapter\_controlnet\_demo\_new.ipynb}. For fine-tuning the IP-Adapter, we reference the \texttt{train\_tutorial.py} script from the IP-Adapter repo that originally only trains on text-image pairs and adapt it to train with ControlNet on our reference image, pose image and target action image triplet pairs.
Appendix \ref{sec:ip-adapter}.

% \subsection{Animate Anyone}
\textbf{Animate Anyone}: We adapt Animate Anyone \citep{hu2024animate} and fine-tune the models for our specific domains. Most of the code is based on \href{https://github.com/MooreThreads/Moore-AnimateAnyone}{Moore-Animate Anyone}, but we made several modifications: (1) rewriting the script for loading our sprite sheet dataset (Figure \ref{fig:dataset}), (2) adding data augmentation by flipping images (Figure \ref{fig:random-flip}), (3) integrating Wandb to track the training process (Figure \ref{fig:wandb}), and (4) rewriting the inference code to generate frames without requiring video format output (Figure \ref{fig:output}). Appendix \ref{sec:animateanyone}.

% \subsection{Experiment Scripts}
\textbf{Experiment Scripts}:
To streamline evaluation and analysis, we implemented two scripts for automated assessment. Each script could process inputs and generate a Markdown file with overall and per-character/motion statistics. The first script (Appendix \ref{code:eval_quality}) evaluates similarity between ground truth and generated frames using SSIM, PSNR, and LPIPS. The second (Appendix \ref{code:eval_consistency}) assesses subject consistency using the Subject Consistency Score from the vbench library \citep{huang2024vbench}.
% To accelerate the evaluation process and facilitate subsequent analysis, we implemented two evaluation scripts to automatically assess our results. Each script takes the required inputs, automatically evaluates the results, and generates a Markdown file documenting the outcomes. These outcomes include both overall statistics and detailed statistics for each character and motion.

% The first script evaluates the similarity between the ground truth and the generated motion frames (see \texttt{experiment/eval\_img\_quality.py}). It incorporates metrics such as SSIM, PSNR, and LPIPS, which were implemented using existing libraries. The second script assesses subject consistency within the generated sequence (see \texttt{eval\_sub\_consistency.py}) by utilizing the Subject Consistency Score from the vbench library \citep{huang2024vbench}.

%%%%%%%%%%
% \section{Timeline}
% ToDO: Ava

%%%%%%%
\section{Research Log and Timeline (Detailed Timeline: Table \ref{tab:timeline})}

\subsection{Task Definition and Dataset Creation, 50 hrs}

We grouped early in the semester to brainstorm our topic of interest. After deciding to work on game character sequence generation, our first challenge was finding a suitable dataset. Websites like SpriteDatabase contained many sprite sheets, but none provided paired action and pose sequences. We wrote labeling pipeline scripts to manually crop action sequences from sprite sheets and annotate poses using pose detection models such as DW-Pose and OpenPose. While these models worked well for humanoid sprites, they performed poorly on characters with exaggerated proportions or non-standard anatomy, conflicting with our goal of building a diverse dataset. To address this, we manually annotated challenging cases, balancing automation with manual correction to ensure quality. (32 hrs)

Later, we discovered GameArt2D, a website that provides pre-cropped action sequences. Although these lacked pose diversity, they were consistent, high-resolution and easier to annotate. For these sprites, since they are all disproportional with human figure and pose detection models do not perform well on them, we wrote a manual annotation script inspired by OpenPose’s outputs. We then supplemented our dataset with several more non-human-like sprites from SpriteDatabase to ensure the dataset’s diversity and representativeness. (18 hrs)

\subsection{Literature Review and Refining Focus, ~20 hrs}

Parallel to dataset creation, we also explored the broader literature on diffusion-based image and video generation. After recognizing the significance of pose conditioning, we narrowed our focus to works on conditional pose generation. Initially, we tested AniPortrait, a facial video generation framework, on our dataset. However, without pretrained weights, training required substantial time and was prone to overfitting. We later discovered that AniPortrait’s codebase was derived from Animate Anyone, which targets character video generation and aligns more closely with our task. This realization prompted us to pivot to Animate Anyone, leveraging its pretrained weights for sprite sequence generation.

Our exploration also revealed layered relationships between image and video generation research. Image generation methods like IP-Adapter focus on single-frame fidelity, while video generation emphasizes temporal coherence. Our task lies somewhere in the middle and can take existing methods from those two lines of work as baselines.

\subsection{Method Implementation, 45 hrs}

We selected IP-Adapter for image generation and Animate Anyone for video generation as our baseline approaches. Each re-implementation presented some unique challenges.

\href{https://github.com/tencent-ailab/IP-Adapter/blob/main/ip_adapter_controlnet_demo_new.ipynb}{IP-Adapter} provided a notebook for testing, where we explored two strategies: generating all frames in an action sequence simultaneously or treating each frame independently. The single-pass approach showed slightly better consistency and was used for reporting. However, its \href{https://github.com/tencent-ailab/IP-Adapter/blob/main/tutorial_train.py}{training logic}, originally designed for style adaptation with text-image pairs, was incompatible with our pose-conditioned task. And repository defines two different classes for the core component \texttt{ip-adaptor}, leading to ambiguity in understanding.  We extensively modified the training script to align with our needs and tested both fine-tuning pre-trained weights and training from scratch. Fine-tuning performed significantly better, likely due to the limited size of our dataset, and was used in the final results. (15 hrs)

For Animate Anyone, the official authors of the paper did not release the source code. Initially, we adapted \href{https://github.com/Zejun-Yang/AniPortrait/tree/main}{AniPortrait}, but later discovered that its codebase originated from \href{https://github.com/MooreThreads/Moore-AnimateAnyone}{Moore-Animate Anyone}, an open-source work that tries to re-implement  Animate Anyone. Upon comparing the two, we found the only difference was in the Pose Guider structure. AniPortrait enhances the Pose Guider by adding more convolution layers, incorporating the reference pose image, and integrating the pose image features into the denoising downsampling process. After evaluating the results, we found that the improved Pose Guider generated images closer to the reference image. We made minimal changes to their code, only adding functions for loading the dataset, data augmentation, and inference. (30 hrs)

\subsection{Evaluation and Experimental Design, 24 hrs}
Defining evaluation metrics was an iterative process due to the novelty of our task, which lacked prior research for reference. We initially used common image generation metrics (SSIM, PSNR, and LPIPS) to evaluate the similarity between ground truth and generated frames. However, these metrics focused solely on frame-to-frame similarity and proved insufficient for assessing subject consistency across frames. Given the limited research in this area, we conducted a thorough literature review and incorporated the Subject Consistency Score proposed by VBench \citep{huang2024vbench} into our evaluation framework. (6 hrs)

For the experimental design, we first created an in-sample and out-sample test set to evaluate the method's performance from two perspectives: seen characters with unseen motion and unseen characters. Our initial results showed significant improvements in the fine-tuned Animate Anyone compared to its vanilla weights, which motivated us to conduct an ablation study (Section \ref{ablation_study}) to analyze the impact of each component and understand the key factors driving the performance gains. The ablation study revealed that the Pose Guider was critical in achieving these improvements, but it also exposed potential overfitting issues in our Stage 2 training. Unfortunately, due to computational constraints, we could not run additional experiments to optimize the Stage 2 training settings. Nevertheless, this highlights a promising direction for future work to improve subject consistency. (18 hrs)

%%%%%%%%%%%%
\section{Conclusion}
In this work, we address a critical yet overlooked task in game development: sprite sheet generation. We creatively adapted the video generation framework Animate Anyone for this task and conducted comprehensive experiments alongside baselines to validate the effectiveness of our fine-tuned model. An ablation study further evaluated the contribution of each component during training. To tackle the lack of datasets in this field, we created a high-quality sprite-domain dataset with 150+ paired reference images, pose sequences, and target action sequences, laying a foundation for future research.

Both quantitative and qualitative results show that our fine-tuned model could effectively generate motion frames aligned with the reference image and the pose structure, which addresses the original Animate Anyone's inability to connect the sprite image with the pose image and overcomes the IP-Adapter-based methods' limitations in transferring image details. However, our model still struggles with overfitting during Stage 2 training and maintaining subject consistency and finer details. Future work will focus on refining the framework and optimizing fine-tuning configuration.

\newpage
\bibliographystyle{plainnat}
\bibliography{reference}

\newpage

\appendix
\section{Visual Illustration}

\begin{figure}[h!]
    \centering
    \begin{minipage}[t]{0.49\textwidth}
        \centering
        \includegraphics[height=3.5cm]{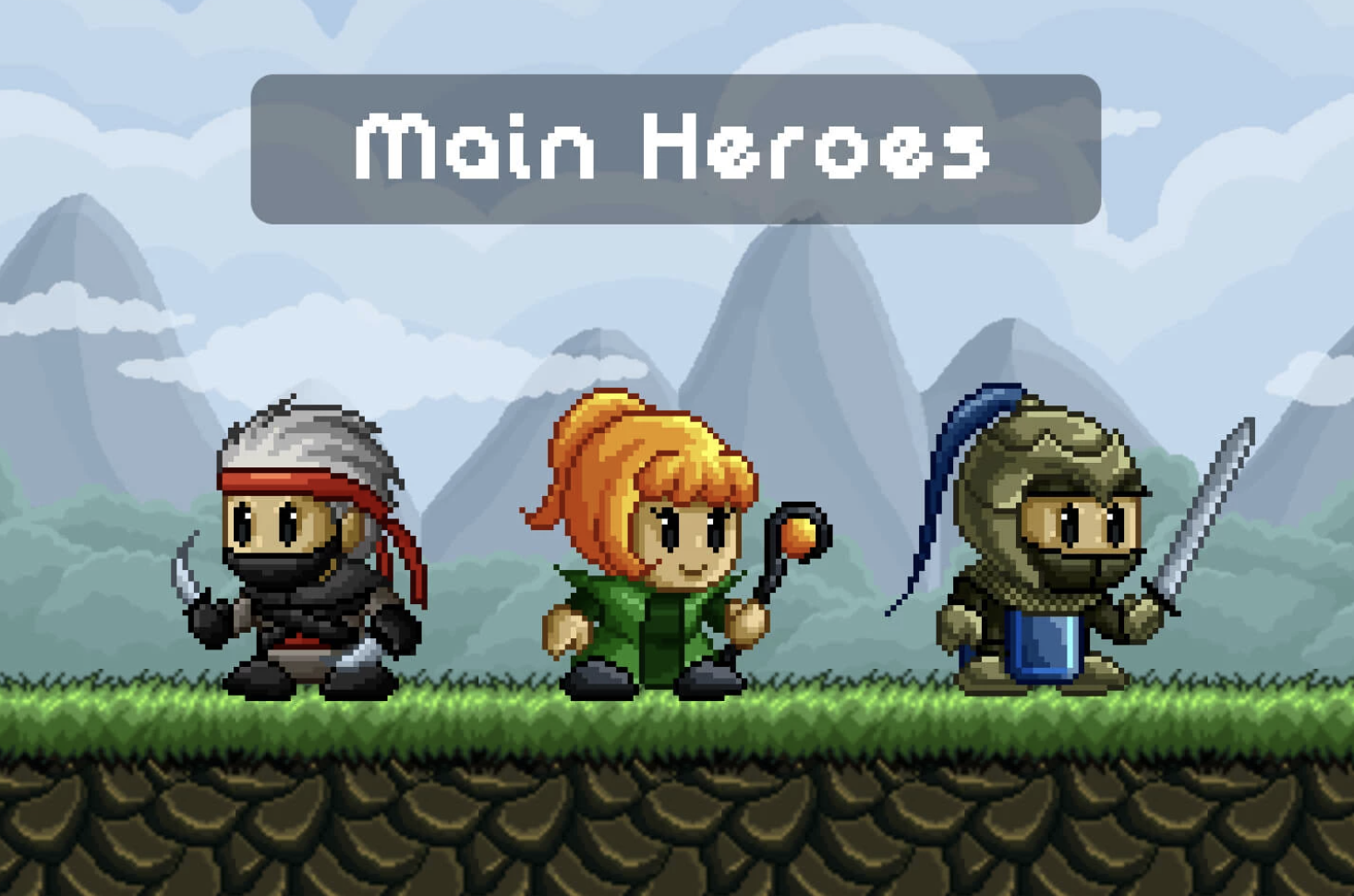}
        % \caption{Caption for the first image.}
        % \label{fig:img1}
    \end{minipage}%
    \hfill
    \begin{minipage}[t]{0.49\textwidth}
        \centering
        \includegraphics[height=3.5cm]{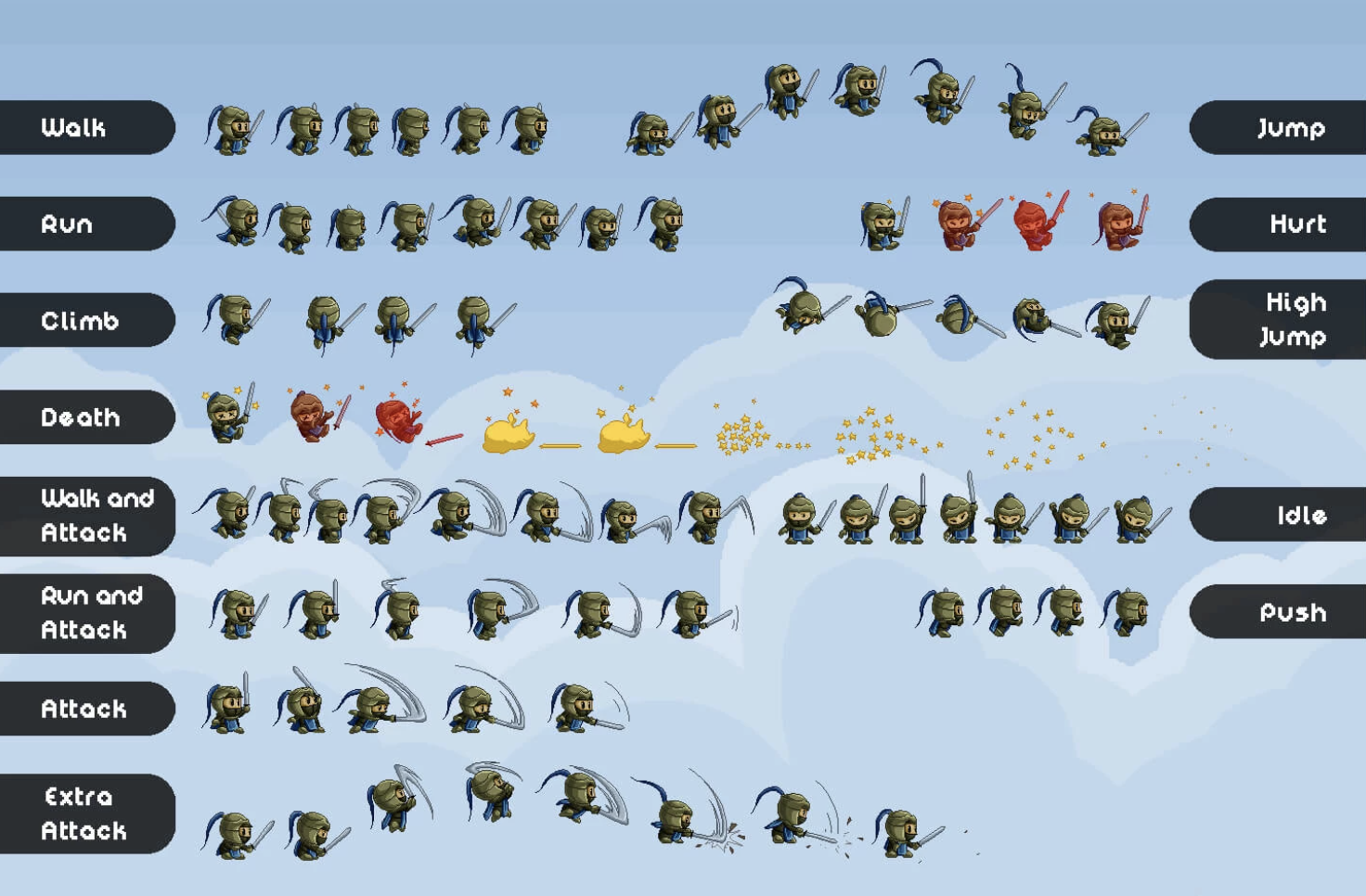}
        % \caption{Caption for the second image.}
        % \label{fig:img2}
    \end{minipage}
    \caption{A sprite in game development is a 2D bitmap graphic representing a character, object, or visual element. A sprite sheet is a single image file containing multiple sprites. The left image shows three game character sprites, while the right displays the sprite sheet of one character's action sequences. Image source: \url{https://craftpix.net/product/pixel-art-characters-for-platformer-games/.}}
    \label{fig:side_by_side}
\end{figure}

% \begin{figure}[h!]
%     \centering
%     % Left image (reference image)
%     \begin{minipage}[c]{0.35\textwidth}
%         \centering
%         \includegraphics[width=\textwidth]{ref_img.png}
%         \caption{Reference Image}
%         \label{fig:ref_img}
%     \end{minipage}%
%     \hfill
%     % Right images (pose and action sequences vertically centered)
%     \begin{minipage}[c]{0.65\textwidth}
%         \centering
%         \begin{subfigure}{\textwidth}
%             \centering
%             \includegraphics[width=0.9\textwidth]{pose_seq.png}
%             \caption{Pose Sequence}
%             \label{fig:pose_seq}
%         \end{subfigure}
%         \vspace{1em} % Adjust this space as needed
%         \begin{subfigure}{\textwidth}
%             \centering
%             \includegraphics[width=0.9\textwidth]{action_seq.png}
%             \caption{Action Sequence}
%             \label{fig:action_seq}
%         \end{subfigure}
%     \end{minipage}
%     % \caption{Reference image with pose and action sequences aligned.}
%     \label{fig:full_layout}
% \end{figure}

\newpage
\section{Dataset}
\begin{figure}[h!]
    \centering
    % First column: Reference image
    \begin{subfigure}[b]{0.3\textwidth}
        \centering
        \vspace{1.5cm}
        \includegraphics[width=\textwidth]{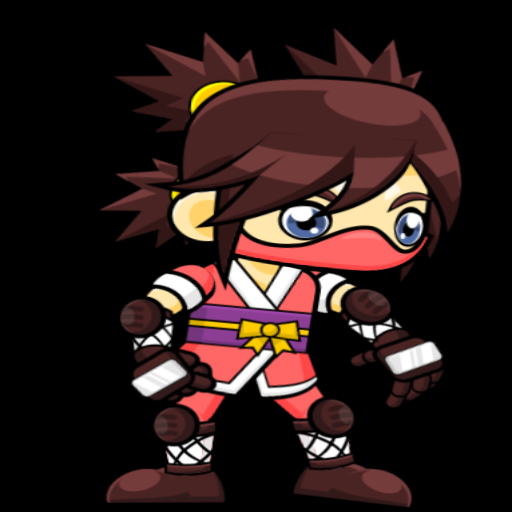}
        \vspace{3.6cm}
        \caption{Reference Image}
        \label{fig:ref_image} % Label for this subfigure
    \end{subfigure}
    \hfill
    % Second column: Pose sequence
    \begin{subfigure}[b]{0.3\textwidth}
        \centering
        \includegraphics[width=0.7\textwidth]{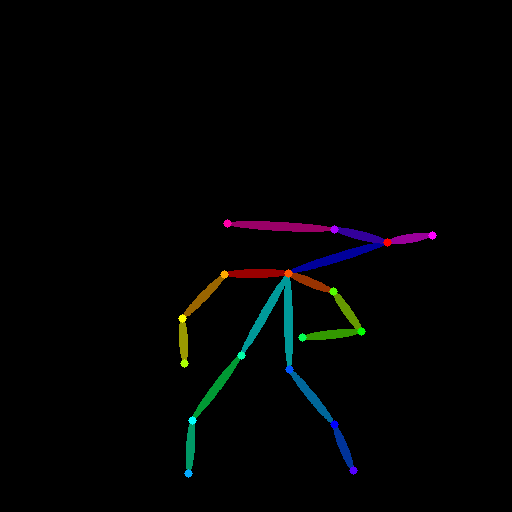} \\[5pt]
        \includegraphics[width=0.7\textwidth]{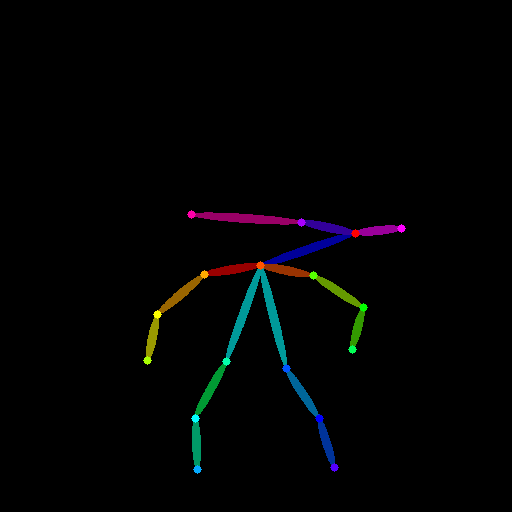} \\[5pt]
        \includegraphics[width=0.7\textwidth]{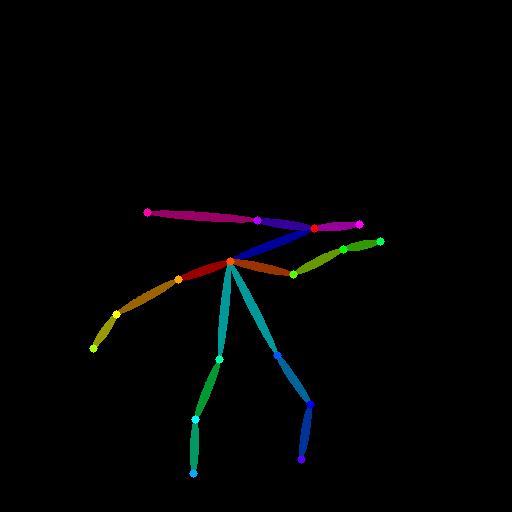} \\[5pt]
        \includegraphics[width=0.7\textwidth]{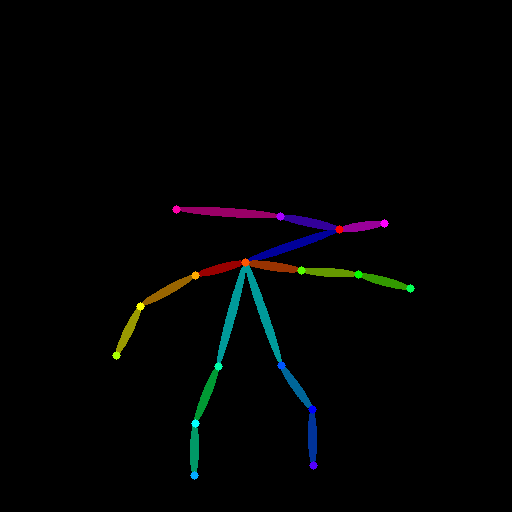}
        \caption{Pose Sequence}
        \label{fig:pose_sequence} % Label for this subfigure
    \end{subfigure}
    \hfill
    % Third column: Target sequence
    \begin{subfigure}[b]{0.3\textwidth}
        \centering
        \includegraphics[width=0.7\textwidth]{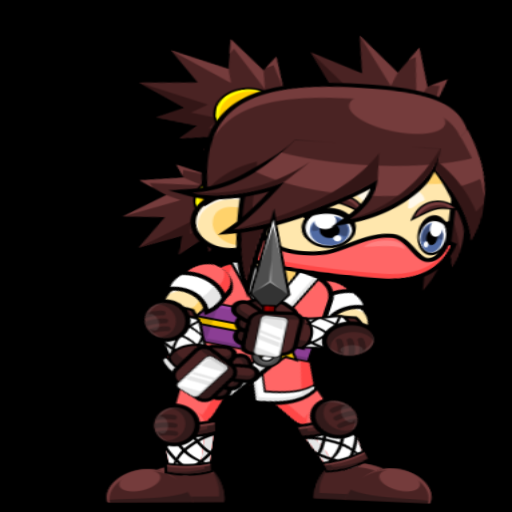} \\[5pt]
        \includegraphics[width=0.7\textwidth]{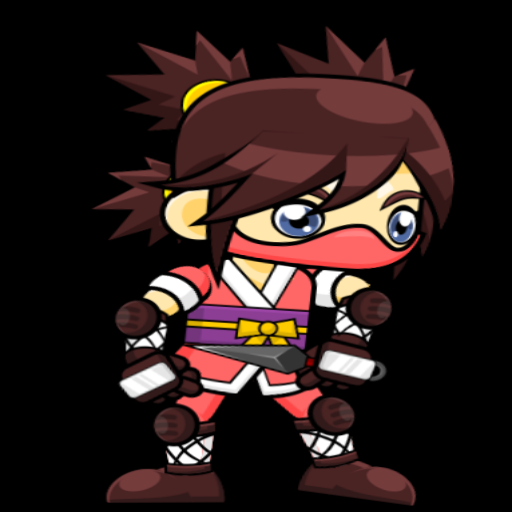} \\[5pt]
        \includegraphics[width=0.7\textwidth]{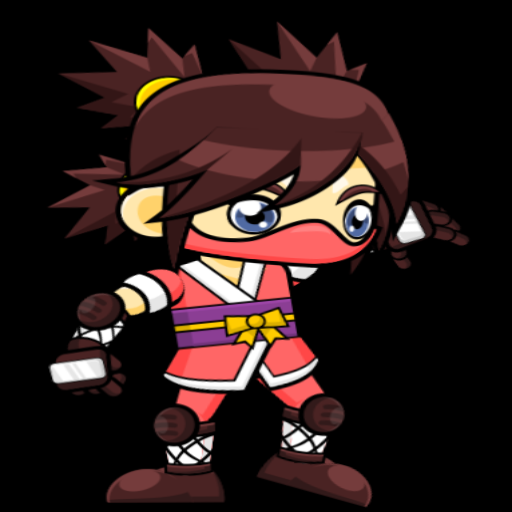} \\[5pt]
        \includegraphics[width=0.7\textwidth]{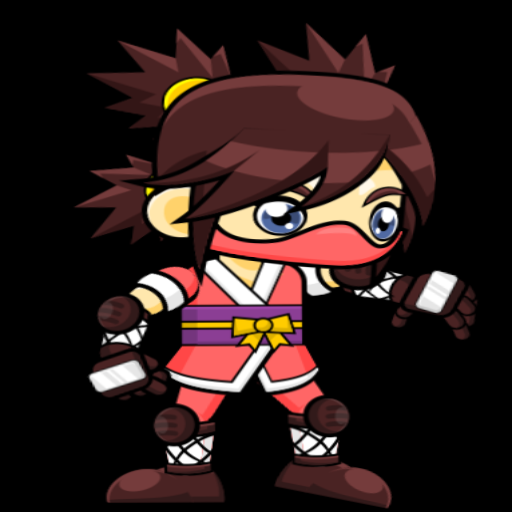}
        \caption{Action Sequence}
        \label{fig:action_sequence} % Label for this subfigure
    \end{subfigure}
    \caption{An example of an action sequence for one character.}
    \label{fig:data_example} % Label for the main figure
\end{figure}

\vspace{4.0em}

\begin{table*}[ht]
\centering
\renewcommand{\arraystretch}{1.2} % 调整行间距
\setlength{\tabcolsep}{15pt} % 调整列间距
\begin{tabular}{lcccc}
\hline
 & \textbf{Train} & \textbf{Val} & \textbf{Test (in-sample)} & \textbf{Test (out-sample)} \\ \hline
\textbf{Character Count} & 28 & 28 & 28 & 12 \\
(GameArt2D) & 12 & 12 & 12 & 4 \\
(SpriteDatabase) & 16 & 16 & 16 & 8 \\ \hline
\textbf{Action Seq Count} & 84 & 28 & 28 & 12 \\ \hline
\textbf{Frame Seq Count} & 533 & 174 & 157 & 52 \\ \hline
\end{tabular}
\vspace{1.0em}
\caption{Dataset Summary.}
\label{tab:dataset_summary}
\end{table*}

% \newpage
% \section{Framework}
% \begin{figure}[h]
% \centering
% \includegraphics[width=0.9\textwidth]{method.png}
% \caption{Framework}
% \label{fig:framework}
% \end{figure}

\newpage
\section{Qualitative Comparison}
\label{app:qualitative}
\subsection{In-Sample}
\begin{figure}[h!]
    \centering
    \scalebox{0.8}{
    \begin{minipage}[t]{0.19\textwidth} % 调整宽度为 0.19\textwidth
        \centering
        \includegraphics[width=\textwidth]{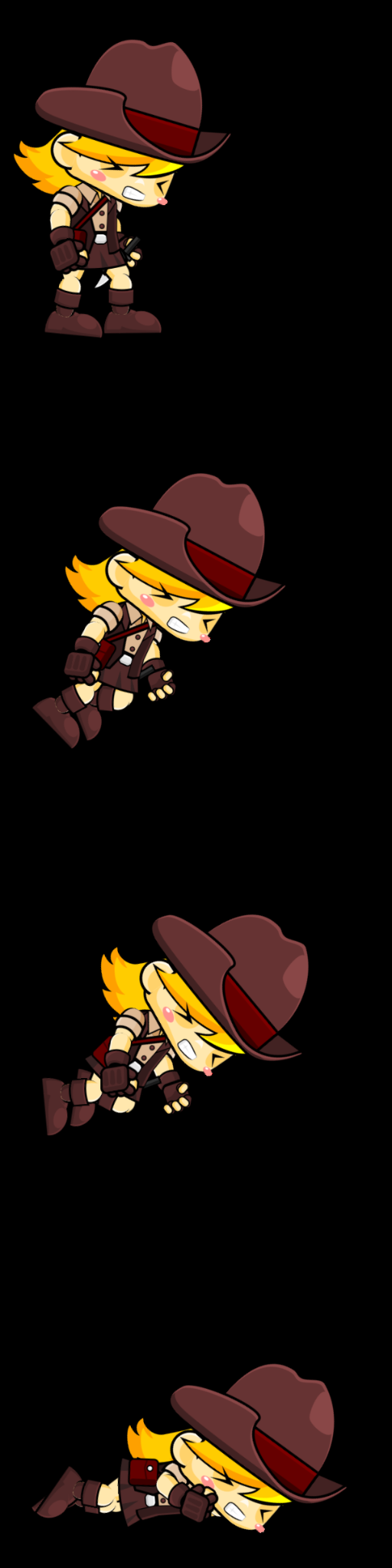}
        \textbf{Ground Truth} 
        % \caption*{Reference Image}
    \end{minipage}
    \hfill
    \begin{minipage}[t]{0.19\textwidth}
        \centering
        \includegraphics[width=\textwidth]{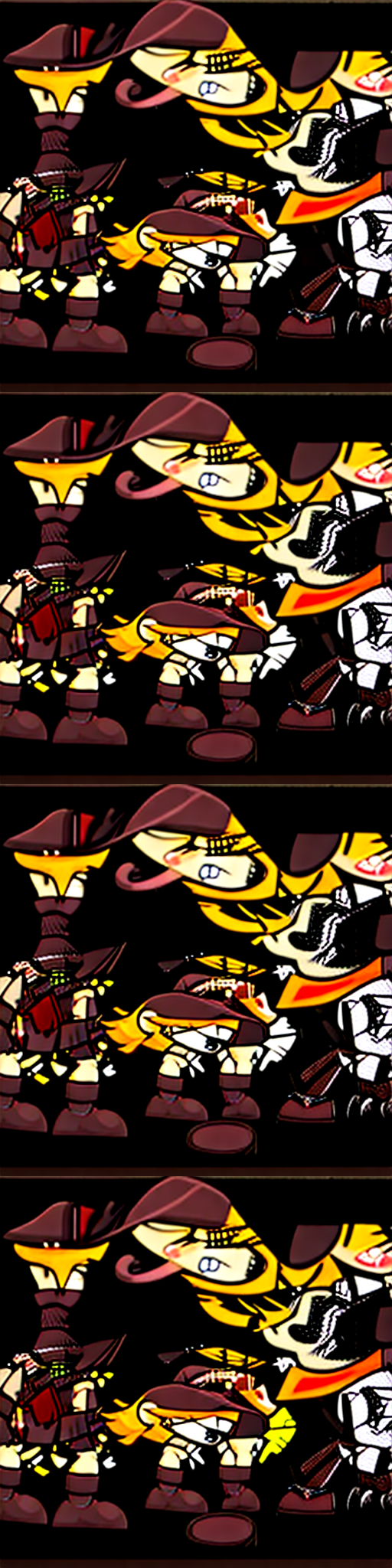}
        \textbf{Animate Anyone}
    \end{minipage}
    \hfill
    \begin{minipage}[t]{0.19\textwidth}
        \centering
        \includegraphics[width=\textwidth]{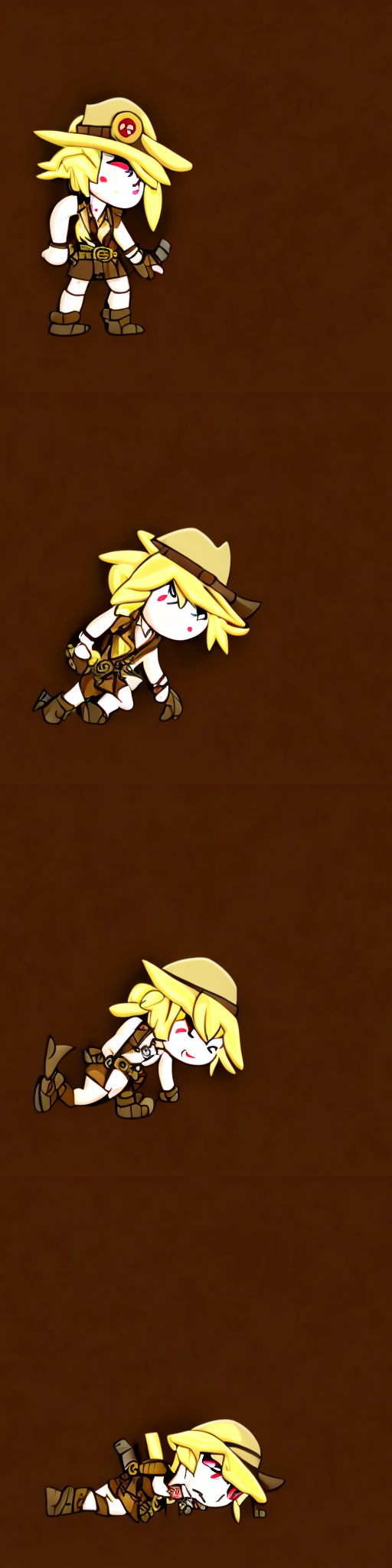}
        \textbf{SD-IPCN}
    \end{minipage}
    \hfill
    \begin{minipage}[t]{0.19\textwidth}
        \centering
        \includegraphics[width=\textwidth]{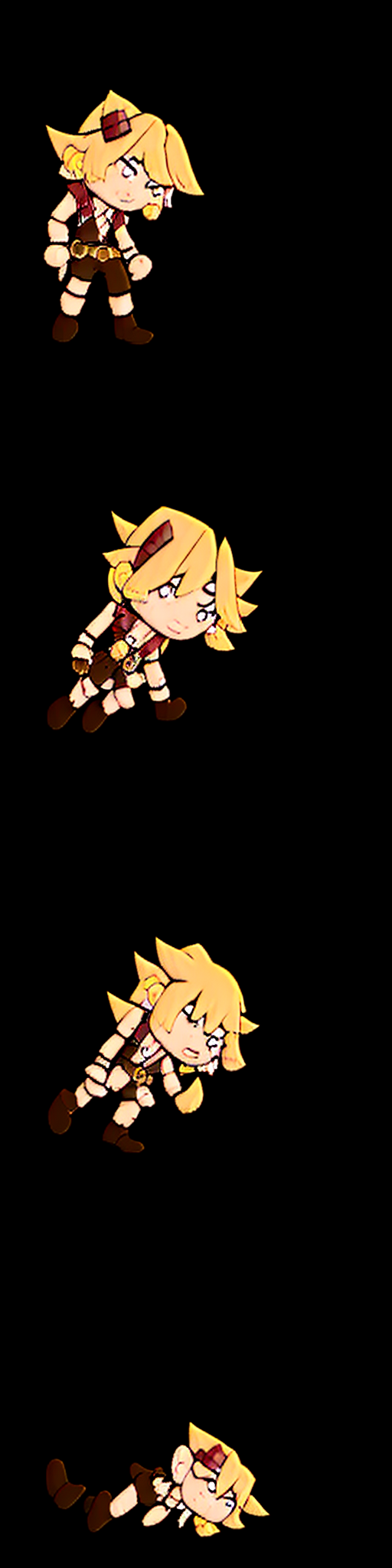}
        \textbf{SD-IPCN (finetuned)}
    \end{minipage}
    \hfill
    \begin{minipage}[t]{0.19\textwidth} % 添加新列
        \centering
        \includegraphics[width=\textwidth]{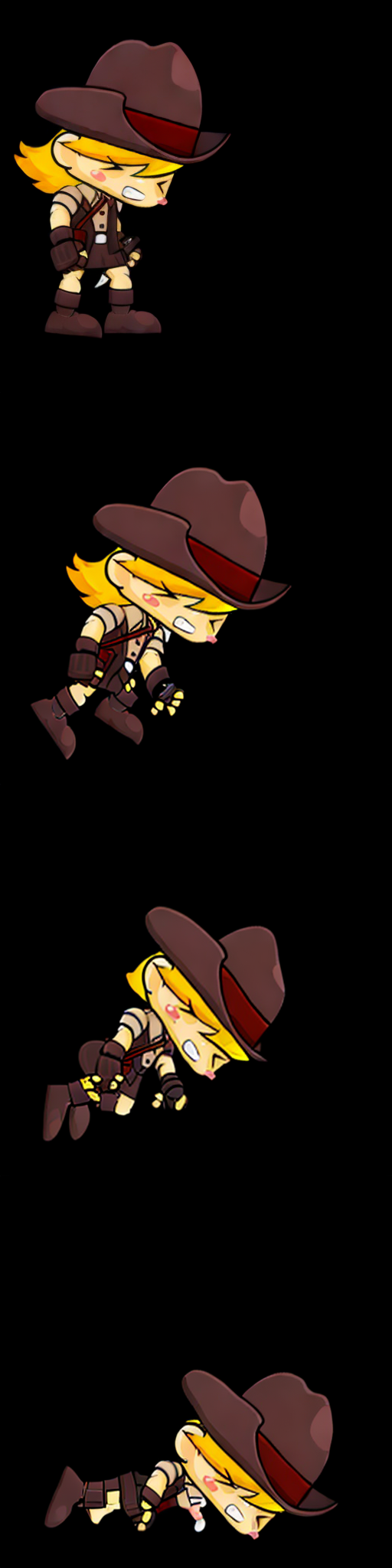}
        \textbf{Animate Anyone (finetuned)} 
    \end{minipage}
    }
    \caption{Qualitative Comparison (In-Sample - Adventure Girl - Dead)}
    \label{fig: QA1}
\end{figure}

\begin{figure}[h!]
    \centering
    \scalebox{0.8}{
    \begin{minipage}[t]{0.19\textwidth} % 调整宽度为 0.19\textwidth
        \centering
        \includegraphics[width=\textwidth]{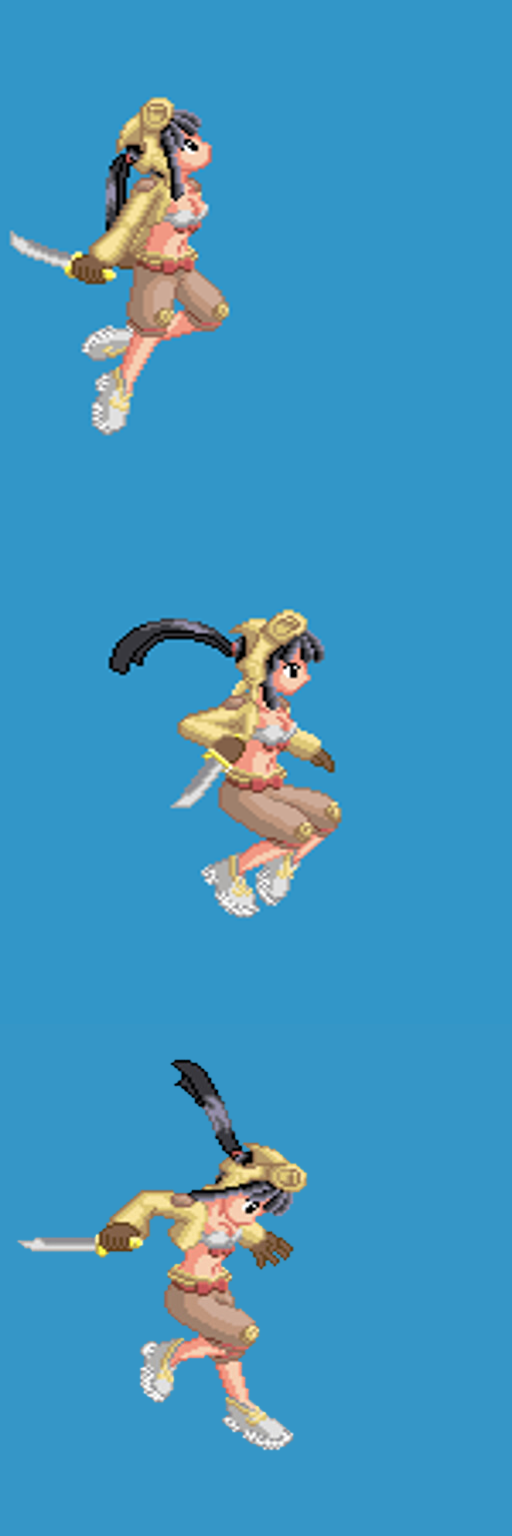}
        \textbf{Ground Truth} 
        % \caption*{Reference Image}
    \end{minipage}
    \hfill
    \begin{minipage}[t]{0.19\textwidth}
        \centering
        \includegraphics[width=\textwidth]{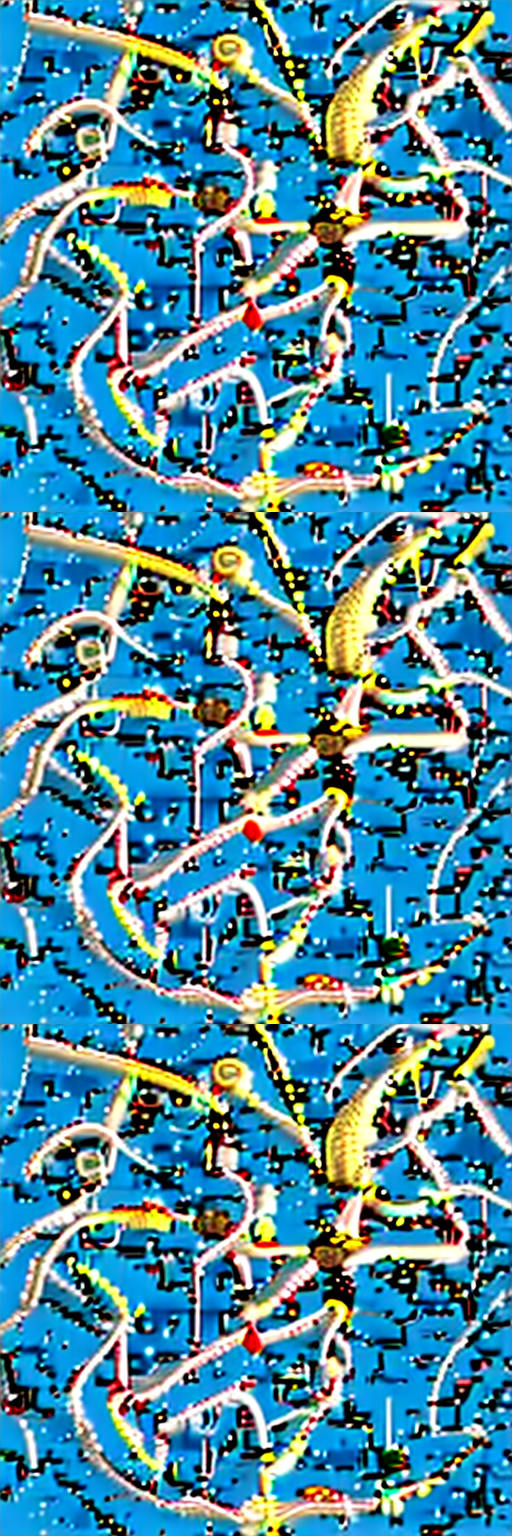}
        \textbf{Animate Anyone}
    \end{minipage}
    \hfill
    \begin{minipage}[t]{0.19\textwidth}
        \centering
        \includegraphics[width=\textwidth]{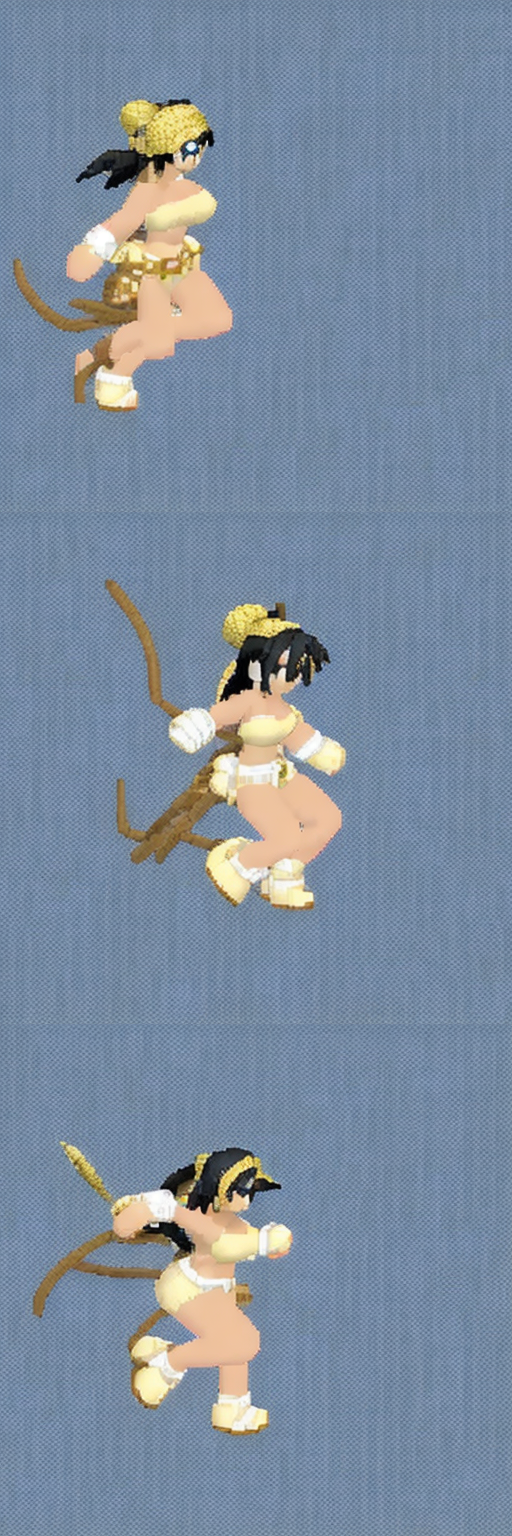}
        \textbf{SD-IPCN}
    \end{minipage}
    \hfill
    \begin{minipage}[t]{0.19\textwidth}
        \centering
        \includegraphics[width=\textwidth]{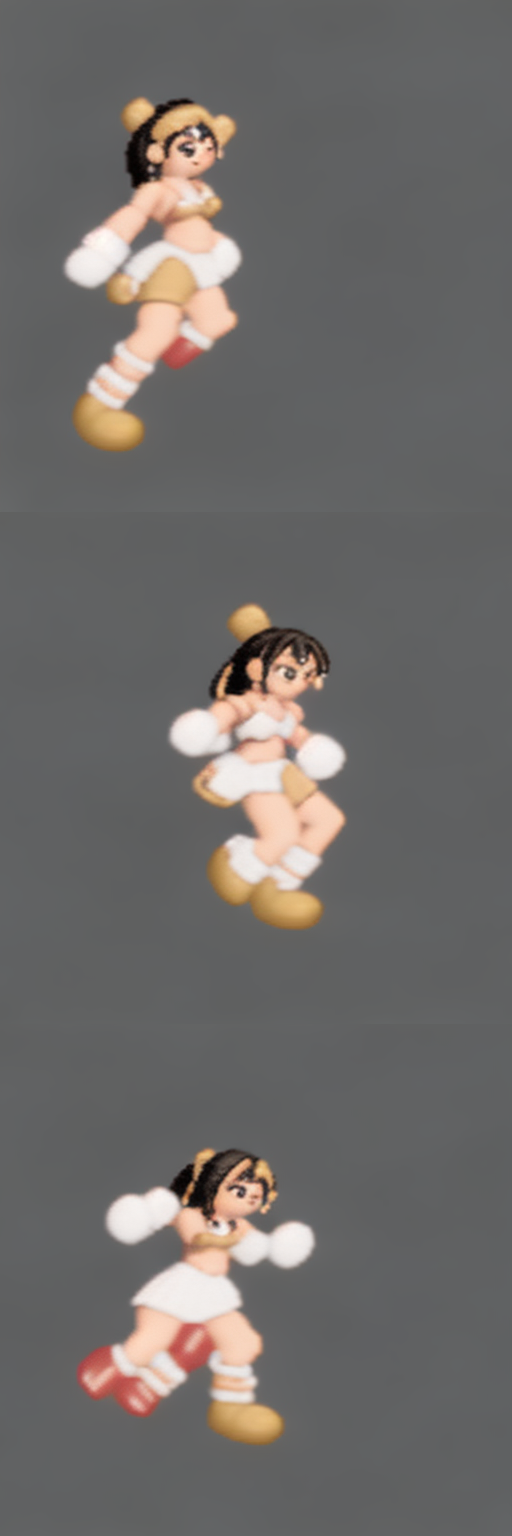}
        \textbf{SD-IPCN (finetuned)}
    \end{minipage}
    \hfill
    \begin{minipage}[t]{0.19\textwidth} % 添加新列
        \centering
        \includegraphics[width=\textwidth]{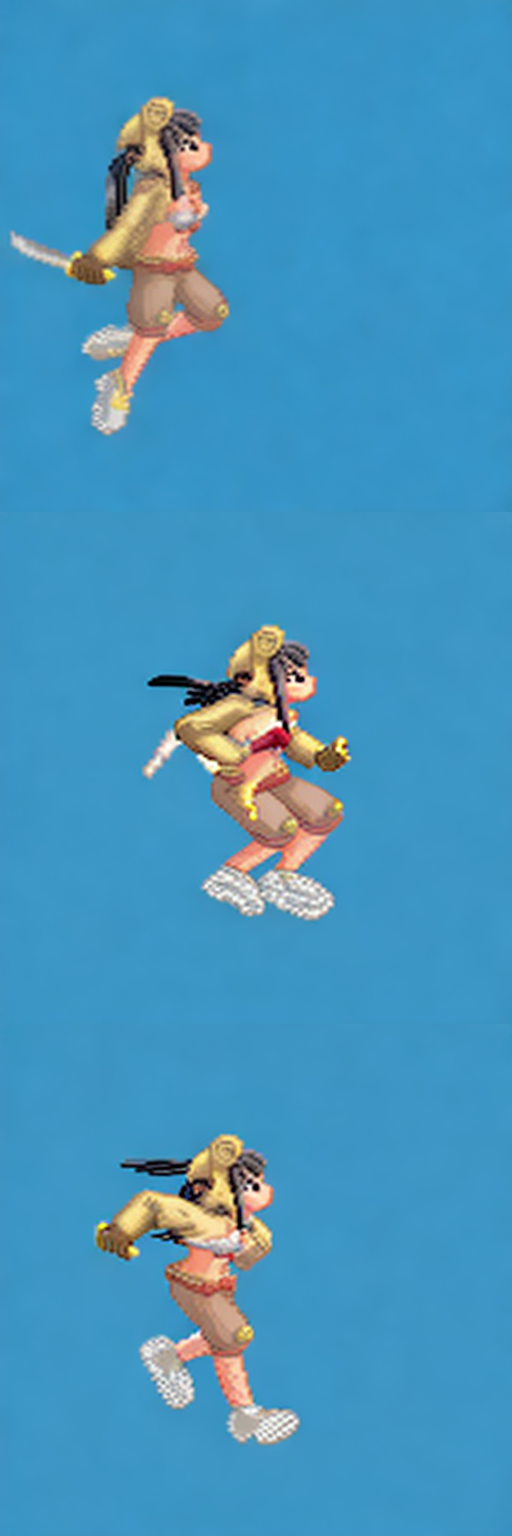}
        \textbf{Animate Anyone (finetuned)} 
    \end{minipage}
    }
    \caption{Qualitative Comparison (In-Sample - Theif - Jump)}
    \label{fig: QA2}
\end{figure}

\newpage
\subsection{Out-Sample}
\begin{figure}[h!]
    \centering
    \scalebox{0.8}{
    \begin{minipage}[t]{0.19\textwidth} % 调整宽度为 0.19\textwidth
        \centering
        \includegraphics[width=\textwidth]{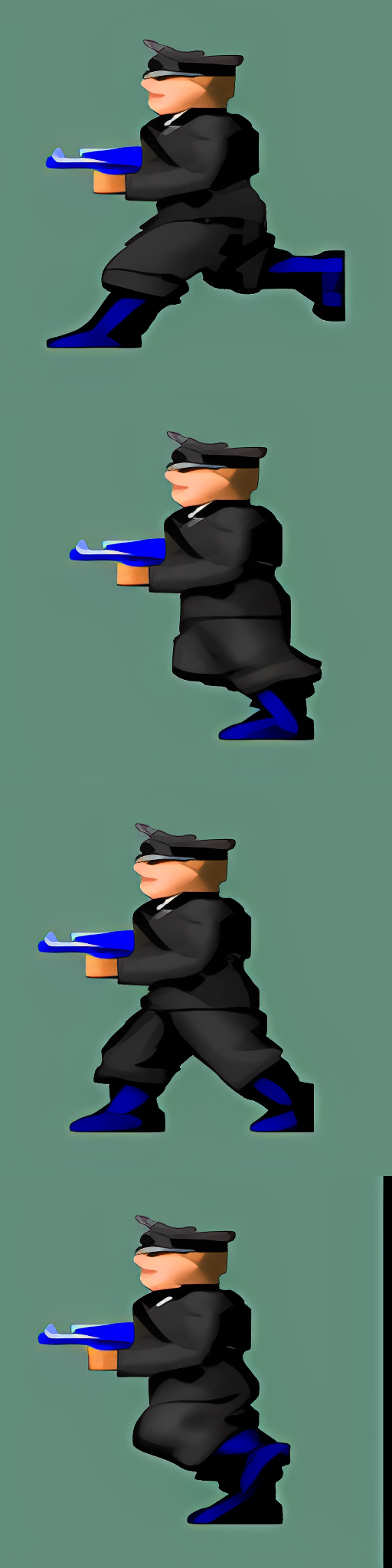}
        \textbf{Ground Truth} 
        % \caption*{Reference Image}
    \end{minipage}
    \hfill
    \begin{minipage}[t]{0.19\textwidth}
        \centering
        \includegraphics[width=\textwidth]{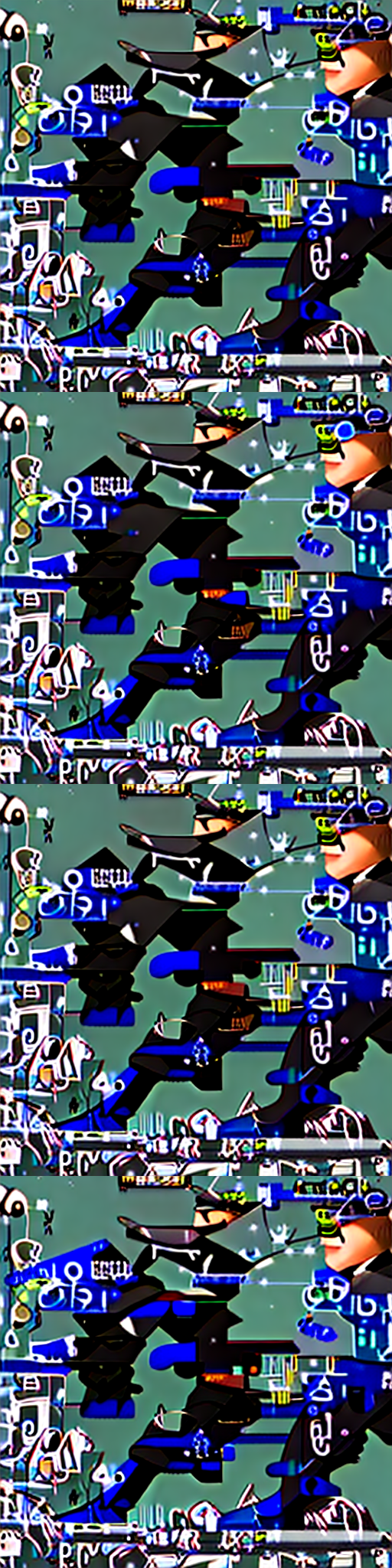}
        \textbf{Animate Anyone}
    \end{minipage}
    \hfill
    \begin{minipage}[t]{0.19\textwidth}
        \centering
        \includegraphics[width=\textwidth]{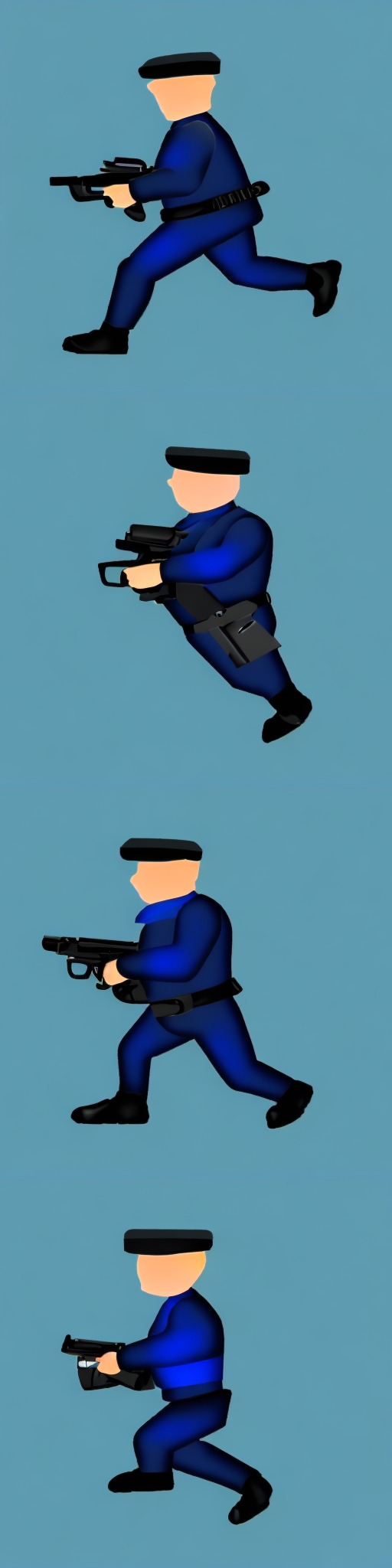}
        \textbf{SD-IPCN}
    \end{minipage}
    \hfill
    \begin{minipage}[t]{0.19\textwidth}
        \centering
        \includegraphics[width=\textwidth]{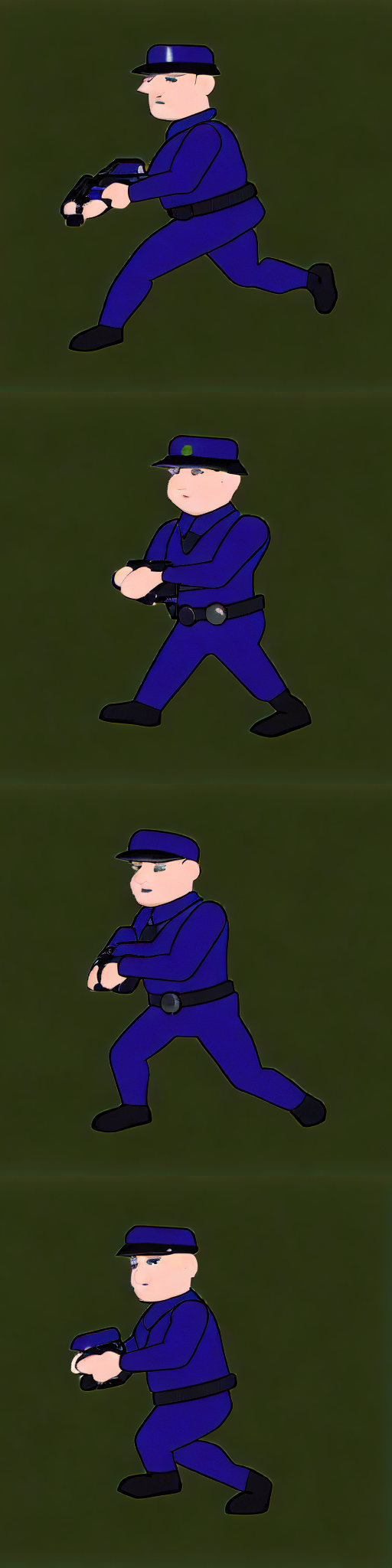}
        \textbf{SD-IPCN (finetuned)}
    \end{minipage}
    \hfill
    \begin{minipage}[t]{0.19\textwidth} % 添加新列
        \centering
        \includegraphics[width=\textwidth]{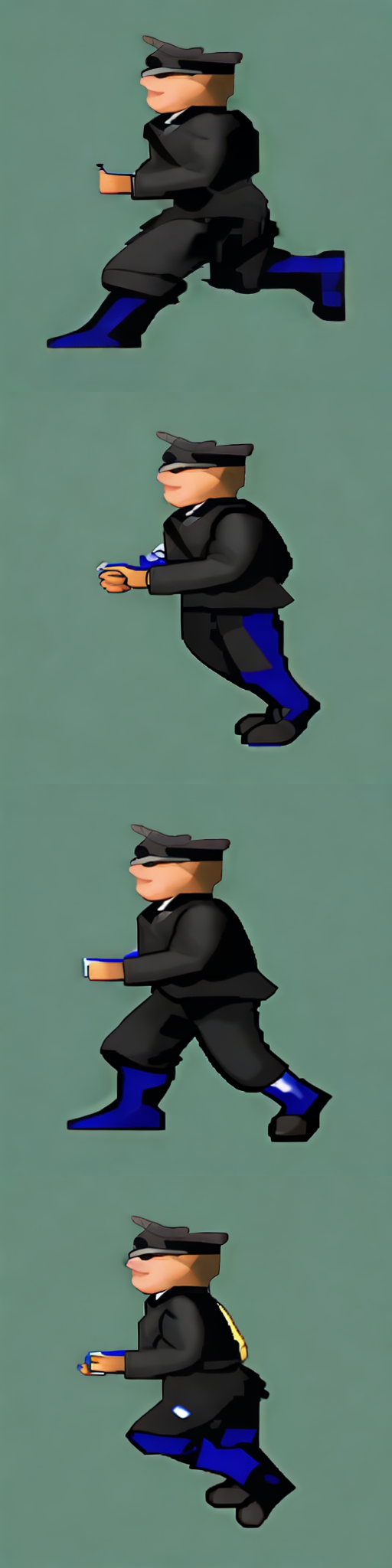}
        \textbf{Animate Anyone (finetuned)} 
    \end{minipage}
    }
    \caption{Qualitative Comparison (Out-Sample - SS - Run)}
    \label{fig: QA3}
\end{figure}

\begin{figure}[h!]
    \centering
    \scalebox{0.8}{
    \begin{minipage}[t]{0.19\textwidth} % 调整宽度为 0.19\textwidth
        \centering
        \includegraphics[width=\textwidth]{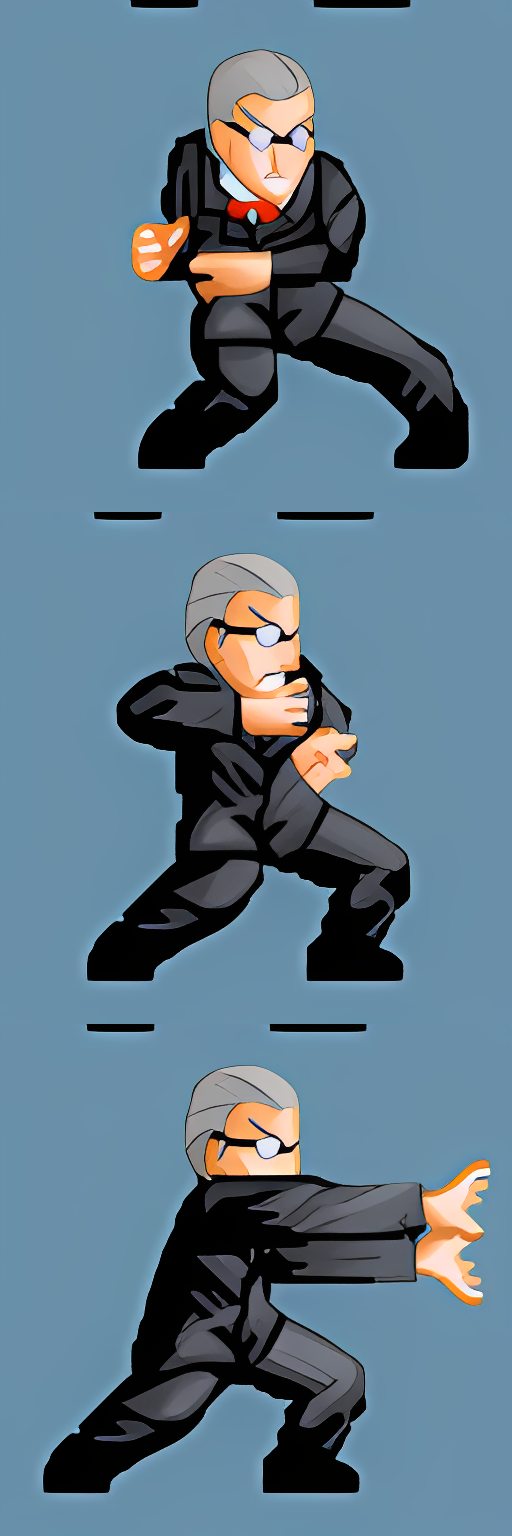}
        \textbf{Ground Truth} 
        % \caption*{Reference Image}
    \end{minipage}
    \hfill
    \begin{minipage}[t]{0.19\textwidth}
        \centering
        \includegraphics[width=\textwidth]{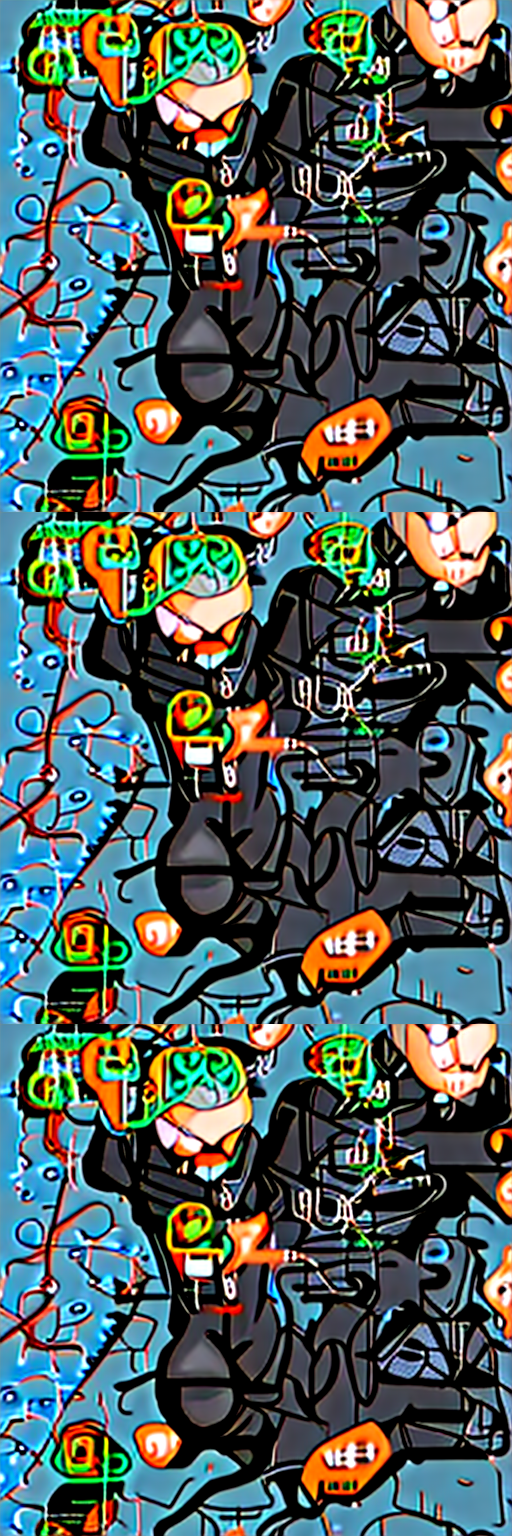}
        \textbf{Animate Anyone}
    \end{minipage}
    \hfill
    \begin{minipage}[t]{0.19\textwidth}
        \centering
        \includegraphics[width=\textwidth]{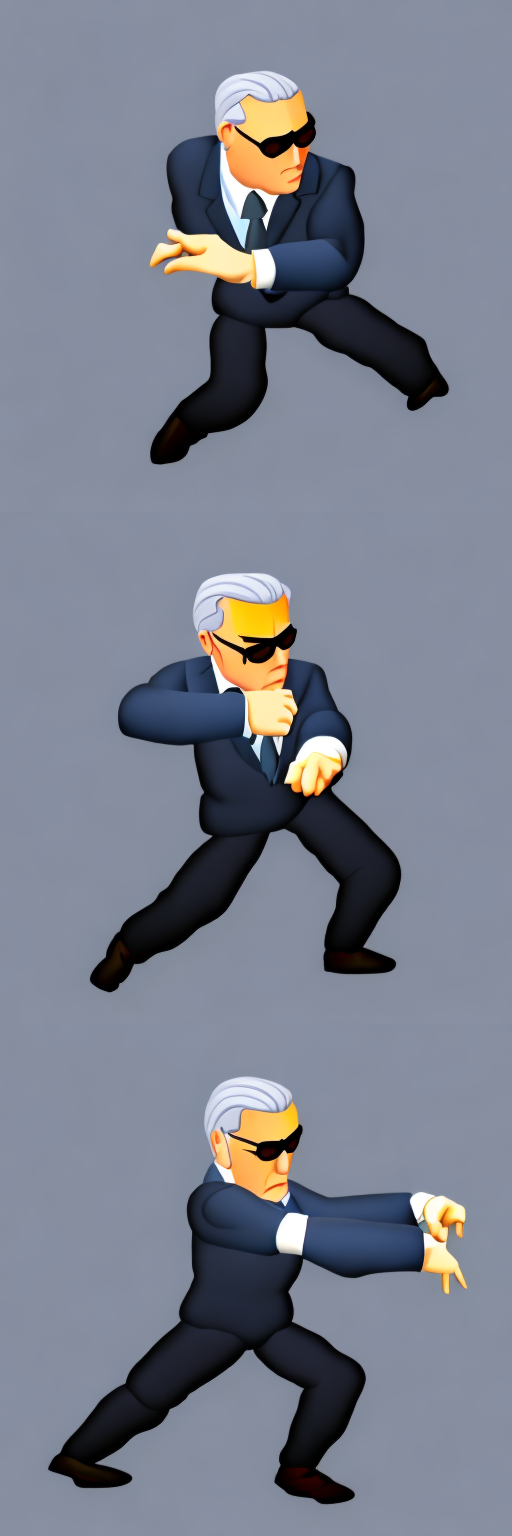}
        \textbf{SD-IPCN}
    \end{minipage}
    \hfill
    \begin{minipage}[t]{0.19\textwidth}
        \centering
        \includegraphics[width=\textwidth]{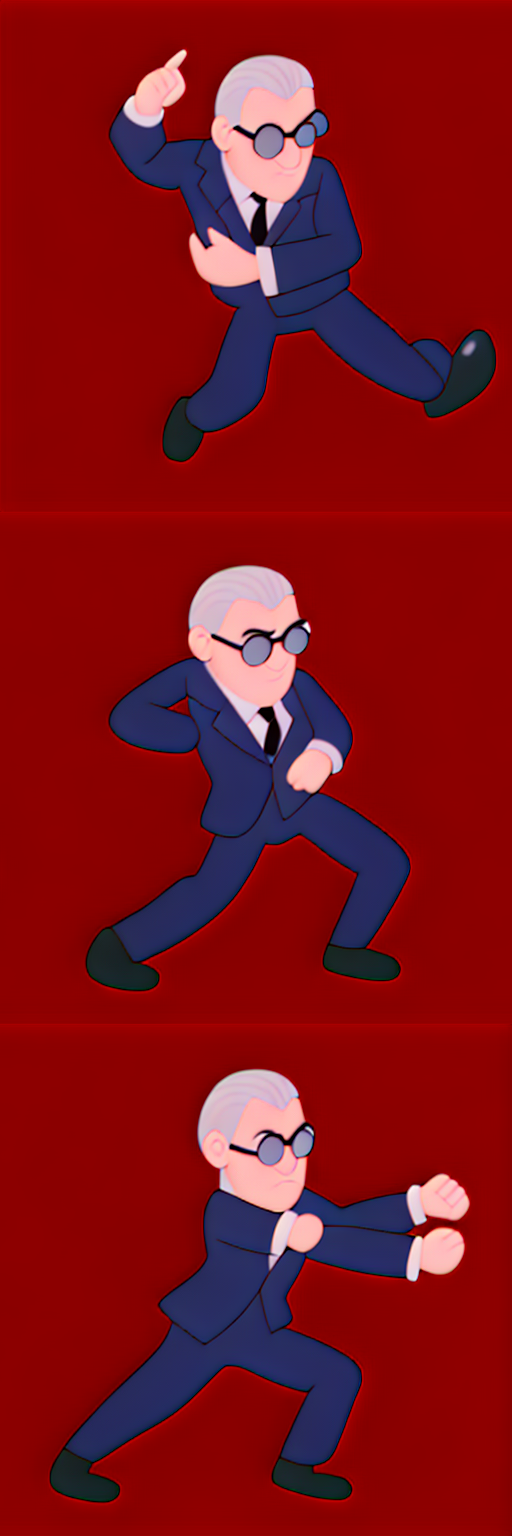}
        \textbf{SD-IPCN (finetuned)}
    \end{minipage}
    \hfill
    \begin{minipage}[t]{0.19\textwidth} % 添加新列
        \centering
        \includegraphics[width=\textwidth]{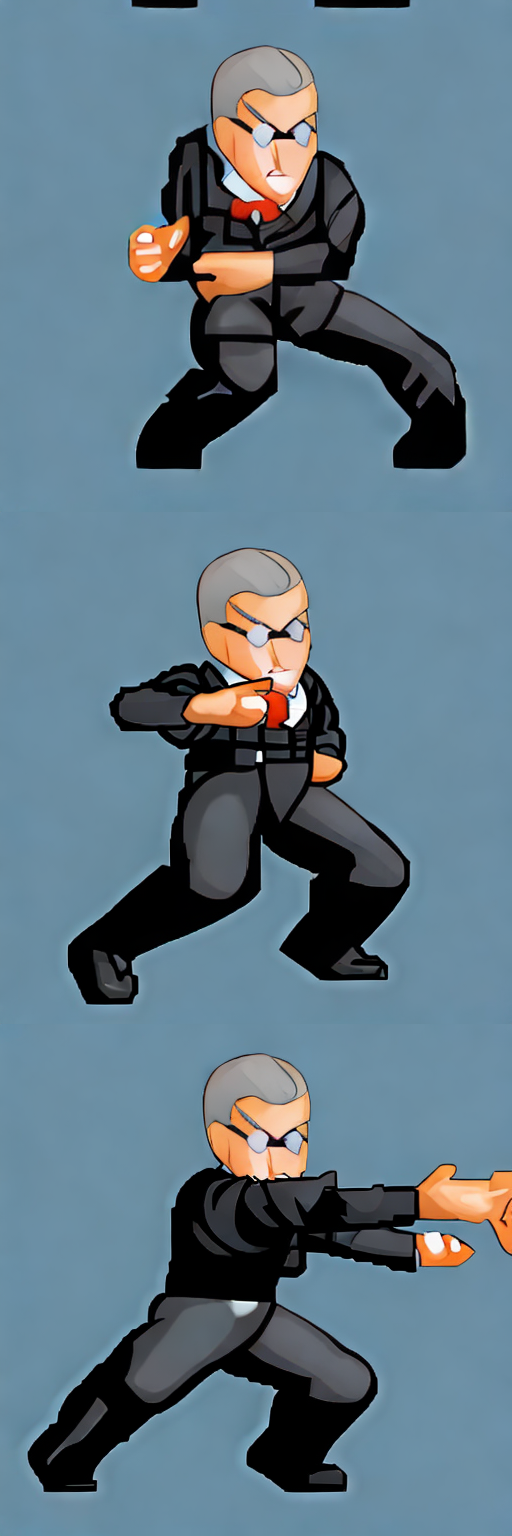}
        \textbf{Animate Anyone (finetuned)} 
    \end{minipage}
    }
    \caption{Qualitative Comparison (Out-Sample - Enemies  Miscellaeous - Attack)}
    \label{fig: QA4}
\end{figure}

\newpage
\section{Ablation Study}
\label{ablation}
\subsection{Quantitative Comparison}
\label{ablation:quantitative}
\begin{table}[!ht]
\centering
\resizebox{\textwidth}{!}{ % Resize to fit within the text width
\begin{tabular}{lcccc}
\toprule
Method & SSIM$ \uparrow$ & PSNR$ \uparrow$ & LPIPS$ \downarrow$ & Subject Consistency$ \uparrow$ \\ 
\midrule
Pose Guider Only & 0.535 ± 0.226 & 15.716 ± 4.503 & 0.211 ± 0.132 & 0.901 ± 0.064 \\
Pose Guider + Unet & 0.606 ± 0.238 & 17.304 ± 5.769 & 0.165 ± 0.114 & 0.914 ± 0.058 \\
Stage 1 Only & 0.619 ± 0.239 & 17.883 ± 6.226 & 0.153 ± 0.108 & \textbf{0.920 ± 0.05}1 \\
Fully Fine-Tuned & \textbf{0.659 ± 0.250} & \textbf{18.405 ± 5.280} & \textbf{0.125 ± 0.088} & 0.901 ± 0.064 \\
\bottomrule
\end{tabular}
}
\caption{Ablation Study Results (In-Sample).}
\label{tab:ablation:in-sample}
\end{table}

\begin{table}[!ht]
\centering
\resizebox{\textwidth}{!}{ % Resize to fit within the text width
\begin{tabular}{lcccc}
\toprule
Method & SSIM$ \uparrow$ & PSNR$ \uparrow$ & LPIPS$ \downarrow$ & Subject Consistency$ \uparrow$ \\ 
\midrule
Pose Guider Only & 0.574 ± 0.182 & 16.966 ± 4.042 & 0.178 ± 0.099 & 0.911 ± 0.038 \\
Pose Guider + Unet & 0.599 ± 0.169 & 18.429 ± 5.314 & 0.158 ± 0.096 & 0.929 ± 0.035 \\
Stage 1 Only & 0.644 ± 0.194 & \textbf{19.491 ± 5.884} & \textbf{0.137 ± 0.091} & \textbf{0.931 ± 0.035} \\
Fully Fine-Tuned & \textbf{0.655 ± 0.195} & 18.809 ± 4.806 & 0.139 ± 0.090 & 0.893 ± 0.038 \\
\bottomrule
\end{tabular}
}
\caption{Ablation Study Results (Out-Sample).}
\label{tab:ablation:out-sample}
\end{table}

\newpage
\subsection{Qualitative Comparison}
\label{ablation:qualitative}
\subsubsection{In-Sample}
\begin{figure}[h!]
    \centering
    \scalebox{0.8}{
    \begin{minipage}[t]{0.19\textwidth} % 调整宽度为 0.19\textwidth
        \centering
        \includegraphics[width=\textwidth]{results/QA_ref_1.png}
        \textbf{Ground Truth} 
        % \caption*{Reference Image}
    \end{minipage}
    \hfill
    \begin{minipage}[t]{0.19\textwidth}
        \centering
        \includegraphics[width=\textwidth]{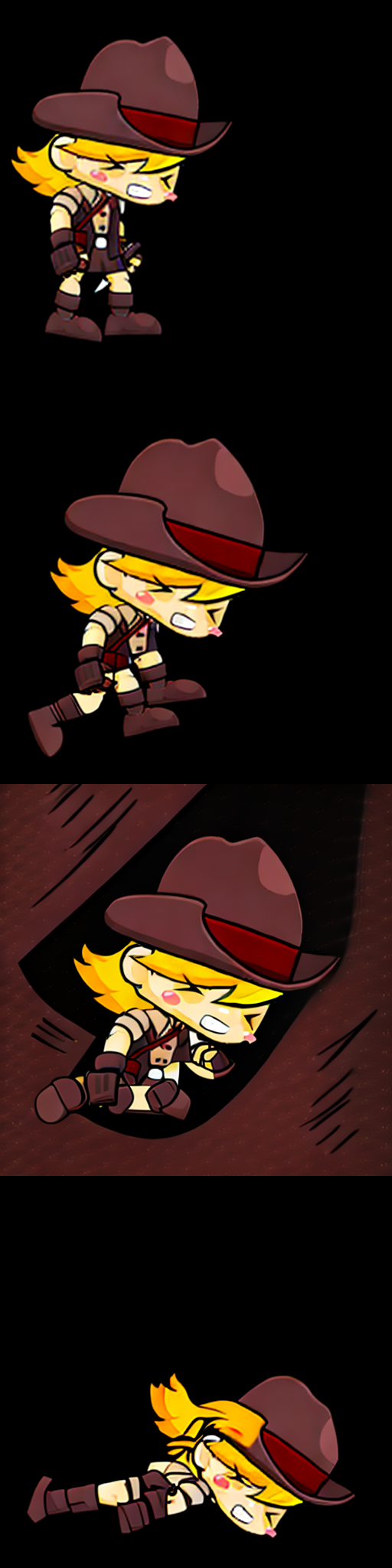}
        \textbf{Pose Guider Only}
    \end{minipage}
    \hfill
    \begin{minipage}[t]{0.19\textwidth}
        \centering
        \includegraphics[width=\textwidth]{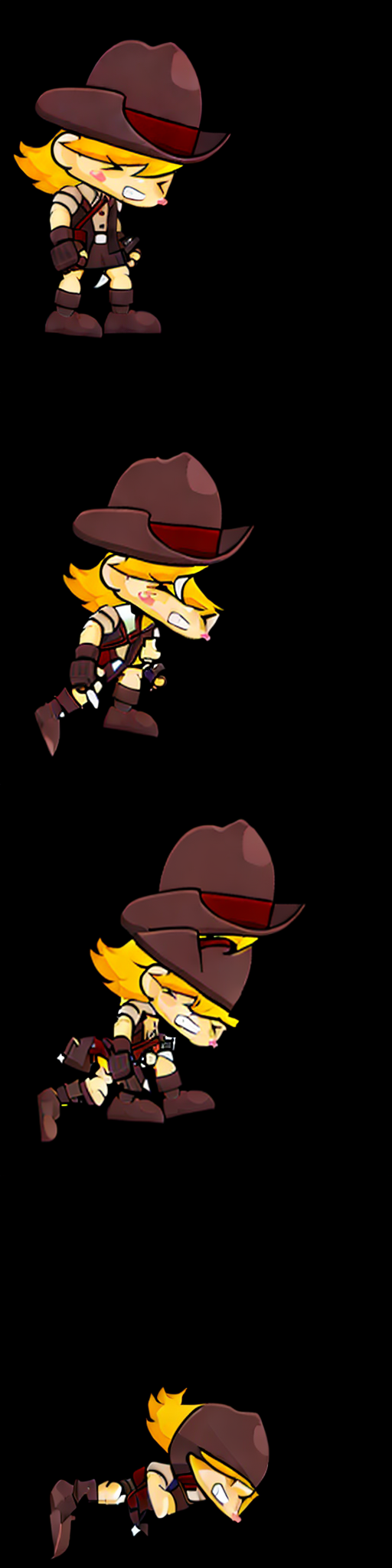}
        \textbf{Pose Guider + UNet}
    \end{minipage}
    \hfill
    \begin{minipage}[t]{0.19\textwidth}
        \centering
        \includegraphics[width=\textwidth]{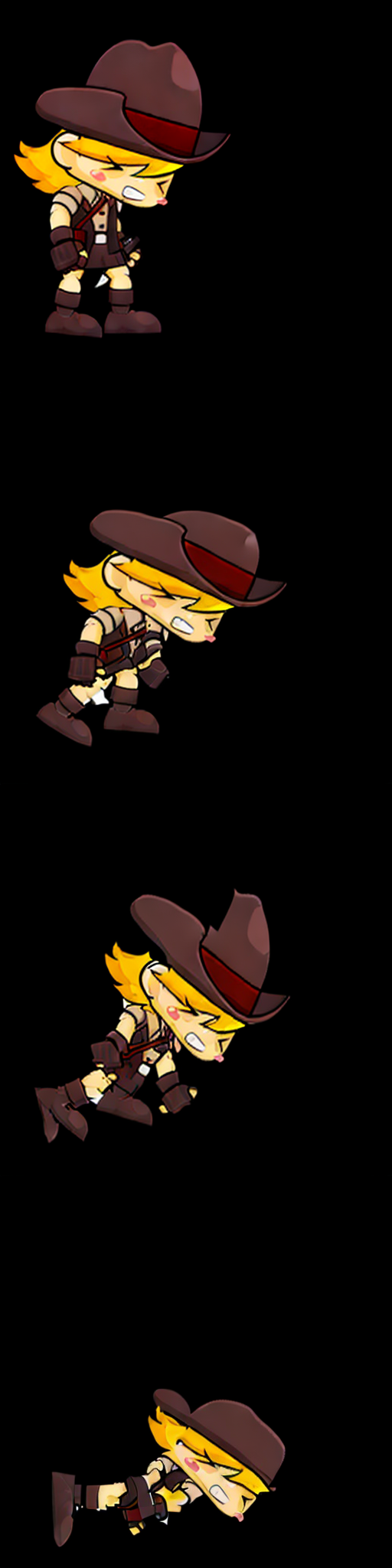}
        \textbf{Stage 1 Only}
    \end{minipage}
    \hfill
    \begin{minipage}[t]{0.19\textwidth} % 添加新列
        \centering
        \includegraphics[width=\textwidth]{results/QA_animate_finetune_1.png}
        \textbf{Fully Fine-Tuned} 
    \end{minipage}
    }
    \caption{Ablation Study Qualitative Comparison (In-Sample - Adventure Girl - Dead)}
    \label{fig-abl: QA1}
\end{figure}

\begin{figure}[h!]
    \centering
    \scalebox{0.8}{
    \begin{minipage}[t]{0.19\textwidth} % 调整宽度为 0.19\textwidth
        \centering
        \includegraphics[width=\textwidth]{results/QA_ref_3.png}
        \textbf{Ground Truth} 
        % \caption*{Reference Image}
    \end{minipage}
    \hfill
    \begin{minipage}[t]{0.19\textwidth}
        \centering
        \includegraphics[width=\textwidth]{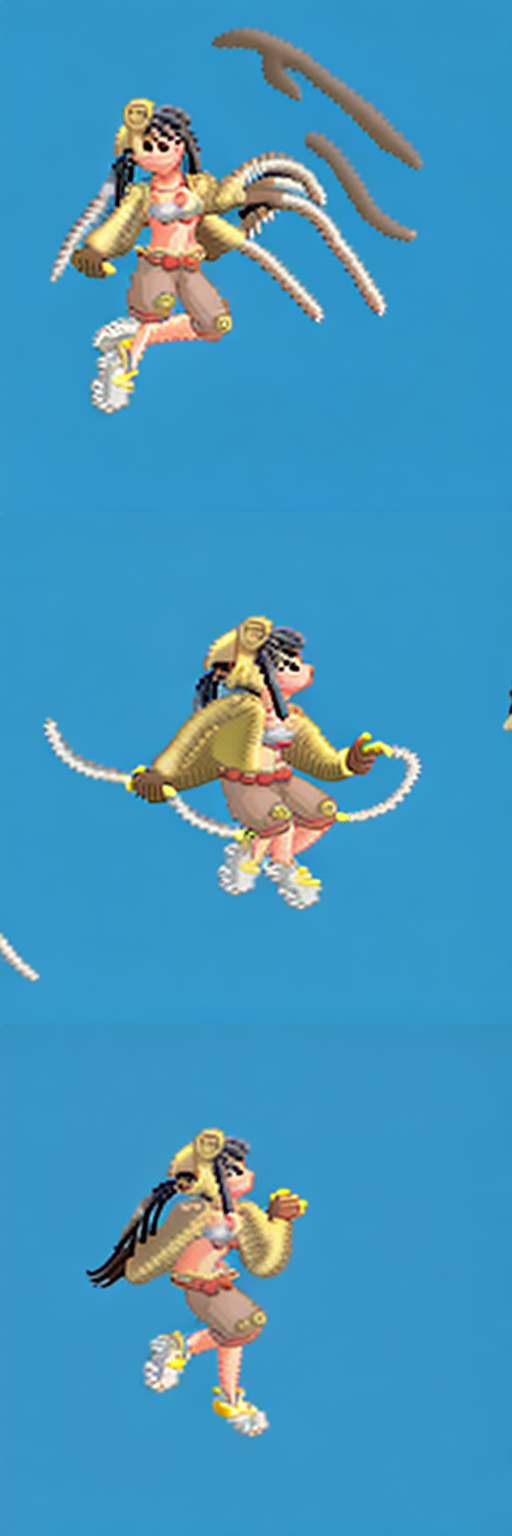}
        \textbf{Pose Guider Only}
    \end{minipage}
    \hfill
    \begin{minipage}[t]{0.19\textwidth}
        \centering
        \includegraphics[width=\textwidth]{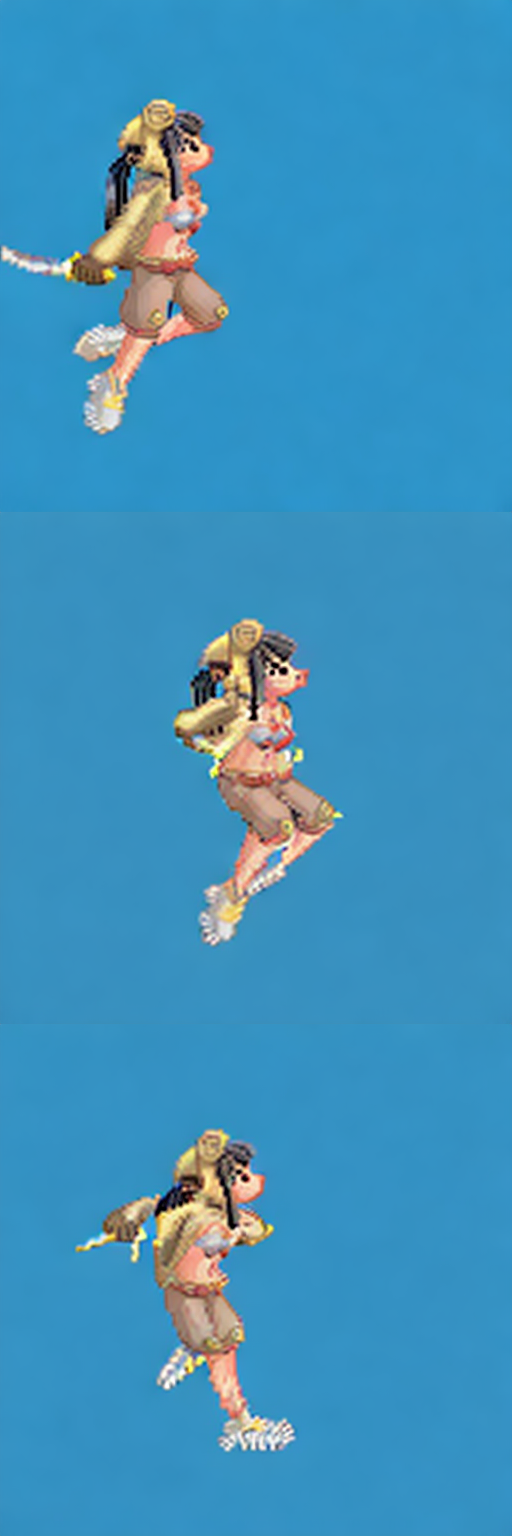}
        \textbf{Pose Guider + UNet}
    \end{minipage}
    \hfill
    \begin{minipage}[t]{0.19\textwidth}
        \centering
        \includegraphics[width=\textwidth]{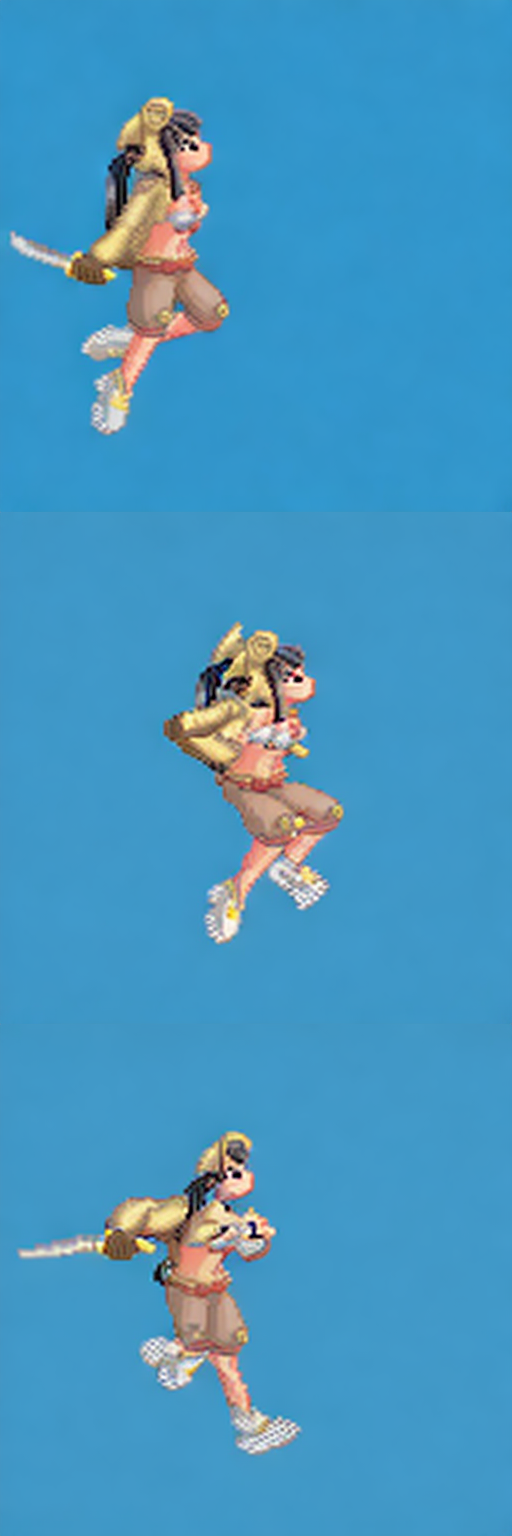}
        \textbf{Stage 1 Only}
    \end{minipage}
    \hfill
    \begin{minipage}[t]{0.19\textwidth} % 添加新列
        \centering
        \includegraphics[width=\textwidth]{results/QA_animate_finetune_3.png}
        \textbf{Fully Fine-Tuned} 
    \end{minipage}
    }
    \caption{Ablation Study Qualitative Comparison (In-Sample - Theif - Jump)}
    \label{fig-abl: QA2}
\end{figure}

\newpage
\subsubsection{Out-Sample}
\begin{figure}[h!]
    \centering
    \scalebox{0.8}{
        \begin{minipage}[t]{0.19\textwidth} % 调整宽度为 0.19\textwidth
        \centering
        \includegraphics[width=\textwidth]{results/QA_ref_2.png}
        \textbf{Ground Truth} 
        % \caption*{Reference Image}
    \end{minipage}
    \hfill
    \begin{minipage}[t]{0.19\textwidth}
        \centering
        \includegraphics[width=\textwidth]{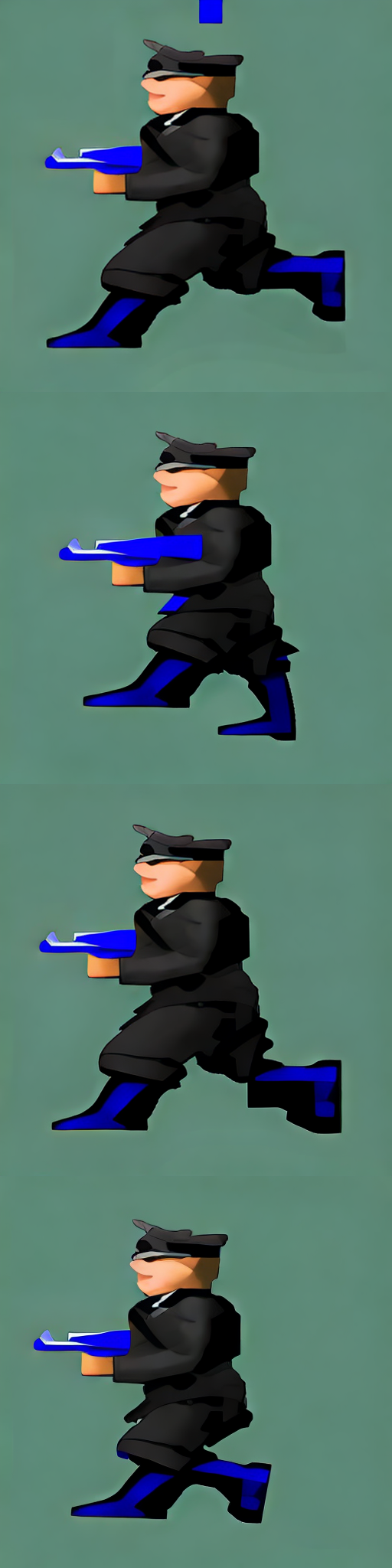}
        \textbf{Pose Guider Only}
    \end{minipage}
    \hfill
    \begin{minipage}[t]{0.19\textwidth}
        \centering
        \includegraphics[width=\textwidth]{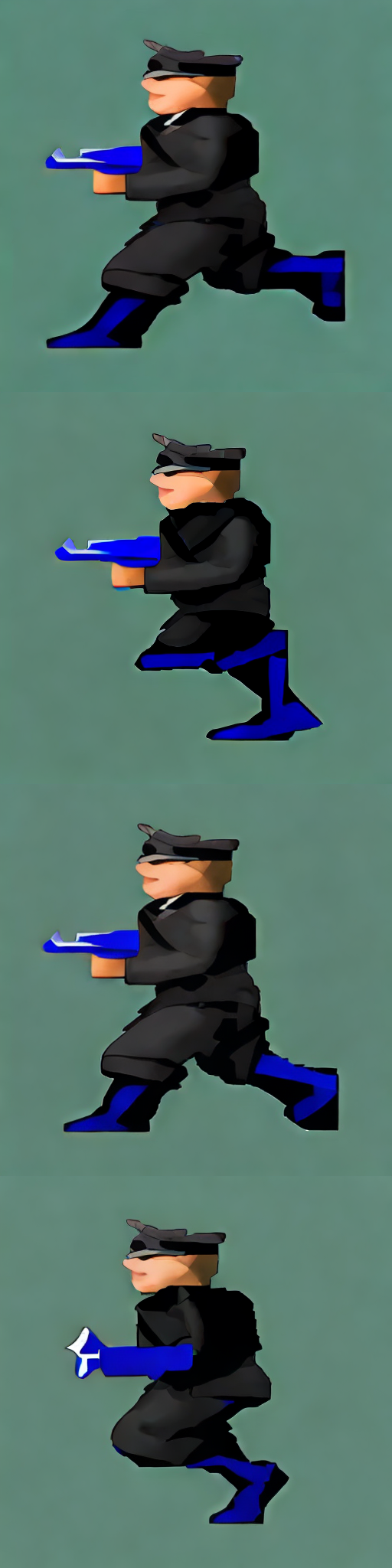}
        \textbf{Pose Guider + UNet}
    \end{minipage}
    \hfill
    \begin{minipage}[t]{0.19\textwidth}
        \centering
        \includegraphics[width=\textwidth]{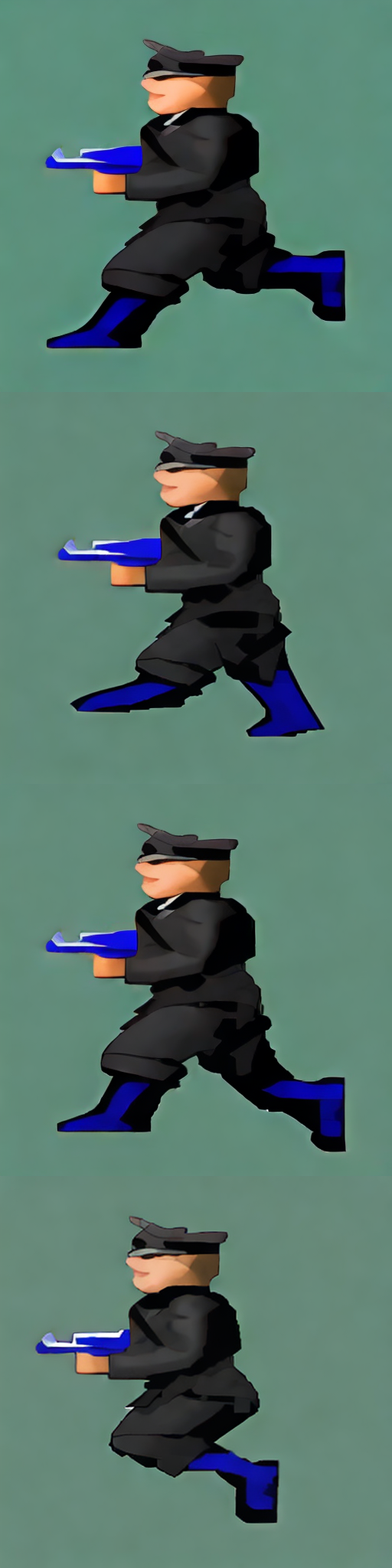}
        \textbf{Stage 1 Only}
    \end{minipage}
    \hfill
    \begin{minipage}[t]{0.19\textwidth} % 添加新列
        \centering
        \includegraphics[width=\textwidth]{results/QA_animate_finetune_2.png}
        \textbf{Fully Fine-Tuned} 
    \end{minipage}
    }
    \caption{Ablation Study Qualitative Comparison (Out-Sample - SS - Run)}
    \label{fig-abl: QA3}
\end{figure}

\begin{figure}[h!]
    \centering
    \scalebox{0.8}{
        \begin{minipage}[t]{0.19\textwidth} % 调整宽度为 0.19\textwidth
        \centering
        \includegraphics[width=\textwidth]{results/QA_ref_4.png}
        \textbf{Ground Truth} 
        % \caption*{Reference Image}
    \end{minipage}
    \hfill
    \begin{minipage}[t]{0.19\textwidth}
        \centering
        \includegraphics[width=\textwidth]{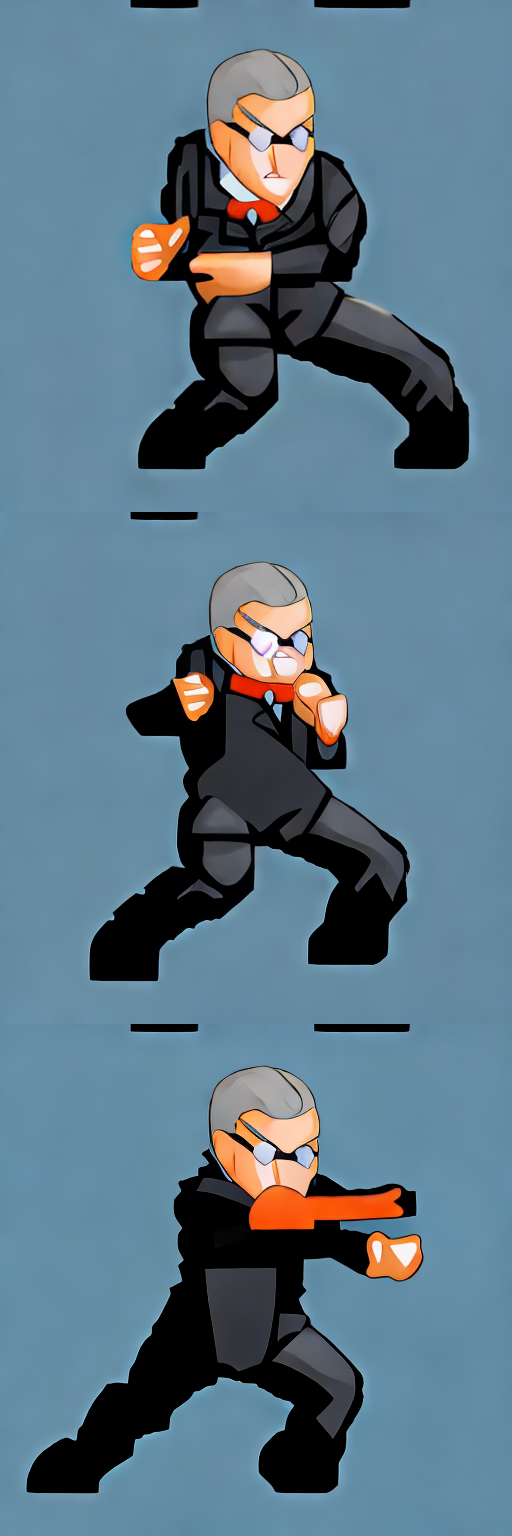}
        \textbf{Pose Guider Only}
    \end{minipage}
    \hfill
    \begin{minipage}[t]{0.19\textwidth}
        \centering
        \includegraphics[width=\textwidth]{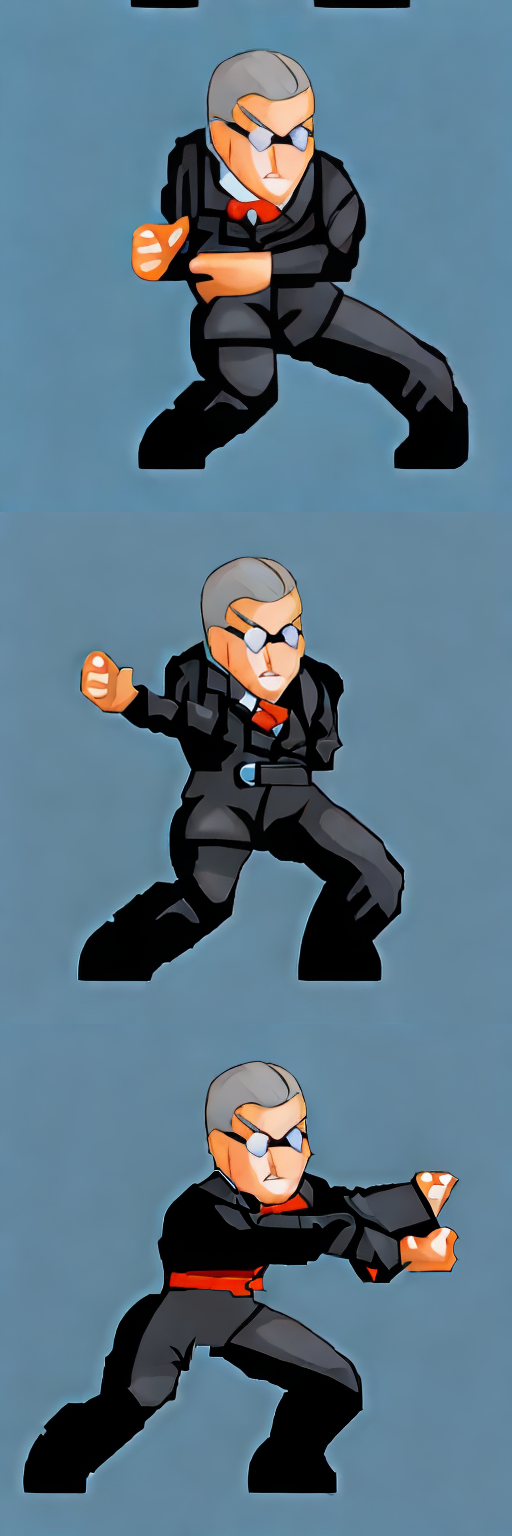}
        \textbf{Pose Guider + UNet}
    \end{minipage}
    \hfill
    \begin{minipage}[t]{0.19\textwidth}
        \centering
        \includegraphics[width=\textwidth]{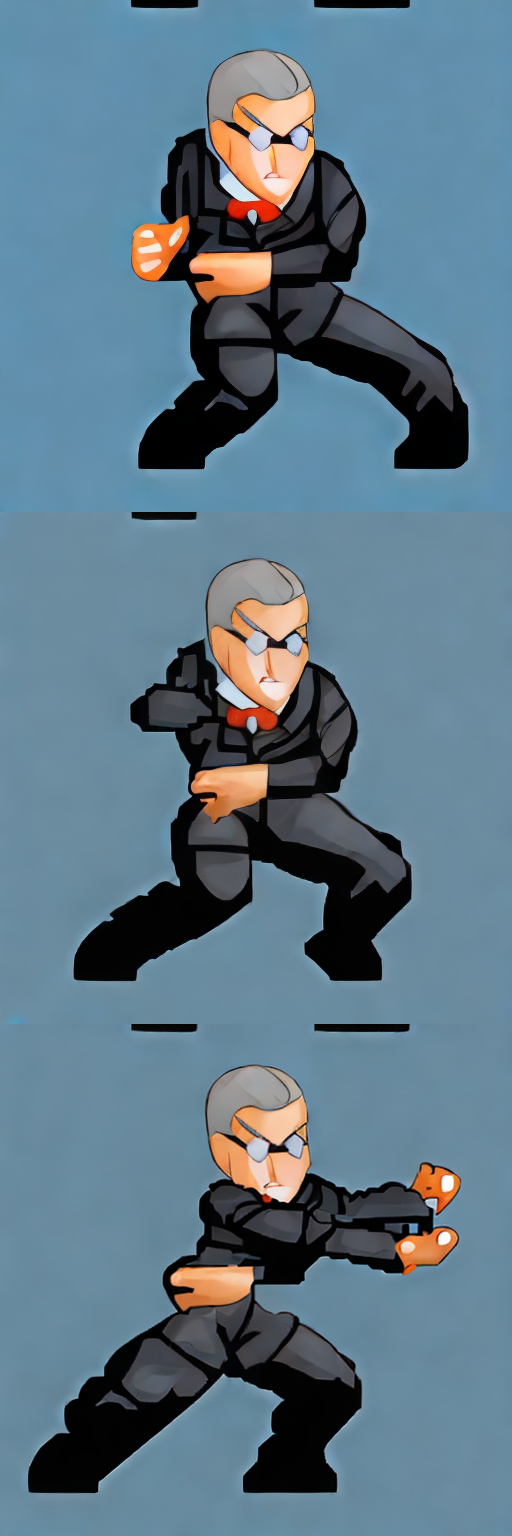}
        \textbf{Stage 1 Only}
    \end{minipage}
    \hfill
    \begin{minipage}[t]{0.19\textwidth} % 添加新列
        \centering
        \includegraphics[width=\textwidth]{results/QA_animate_finetune_4.png}
        \textbf{Fully Fine-Tuned} 
    \end{minipage}
    }
    \caption{Ablation Study Qualitative Comparison (Out-Sample - Enemies  Miscellaeous - Attack)}
    \label{fig-abl: QA4}
\end{figure}

\newpage
\section{Code Overview}
\subsection{Labeling Pipeline}
\label{sec:pipeline}
\begin{figure}[!ht]
    \centering
    \includegraphics[width=0.8\linewidth]{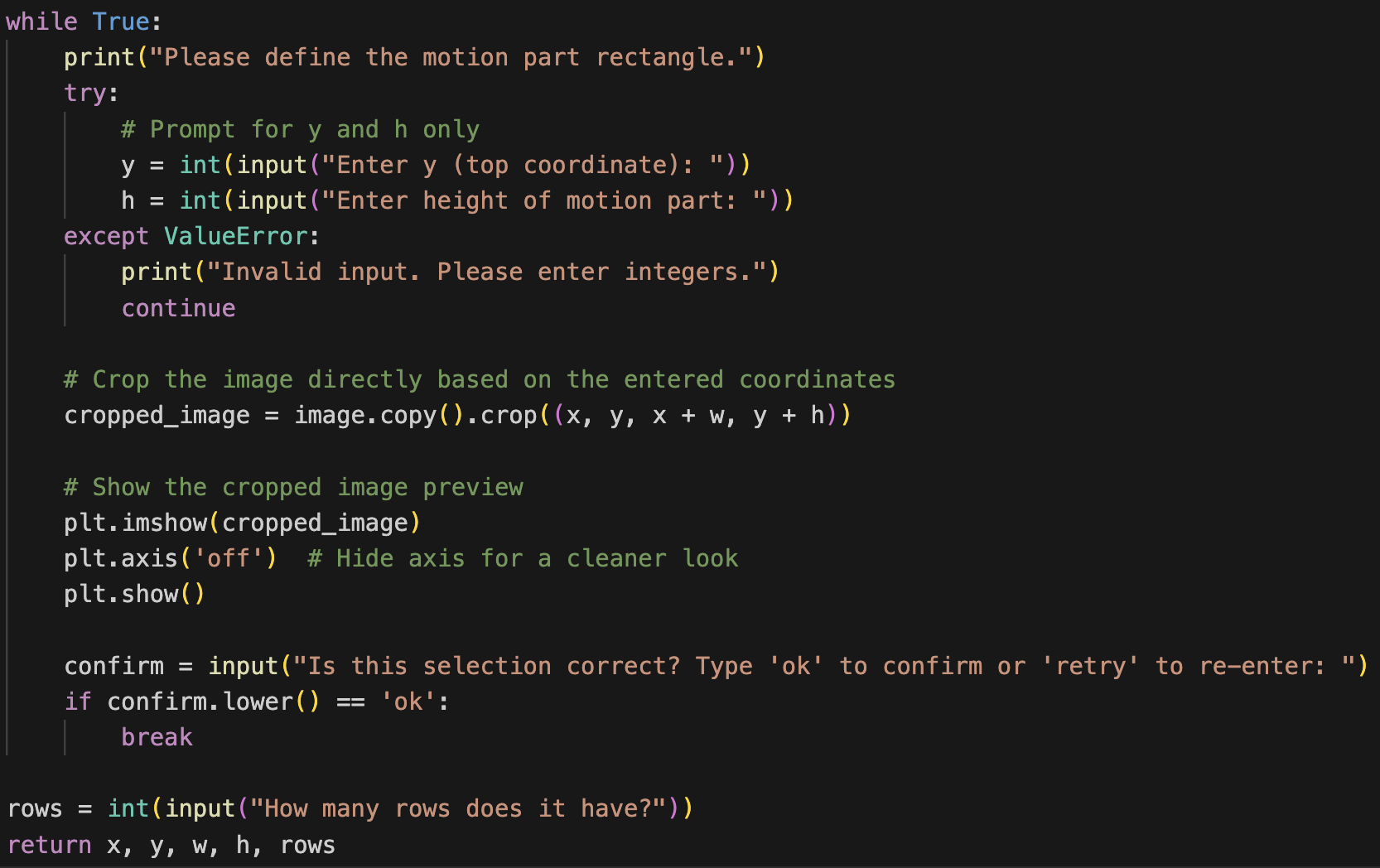}
    \caption{Split sprite sheet into different rows}
    % \label{fig:enter-label}
\end{figure}

\begin{figure}[!ht]
    \centering
    \includegraphics[width=0.8\linewidth]{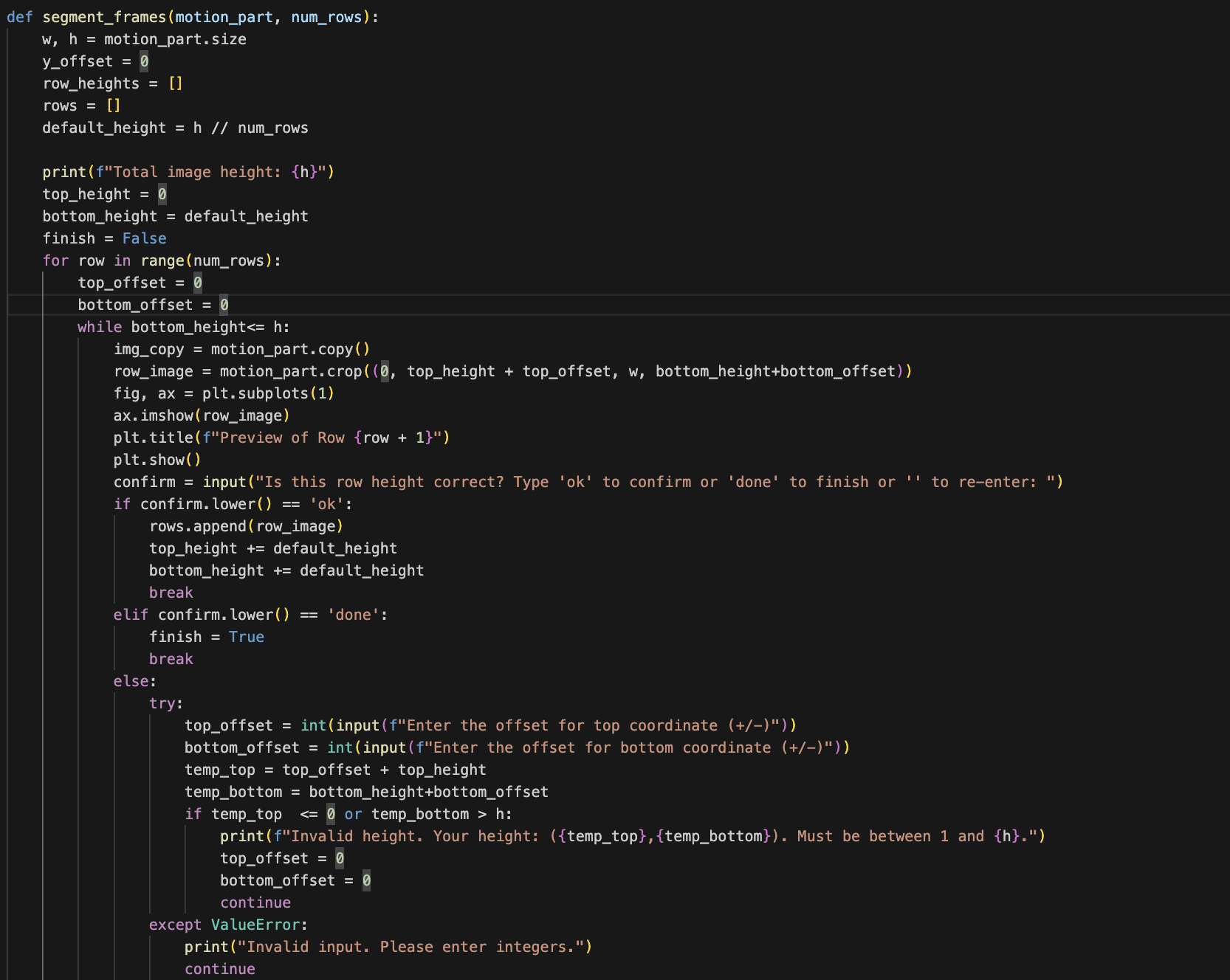}
    \caption{Adjust height and segment each frame}
    % \label{fig:enter-label}
\end{figure}

\begin{figure}[!ht]
    \centering
    \includegraphics[width=0.8\linewidth]{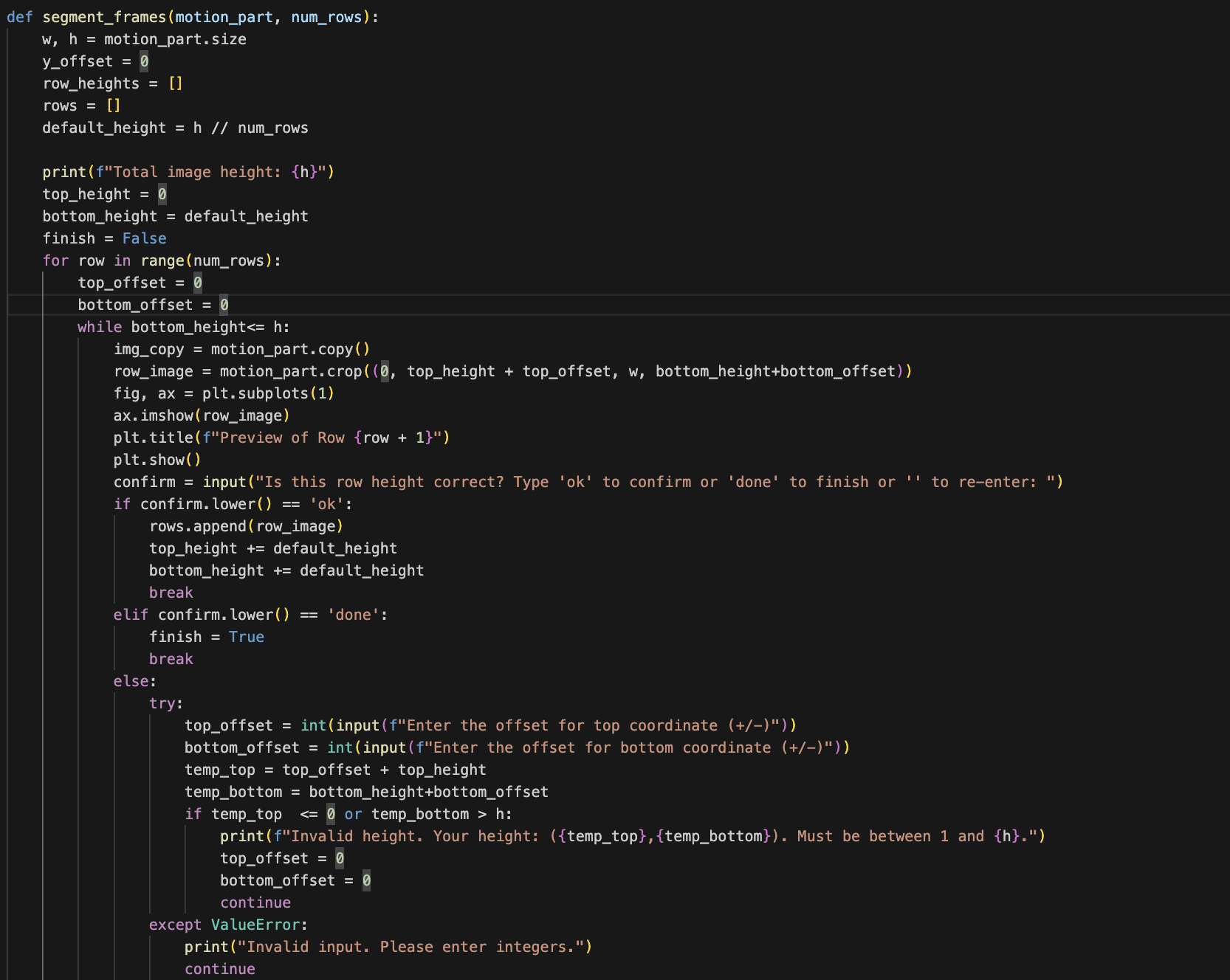}
    \caption{Crop images and label the human pose for each frame}
    % \label{fig:enter-label}
\end{figure} 

\begin{figure}[!ht]
    \centering
    \includegraphics[width=0.8\linewidth]{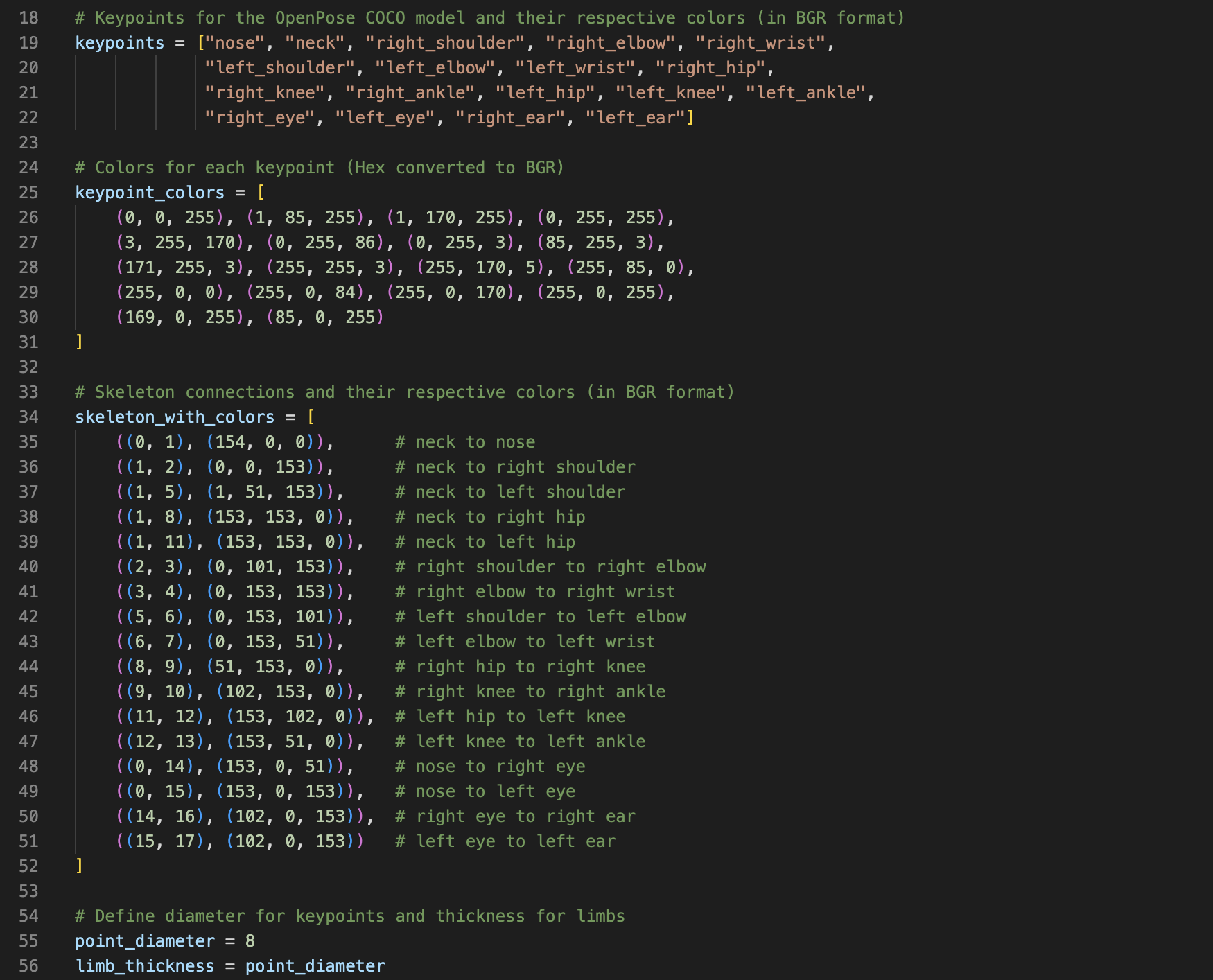}
    \caption{Follow openpose output's format}
    % \label{fig:enter-label}
\end{figure} 

\begin{figure}[!ht]
    \centering
    \includegraphics[width=0.8\linewidth]{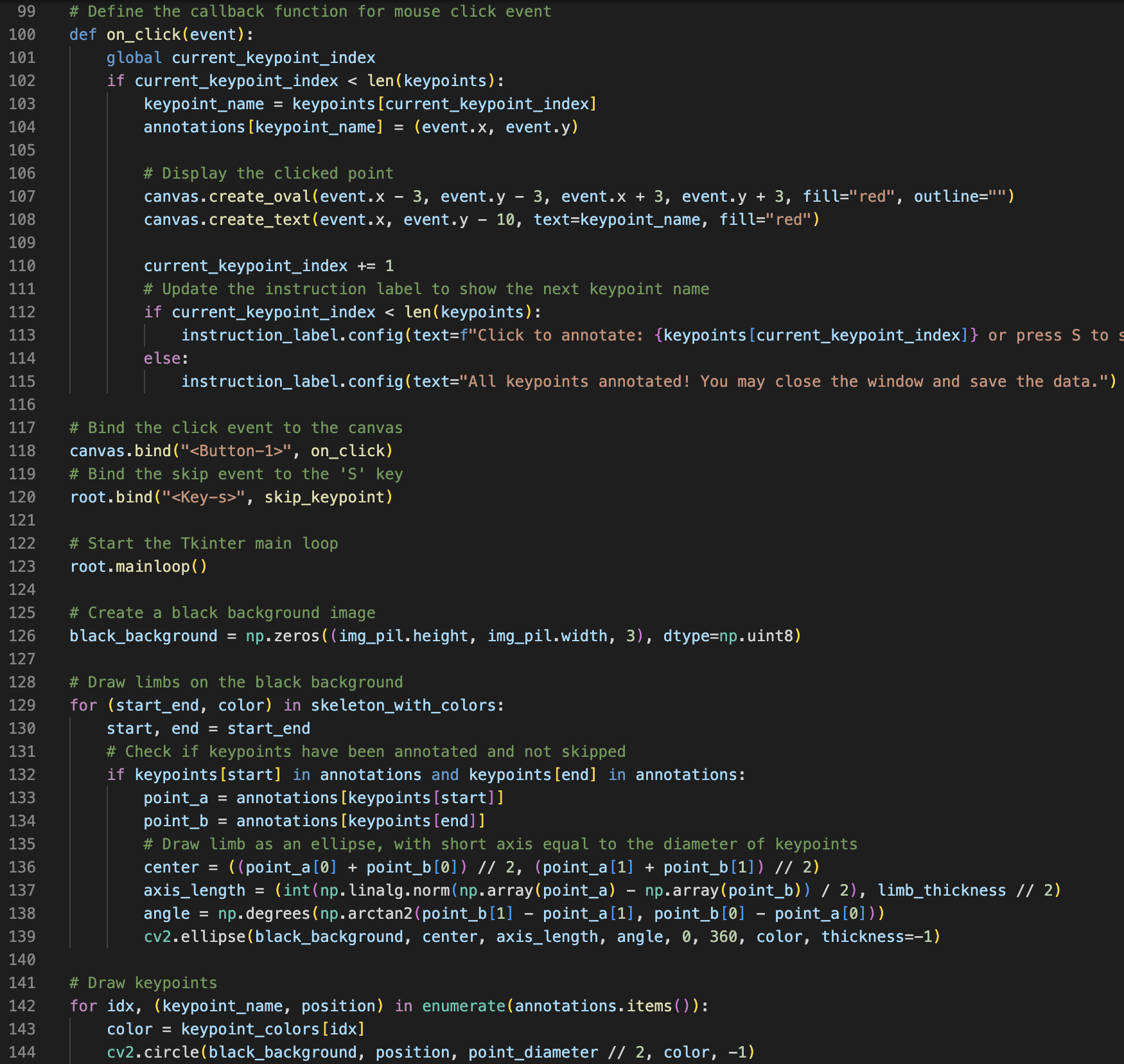}
    \caption{Use tinker library to hand-draw keypoints and limbs of the pose}
    % \label{fig:enter-label}
\end{figure} 
\clearpage
\newpage
\subsection{IP-Adpater}
\label{sec:ip-adapter}
\begin{figure}[!ht]
    \centering
    \includegraphics[width=0.8\linewidth]{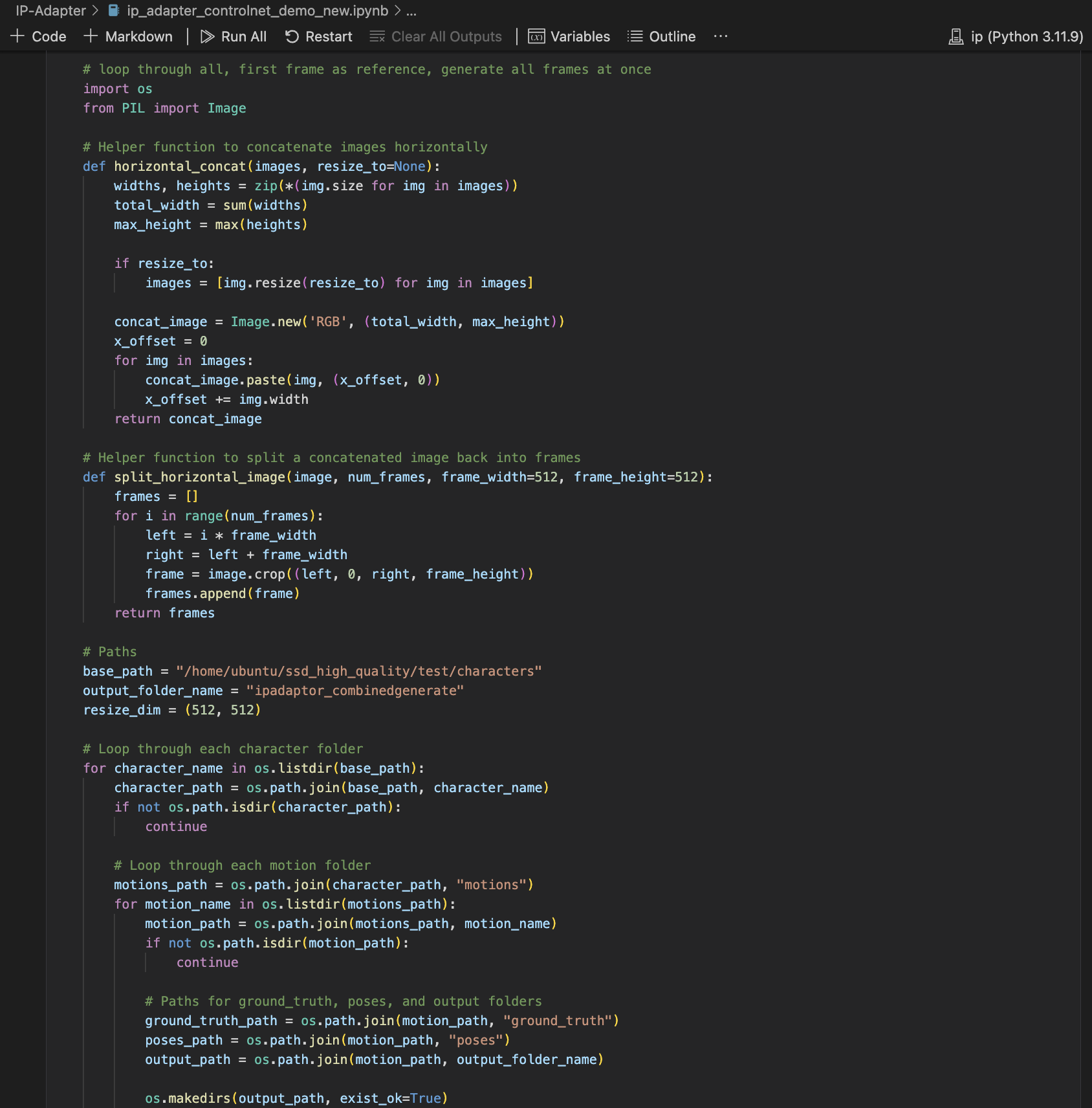}
\end{figure}

\begin{figure}[!ht]
    \centering
    \includegraphics[width=0.8\linewidth]{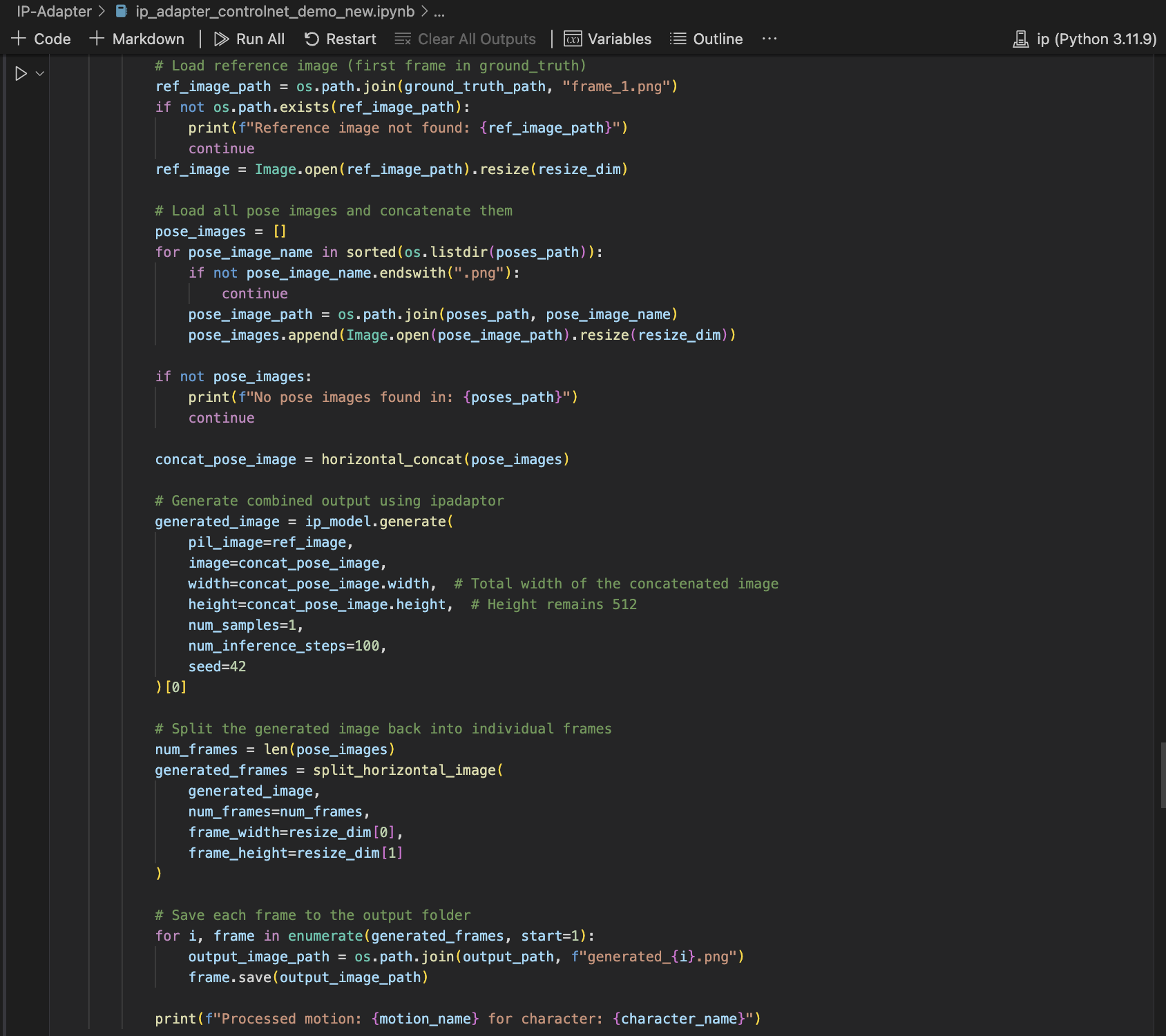}
    \caption{Use Ip-Adapter repo SD-IPCN notebook for testing with out customed dataset}
\end{figure}

\begin{figure}[!ht]
    \centering
    \includegraphics[width=0.8\linewidth]{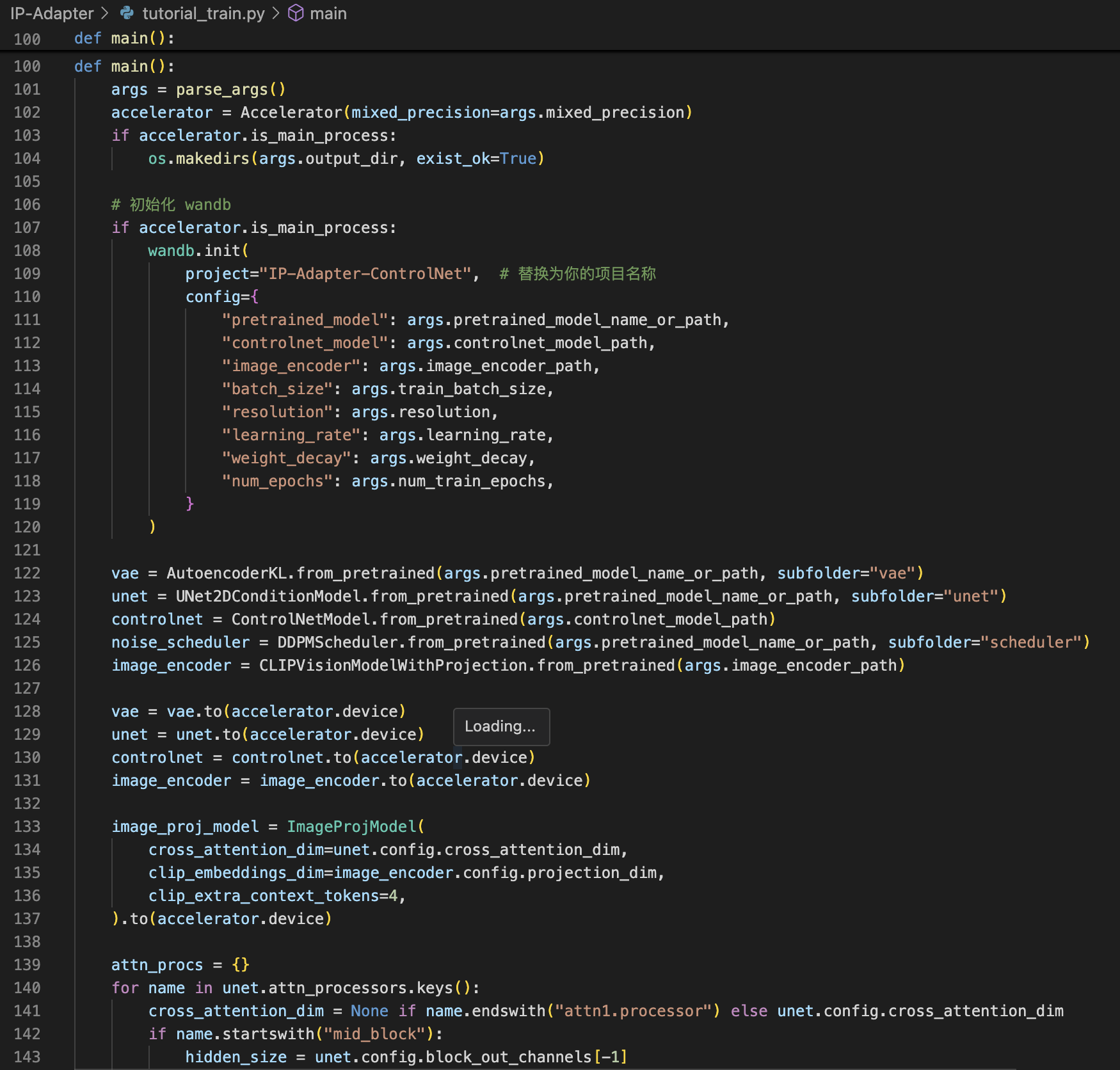}
\end{figure}

\begin{figure}[!ht]
    \centering
    \includegraphics[width=0.8\linewidth]{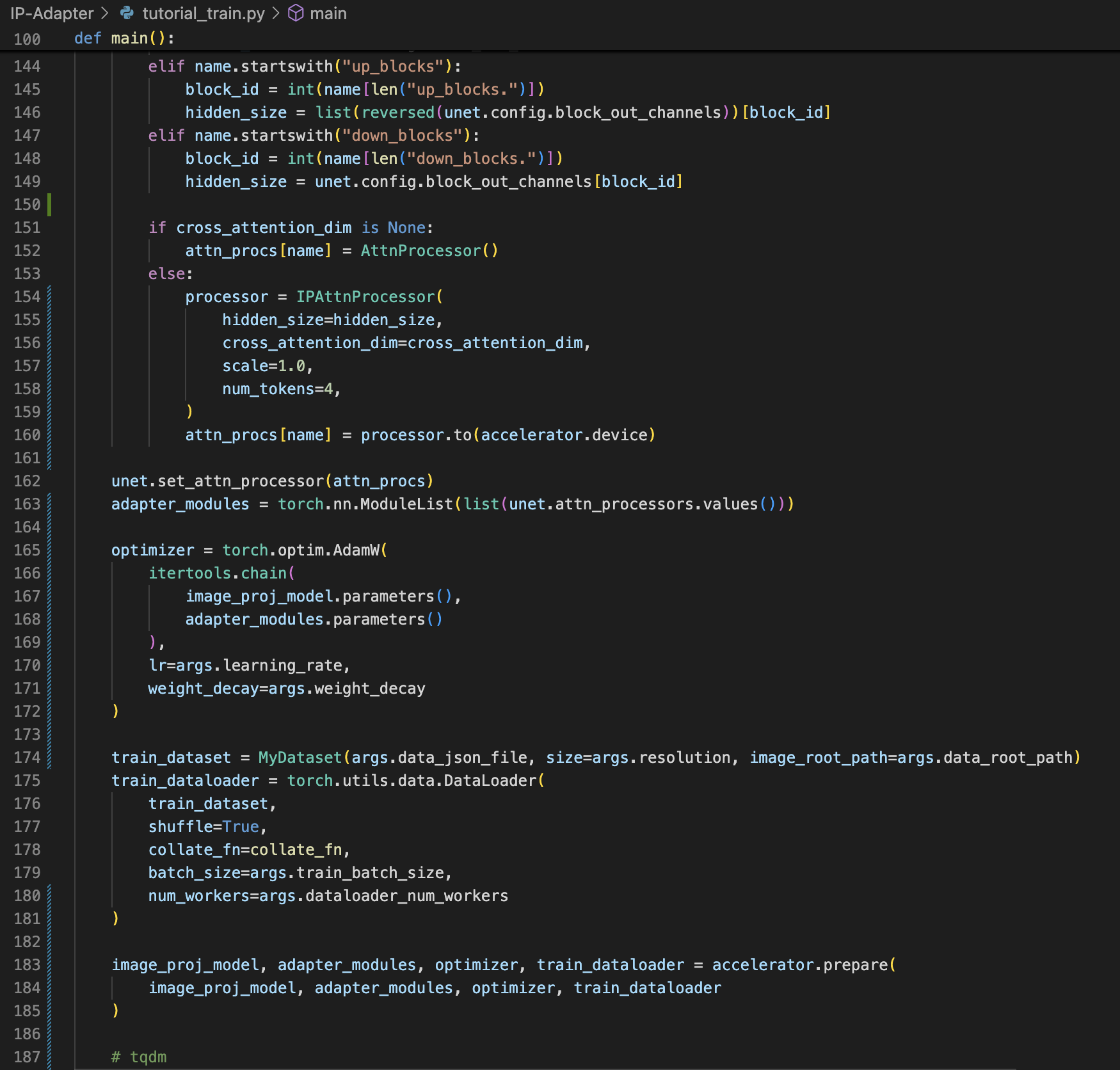}
\end{figure}

\begin{figure}[!ht]
    \centering
    \includegraphics[width=0.8\linewidth]{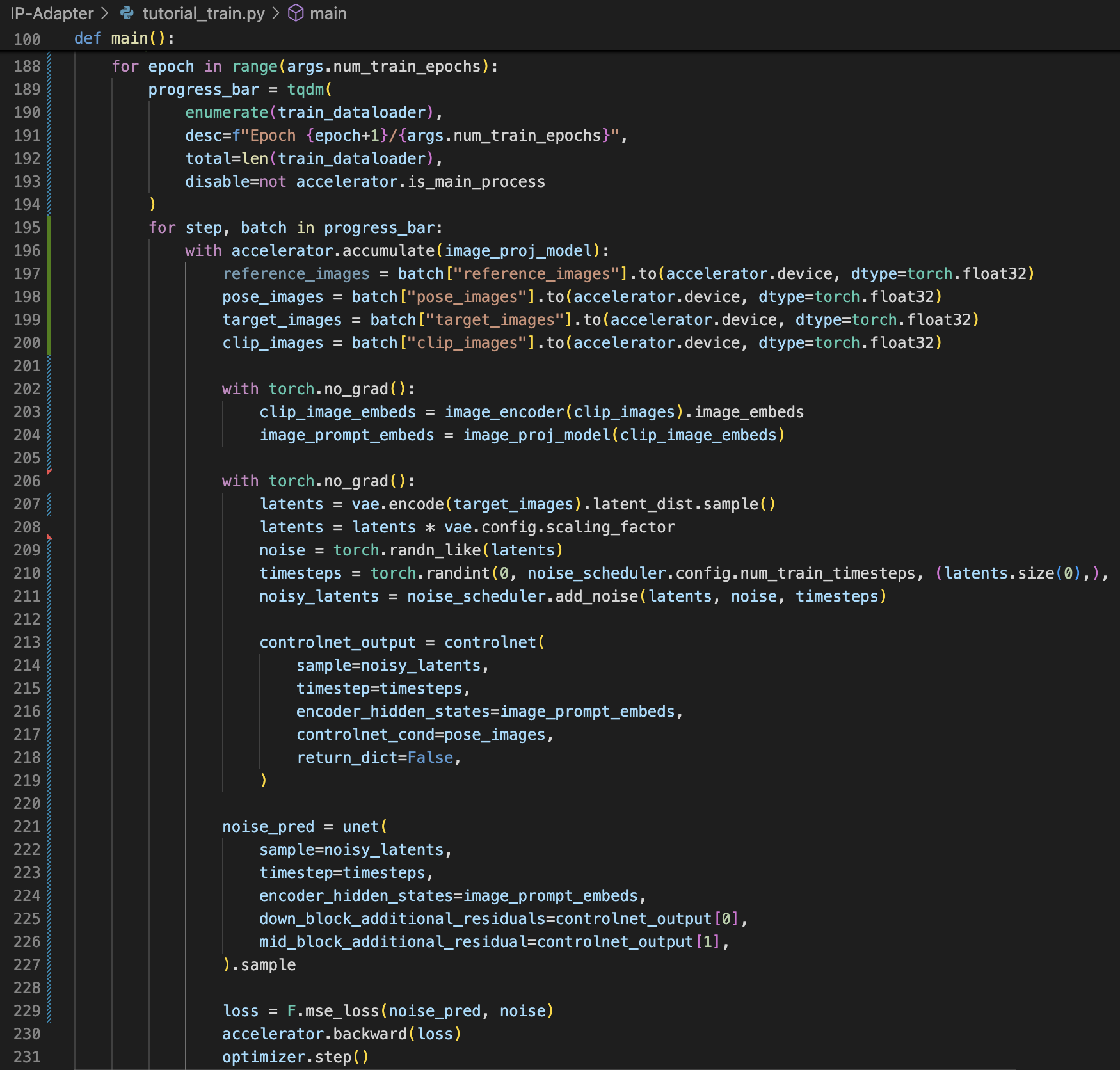}
\end{figure}

\begin{figure}[!ht]
    \centering
    \includegraphics[width=0.8\linewidth]{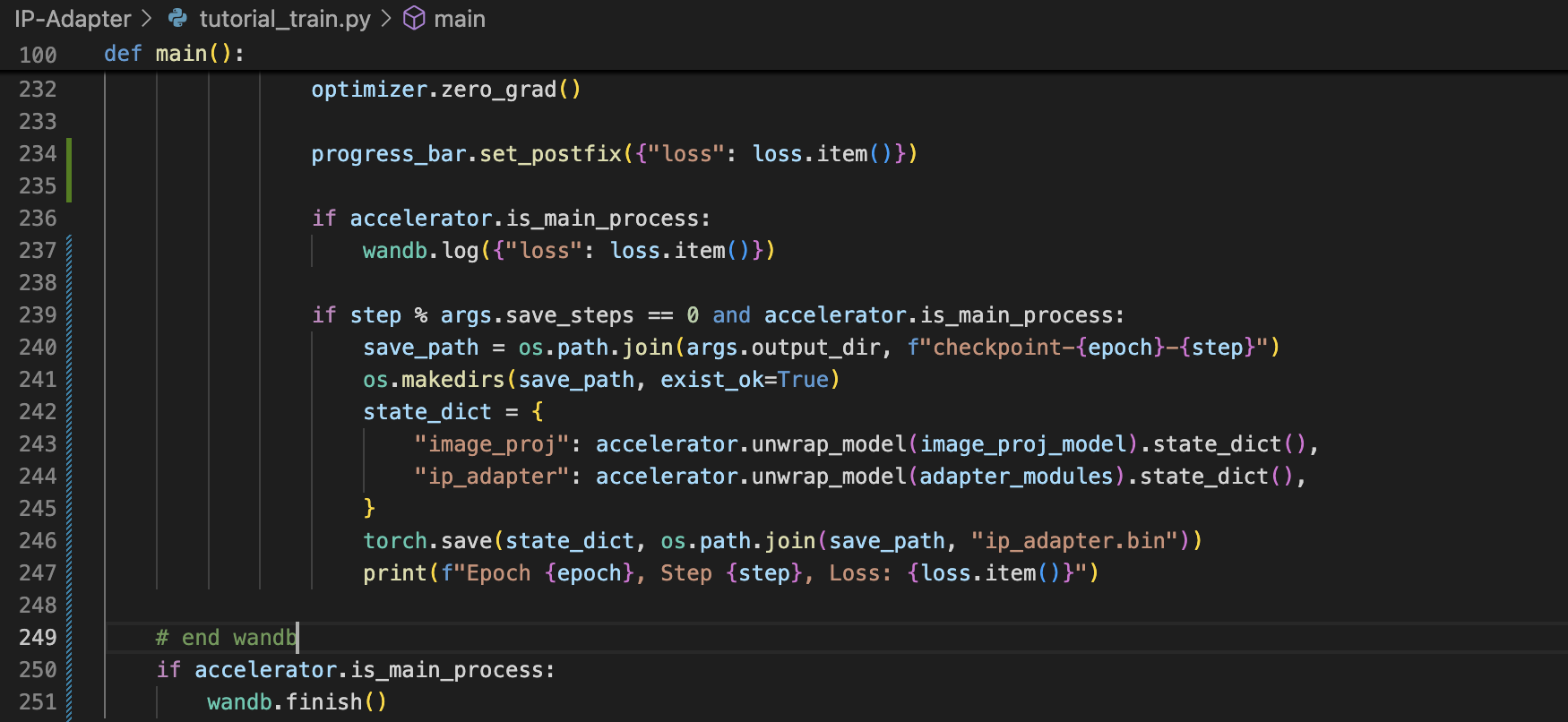}
    \caption{Adapt Ip-Adapter repo \texttt{train\_tutorial.py} to finetune with Controlnet}
\end{figure}

\clearpage
\newpage
\subsection{Animate Anyone}
\label{sec:animateanyone}
\begin{figure}[!ht]
    \centering
    \includegraphics[width=0.8\linewidth]{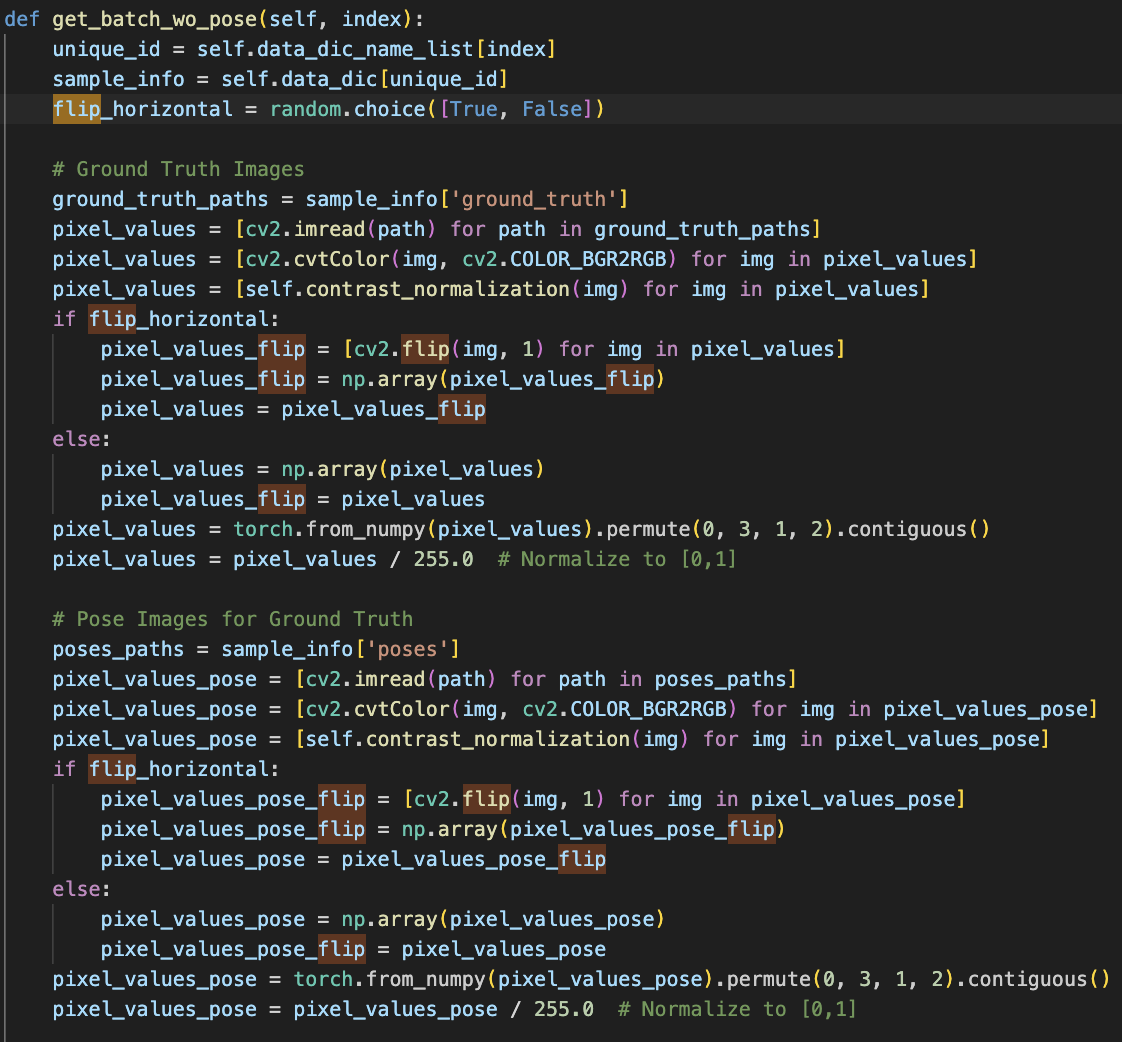}
    \caption{Randomly flip images}
    \label{fig:random-flip}
\end{figure}

\begin{figure}[!ht]
    \centering
    \includegraphics[width=0.8\linewidth]{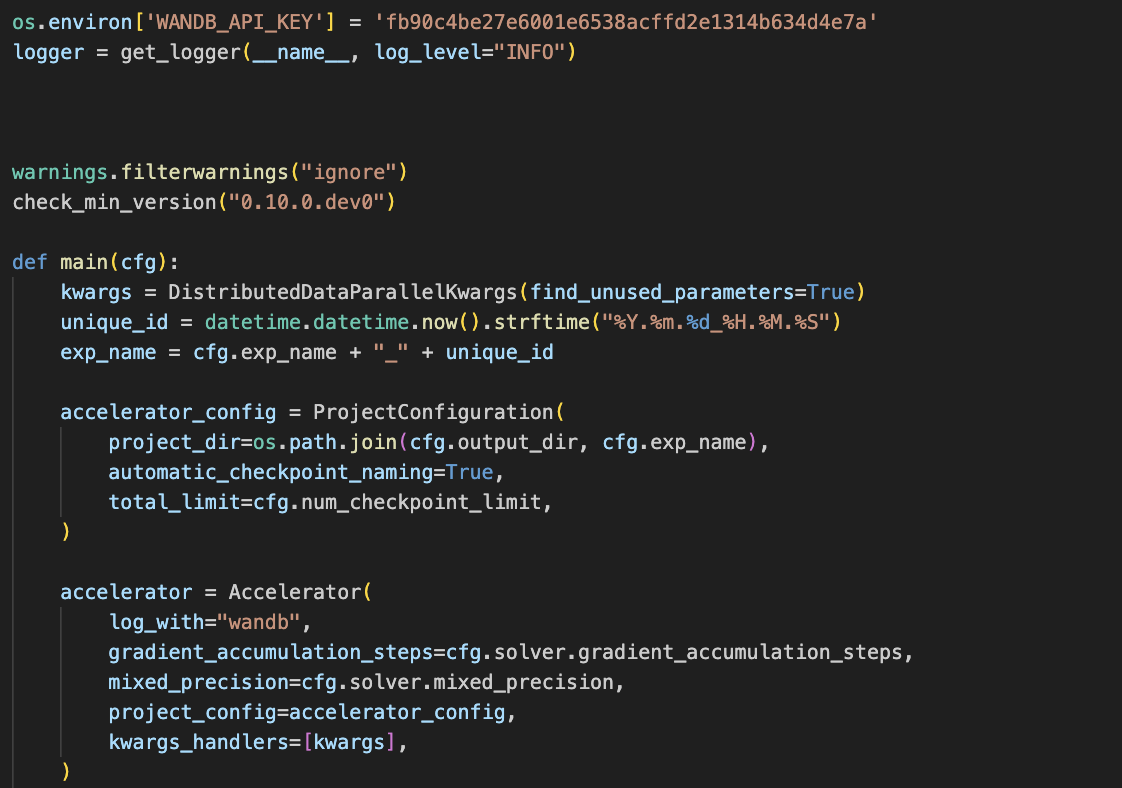}
    \caption{Wandb as the logging method}
    \label{fig:wandb}
\end{figure}

\begin{figure}[!ht]
    \centering
    \includegraphics[width=0.8\linewidth]{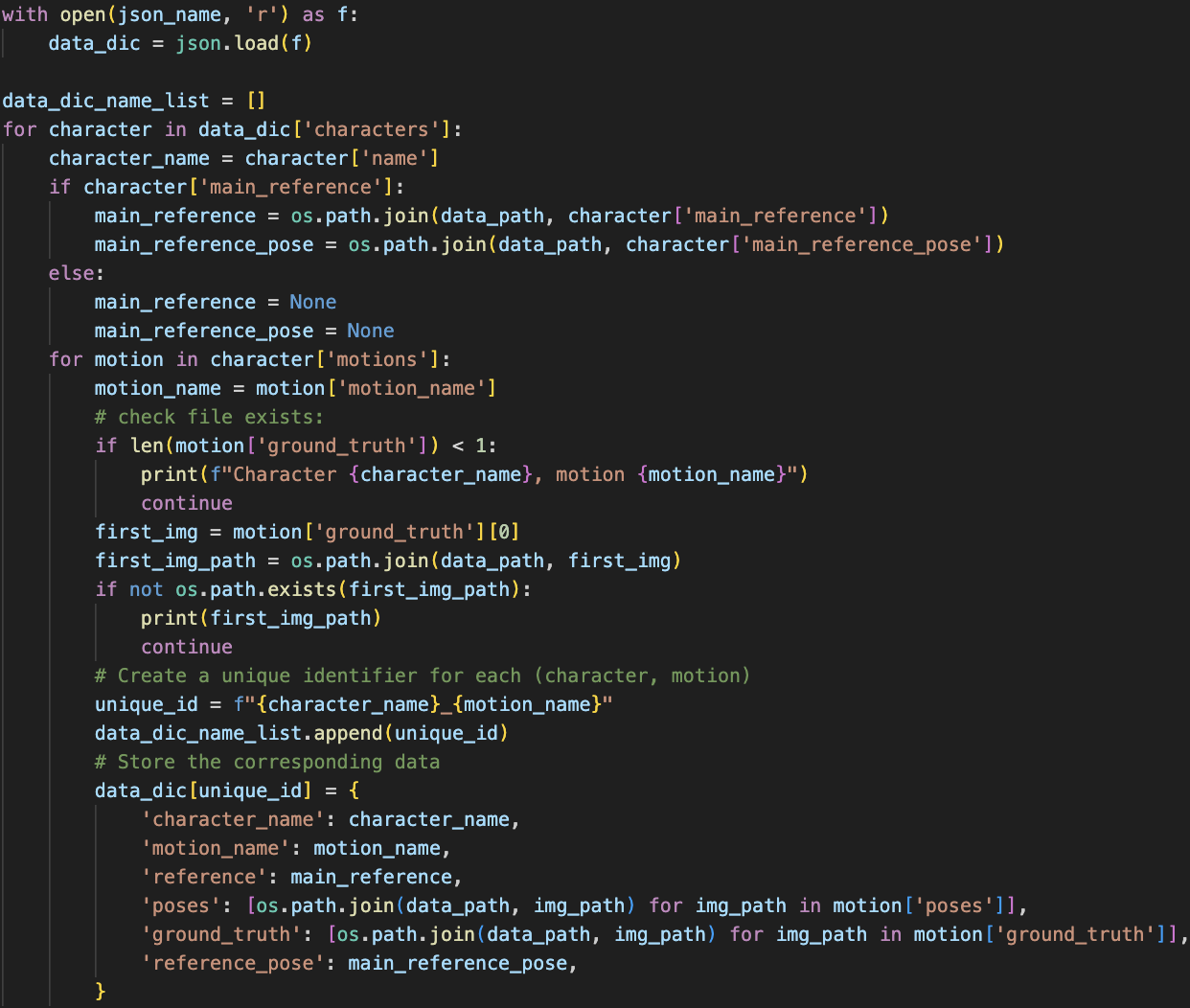}
    \caption{Load our dataset}
    \label{fig:dataset}
\end{figure}

\begin{figure}[!ht]
    \centering
    \includegraphics[width=0.8\linewidth]{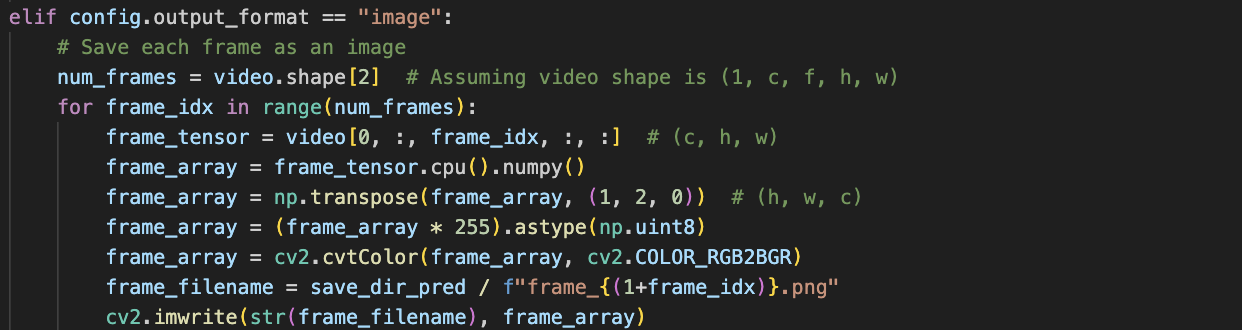}
    \caption{Output the result}
    \label{fig:output}
\end{figure}

\clearpage
\newpage
\subsection{Evaluation and Experimental Design}
\subsubsection{\texttt{experiment/eval\_img\_quality.py}}
\label{code:eval_quality}
\begin{figure}[!ht]
    \centering
    \includegraphics[width=0.8\linewidth]{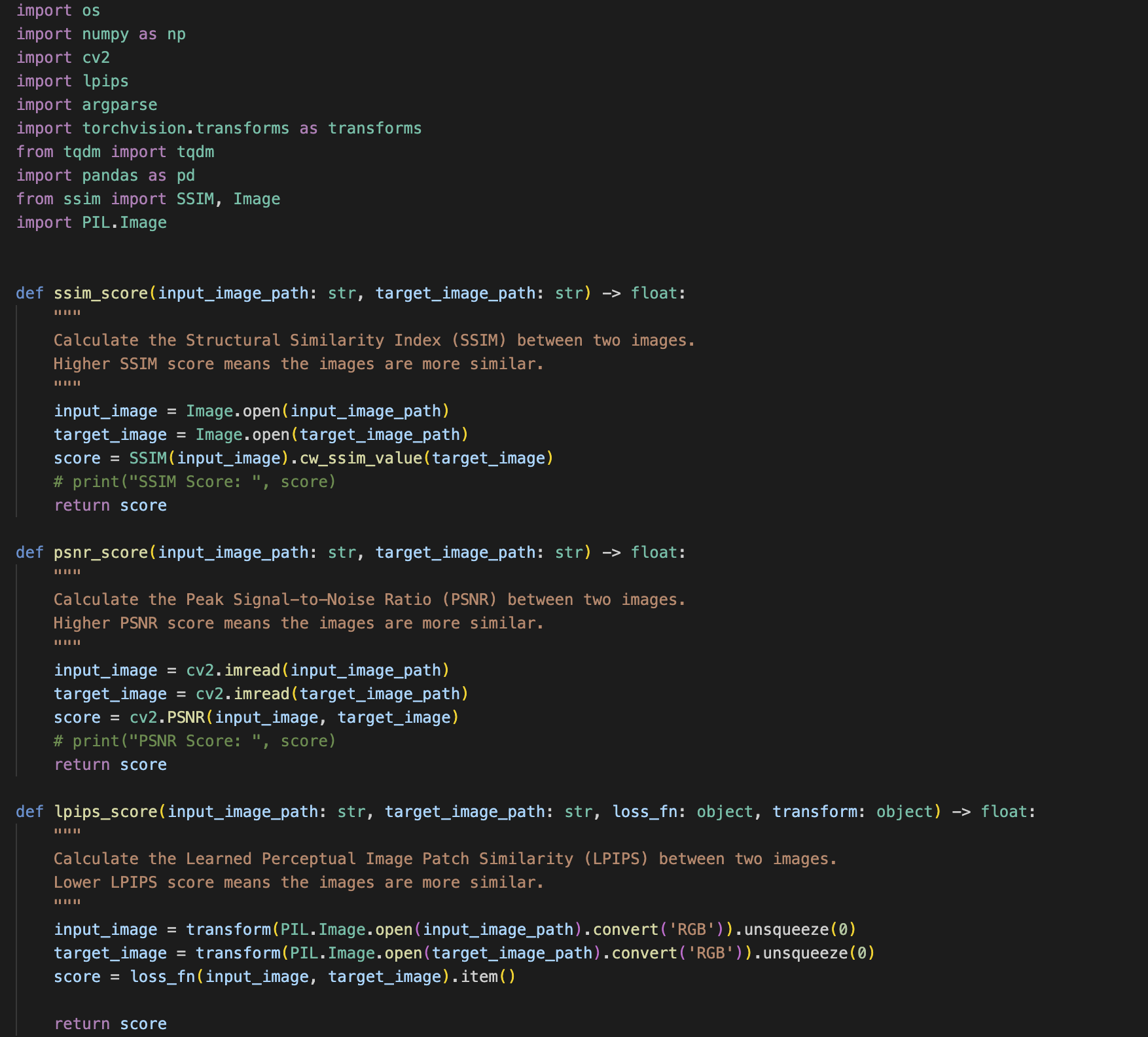}
    \caption{\texttt{eval\_img\_quality.py} - Part 1}
\end{figure}

\begin{figure}[!ht]
    \centering
    \includegraphics[width=0.8\linewidth]{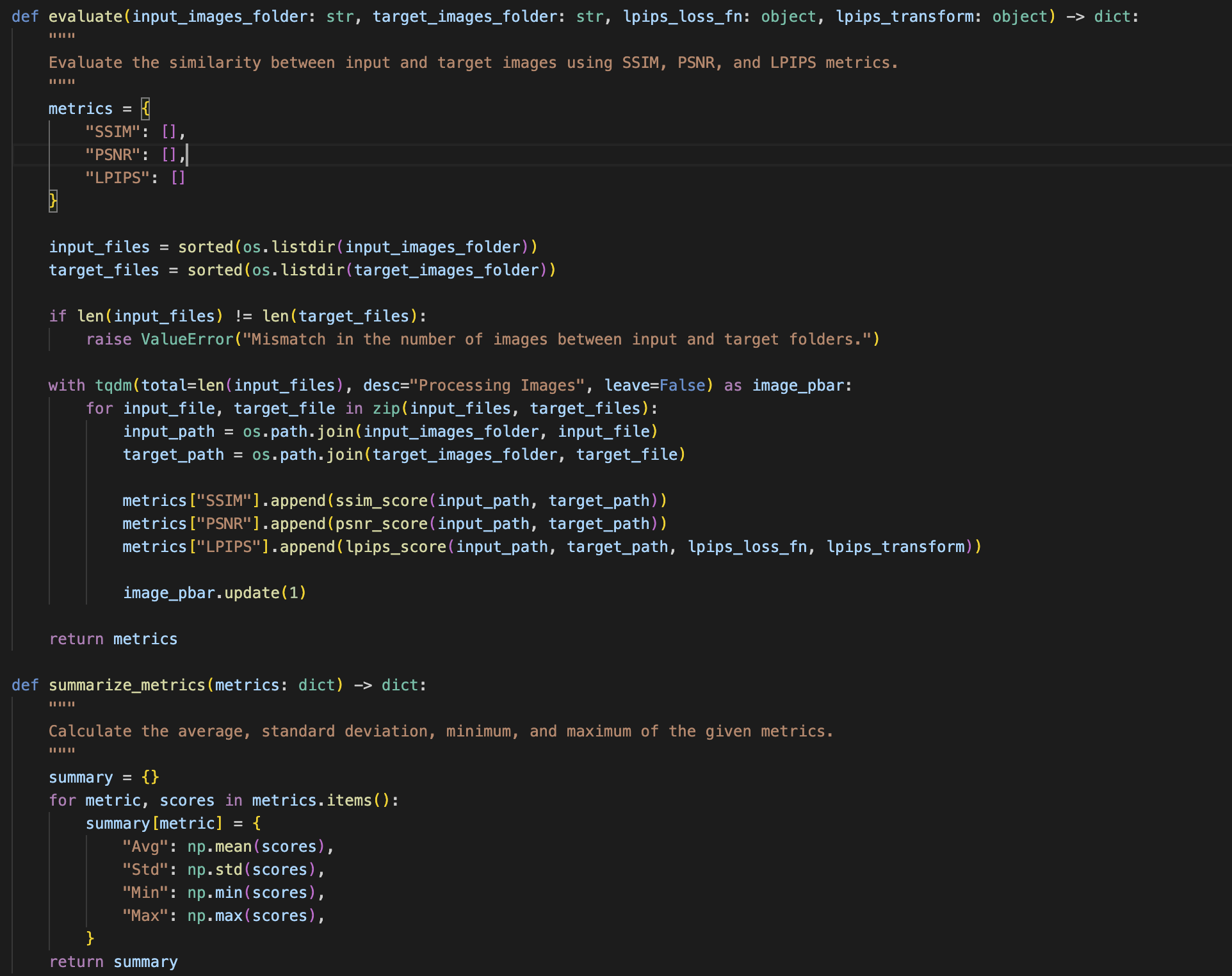}
    \caption{\texttt{eval\_img\_quality.py} - Part 2}
\end{figure}

\begin{figure}[!ht]
    \centering
    \includegraphics[width=0.8\linewidth]{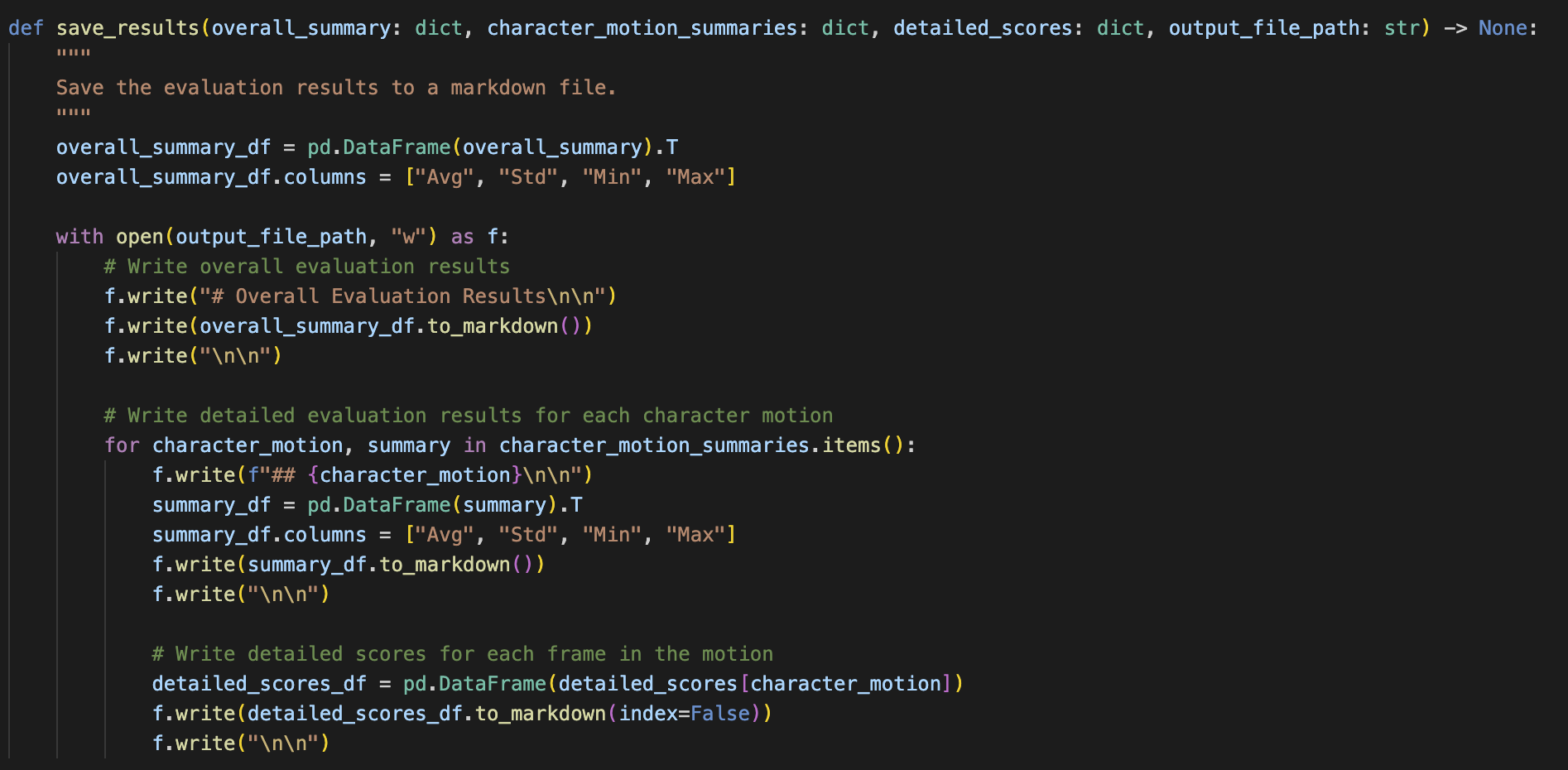}
    \caption{\texttt{eval\_img\_quality.py} - Part 3}
\end{figure}

\begin{figure}[!ht]
    \centering
    \includegraphics[width=0.8\linewidth]{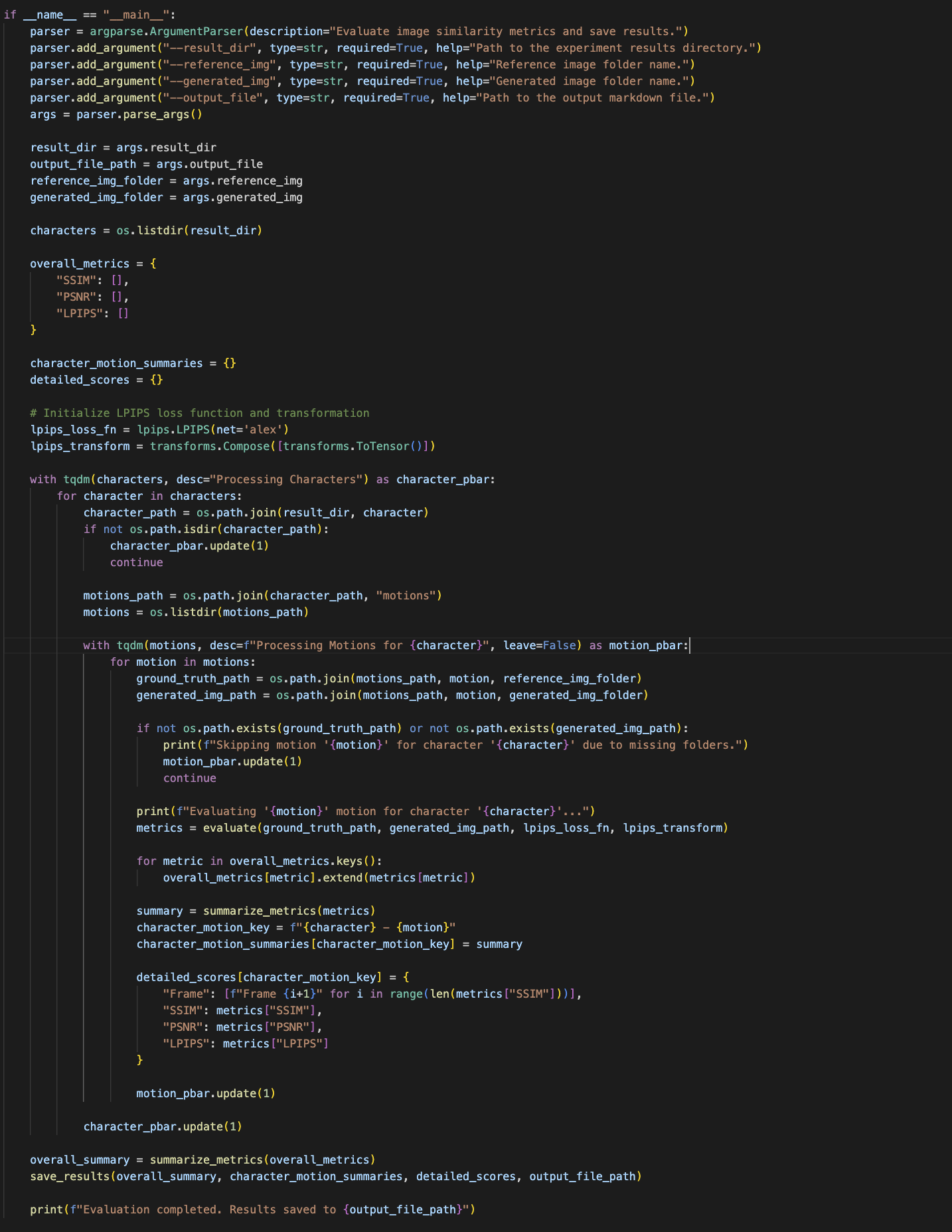}
    \caption{\texttt{eval\_img\_quality.py} - Part 4}
\end{figure}

\clearpage
\newpage
\subsubsection{\texttt{experiment/eval\_sub\_consistency.py}}
\label{code:eval_consistency}
\begin{figure}[!ht]
    \centering
    \includegraphics[width=0.8\linewidth]{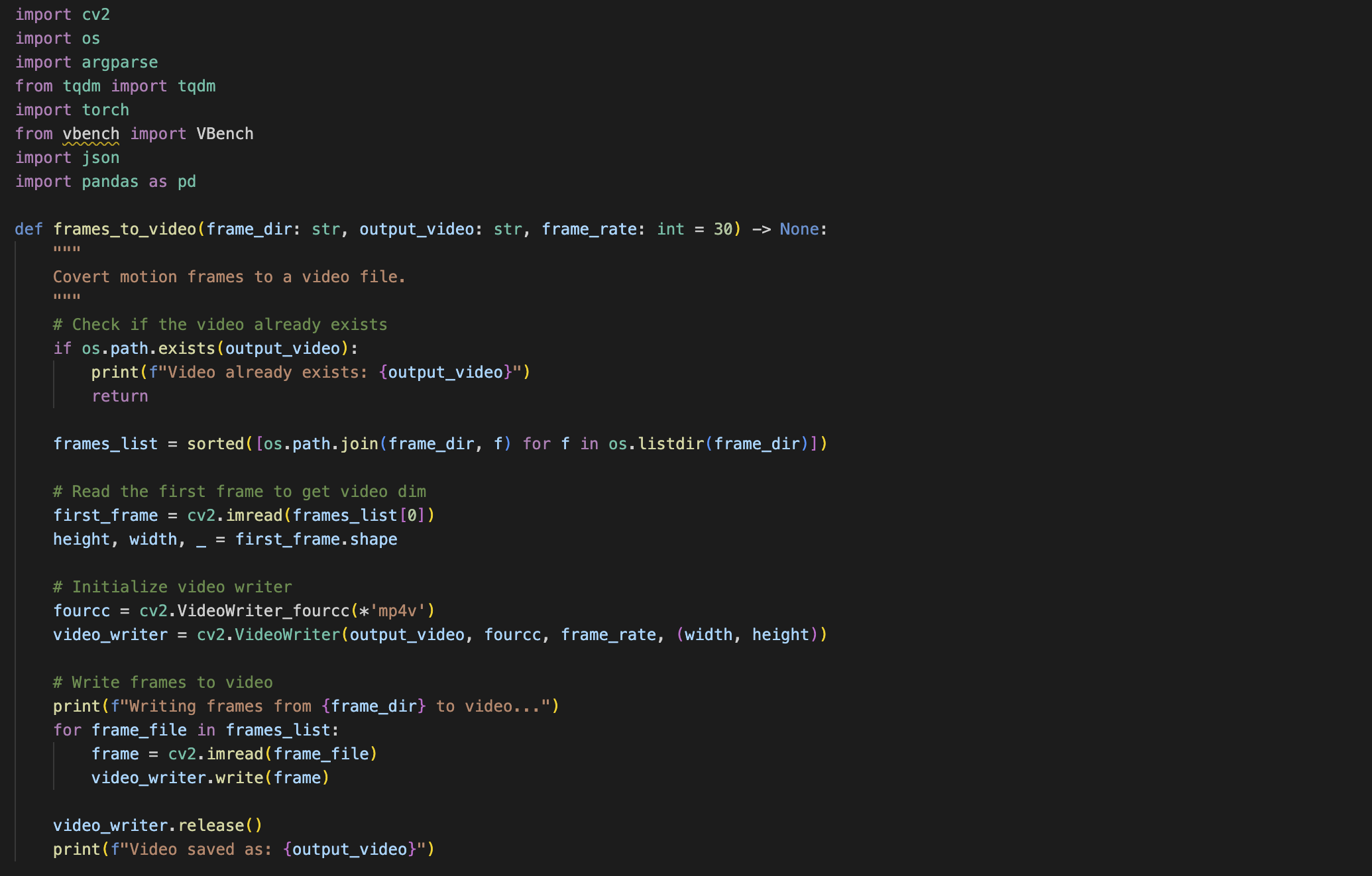}
    \caption{\texttt{eval\_sub\_consistency.py} - Part 1}
\end{figure}

\begin{figure}[!ht]
    \centering
    \includegraphics[width=0.8\linewidth]{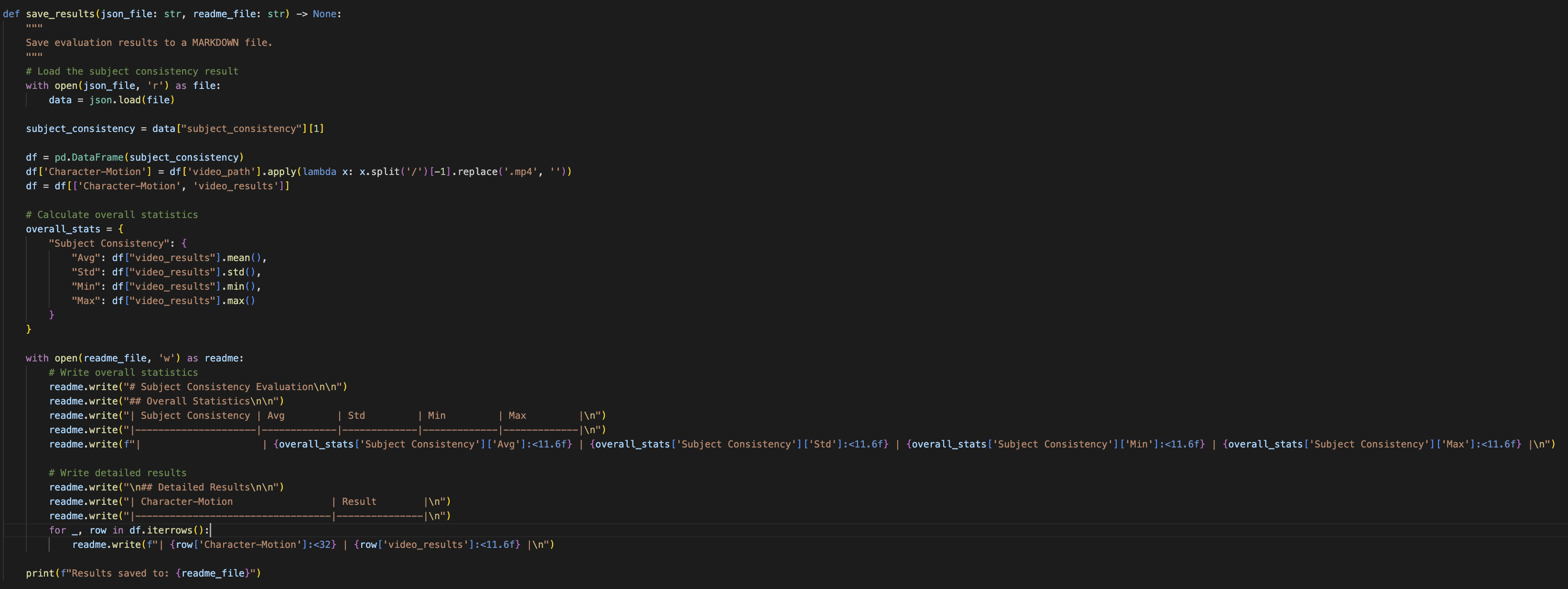}
    \caption{\texttt{eval\_sub\_consistency.py} - Part 2}
\end{figure}

\begin{figure}[!ht]
    \centering
    \includegraphics[width=0.8\linewidth]{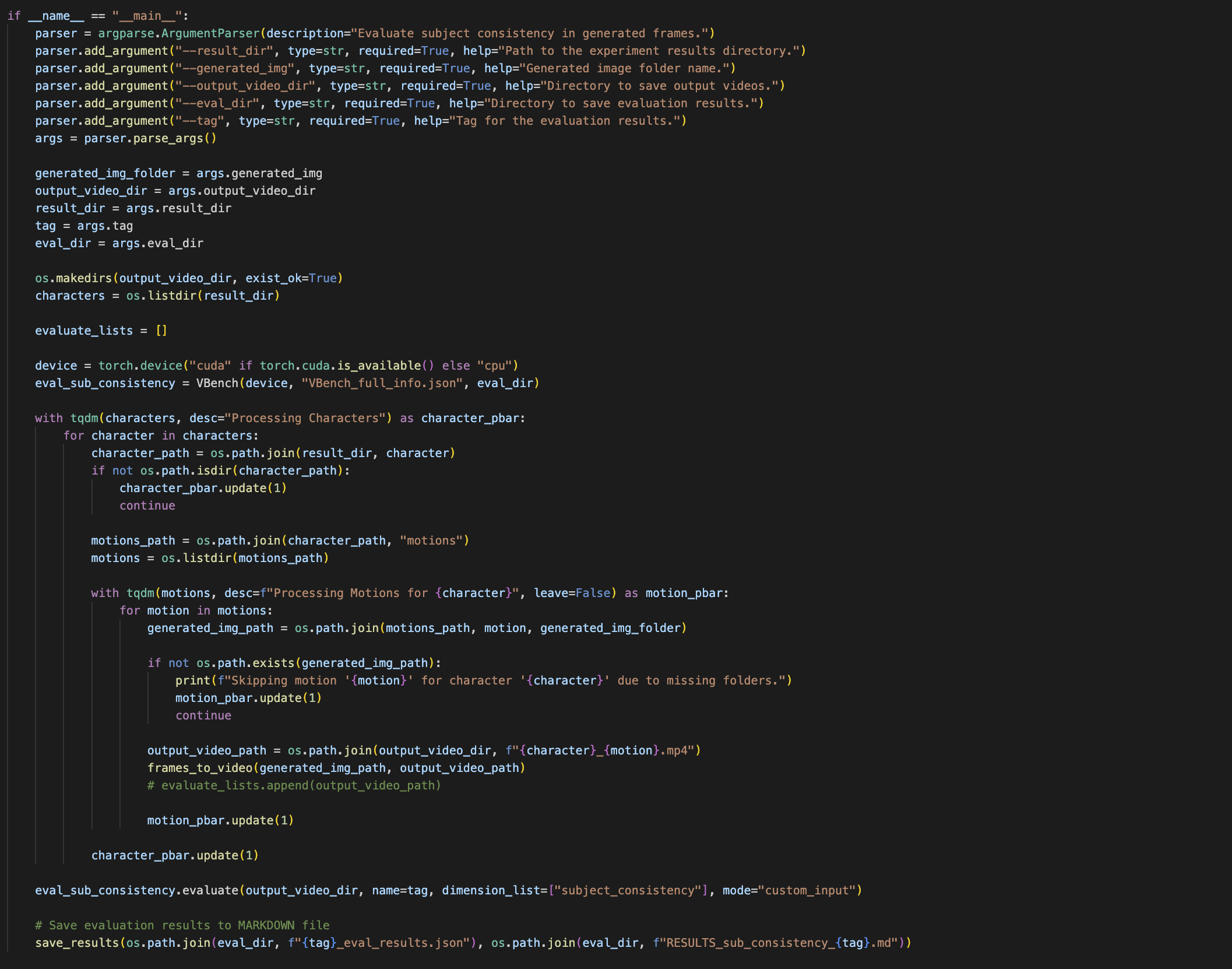}
    \caption{\texttt{eval\_sub\_consistency.py} - Part 3}
\end{figure}

\clearpage
\newpage
\section{Appendix: Detailed Timeline and Task Breakdown}

\begin{table}[h!]
\centering
\renewcommand{\arraystretch}{1.2}
\setlength{\tabcolsep}{8pt}
\begin{tabular}{|p{3cm}|p{10cm}|p{2cm}|}
\hline
\textbf{Task Category}                   & \textbf{Task Description}                 & \textbf{Time Spent (hrs)} \\ \hline
Reading Papers/Dataset Websites          & - Exploring literature on diffusion-based image and video generation.  \newline - Researching existing datasets (SpriteDatabase, GameArt2D). & 12 \\ \hline
Reading Code Documentation               & - Understanding IP-Adapter's code \newline - Reviewing AniPortrait and Animate Anyone repositories. & 12 \\ \hline
Dataset Creation                         & - Writing labeling pipeline for action sequences from sprite sheets.  \newline - Annotating poses using DW-Pose and OpenPose, and manually annotating challenging cases.  \newline - Annotating GameArt2D sprites and supplementing dataset with non-human sprites. & 50 \\ \hline
Understanding Existing Code              & - Studying IP-Adapter's testing notebook and training logic.  \newline - Analyzing Pose Guider modifications in AniPortrait and Moore-Animate Anyone. & 14 \\ \hline
Modifying Existing Code                  & - Adapting IP-Adapter's training script for pose-conditioned tasks.  \newline - Adding dataset loading, augmentation, and inference functions to Animate Anyone. \newline - Run training experiments & 30 \\ \hline
Writing Scripts for Experiments          & - Writing scripts for training and evaluation of IP-Adapter and Animate Anyone.  \newline - Implementing dataset augmentation and evaluation pipelines. & 8 \\ \hline
Running Experiments                      & - Testing frame generation strategies with IP-Adapter.   \newline - Conducting ablation studies. & 12 \\ \hline
Compiling Results                        & - Analyzing and visualizing evaluation metrics (SSIM, PSNR, LPIPS).  \newline - Calculating Subject Consistency Scores.  \newline - Summarizing results for ablation study. & 12 \\ \hline
Writing Report                           & - Documenting project phases, methodology, and findings.  \newline - Preparing detailed timeline and writing key sections. & 18 \\ \hline
Preparing Presentation and Paper         & - Designing and writing the project paper.  \newline - Preparing slides and rehearsing presentation. & 15 \\ \hline
\textbf{Total Hours}                     & \multicolumn{2}{c|}{183} \\ \hline
\end{tabular}
\caption{Detailed timeline and task breakdown.}
\label{tab:timeline}
\end{table}

\end{document}